\documentclass[fleqn,usenatbib,useAMS]{mnras}

\usepackage[T1]{fontenc}    
\usepackage{ae,aecompl}     

\usepackage{graphicx}	
\usepackage{amsmath}	
\usepackage{amssymb}	
\usepackage{multicol}   
\usepackage{bm}		    
\usepackage{pdflscape}	
\usepackage{mathtools}
\usepackage{silence}
\usepackage{xspace}
\DeclareMathOperator\arcsinh{arcsinh}

\newcommand*{\about}{\mathord\sim}
\newcommand*{\Sersic}{S\'{e}rsic\xspace}
\newcommand*{\SExtractor}{\textsc{Source Extractor}\xspace}
\newcommand*{\Gnuastro}{\textsc{Gnuastro}\xspace}
\newcommand*{\NoiseChisel}{\textsc{NoiseChisel}\xspace}
\newcommand*{\Segment}{\textsc{Segment}\xspace}
\newcommand*{\MakeCatalog}{\textsc{MakeCatalog}\xspace}
\newcommand*{\PSFEx}{\textsc{PSFEx}\xspace}
\newcommand*{\GalSim}{\textsc{GalSim}\xspace}
\newcommand*{\LSSTSPs}{\textsc{LSST Science Pipelines}\xspace}
\newcommand*{\LSSTPs}{\textsc{LSST Pipelines}\xspace}
\newcommand*{\DMA}{\textsc{P6}\xspace}
\newcommand*{\DMB}{\textsc{S128}\xspace}
\newcommand*{\VRO}{Vera C. Rubin Observatory\xspace}
\newcommand*{\RO}{Rubin Observatory\xspace}
\newcommand*{\LSST}{Legacy Survey of Space and Time\xspace}

\title[LSB Sky Subtraction]{Sky Subtraction in an Era of Low Surface Brightness Astronomy}
\author[L.~S.~Kelvin et al.]{
\parbox{\textwidth}{
\raggedright
Lee~S.~Kelvin,$^{1,2,3}$
Imran~Hasan$^{2}$
and J.~Anthony~Tyson$^{2}$
}\vspace{0.5cm}\\
\parbox{\textwidth}{
$^{1}$Department of Astrophysical Sciences, Princeton University, 4 Ivy Lane, Princeton, NJ 08544, USA\\
$^{2}$Department of Physics, University of California, One Shields Ave., Davis, CA 95616, USA\\
$^{3}$Astrophysics Research Institute, Liverpool John Moores University, IC2, LSP, 146 Brownlow Hill, Liverpool, L3 5RF, UK\\
}
\vspace{-0.75cm}
}
\date{Accepted XXX. Received YYY; in original form ZZZ}
\pubyear{2021}

\begin{document}
\label{firstpage}
\pagerange{\pageref{firstpage}--\pageref{lastpage}}
\maketitle



\begin{abstract}
The \VRO Wide-Fast Deep (WFD) sky survey will reach unprecedented surface brightness depths over tens of thousands of square degrees. Surface brightness photometry has traditionally been a challenge. Current algorithms which combine object detection with sky estimation systematically over-subtract the sky, biasing surface brightness measurements at the faint end and destroying or severely compromising low surface brightness light. While it has recently been shown that properly accounting for undetected faint galaxies and the wings of brighter objects can in principle recover a more accurate sky estimate, this has not yet been demonstrated in practice. Obtaining a consistent spatially smooth underlying sky estimate is particularly challenging in the presence of representative distributions of bright and faint objects. In this paper we use simulations of crowded and uncrowded fields designed to mimic Hyper Suprime-Cam data to perform a series of tests on the accuracy of the recovered sky. Dependence on field density, galaxy type and limiting flux for detection are all considered. Several photometry packages are utilised: \SExtractor, \Gnuastro, and the \LSSTSPs. Each is configured in various modes, and their performance at extreme low surface brightness analysed. We find that the combination of the \SExtractor software package with novel source model masking techniques consistently produce extremely faint output sky estimates, by up to an order of magnitude, as well as returning high fidelity output science catalogues.
\end{abstract}




\section{Introduction}
\label{sec:introduction}

The human eye is unique in its capacity to look up at the night sky and easily separate out a wide variety of astronomical objects from the all encompassing celestial sphere upon which they lay. A computer, however, finds such a task decidedly non-trivial, often requiring extensive training, testing, and the use of a bespoke data processing pipeline. Tasks such as object detection, segmentation, point source estimation, and, crucially, sky estimation are all important components of many professional image processing pipelines. Deriving algorithms to accurately and rapidly perform such tasks has been an active area of research over the last several decades, with the issue becoming more pressing in recent years as we rapidly enter an era of big data astronomy which combines wide surveyed areas with unprecedented depth.

Contemporary astronomical observatories are located at sites with optimal seeing, photometric stability, and dark skies. Extremely dark ground-based sites with little to no light pollution are able to achieve sky brightness levels down to $\mu_V\sim22\;\mathrm{mag\;arcsec}^{-2}$ \citep{Garstang1989}. Without the impediment of an atmosphere, space-based facilities such as the Hubble Space Telescope (HST) are able to probe $\sim1-2\;\mathrm{mag\;arcsec}^{-2}$ deeper still \citep[see, e.g.,][and references therein]{Knapen2017}. Taking full advantage of such sites in combination with a careful observing strategy and a minimally aggressive sky subtraction technique now allows us to probe beyond the $\mu=30\;\mathrm{mag\;arcsec}^{-2}$ threshold on an increasingly regular basis \citep{Iodice2016,Trujillo2016}. Such methodologies allow for formerly hidden galaxy structures to be seen and catalogued for the first time \citep[e.g.,][]{Kelvin2018}. The previously hidden low surface brightness (LSB) Universe continues to unveil a wealth of information across multiple scales, from LSB-type dwarfs \citep{vanDokkum2015,Williams2016} to circumgalactic stellar haloes \citep{D'Souza2014,Wang2019} and the mass-merging remnant intracluster light \citep{Montes2014,Montes2018,Montes2019}.

Optimal observing conditions notwithstanding, non-signal features of any astronomical image must still be characterised to enable further scientific analyses. Such features typically consist of three primary components: the instrumental background produced by the observing hardware, such as CCD edge effects \citep{Goldstein2015}, scattered light from optics within the telescope \citep{Karabal2017}, and saturation features \citep{Desai2016}; the temporal background from terrestrial effects, such as zodiacal light, airglow and cosmic rays \citep{Fixsen2000}, and; the astrophysical background from sources which remain fixed in the sky, such as diffuse stellar light from bright sources and galactic cirrus. Additional contaminant flux such as satellite trails \citep{Cheselka1999,Vandame2001,Storkey2004} must also be accounted for as part of the background estimation and handling process.

Many contemporary astronomy image reduction pipelines do not accurately preserve LSB flux, destroying this information at the sky subtraction phase. Furthermore, information on the undetected extragalactic background light (EBL), primarily due to fainter galaxies, is often not taken into account. Such pipelines tend to operate top-down: first detecting objects above a running biased initial sky estimate, then iteratively estimating object and sky fluxes. This produces good photometry at higher surface brightnesses, but is subject to sky bias due to very faint undetected galaxies. A complementary approach is to work bottom-up: first estimating sky, building a sky model, and only then detecting and measuring individual objects in a final step. The risk here is mistakenly assigning noise or contaminant flux to such sources, thereby overestimating their total flux. The potential for a bottom-up source detection algorithm to link several astrophysical sources together via a `flux-bridge' is significant, for example, in densely packed regions of the sky, for observations whereby the source encompasses a large fraction of the detector field of view, or in regions whereby some other such flux contaminant is evident such as galactic cirrus. The key towards facilitating science at ultra low surface brightnesses is in producing accurate and robust sky estimations across a wide variety of potentially contaminated source flux.

The process begins with automated photometry and the detection of individual objects. This is a multi-step process which becomes more challenging at faint surface brightnesses. One of the first such software packages to address this task across a wide dynamic range and for faint galaxies was the Faint Object Classification and Analysis System (\textsc{FOCAS})~\cite{Jarvis1981}. Using non-parametric statistical pattern recognition, it was initially driven by the need for surface photometry on faint co-added photographic plates, and later was extensively used on fainter CCD data. The output of a Bayesian detection algorithm with an optimal filter and adjustable detection threshold is fed to an area assembly stage, followed by spatial segmentation, iterative sky level estimation, object photometry, a hypersurface classifier, and output catalogue containing the segmentation history of each object. Recursive \textsc{FOCAS} runs on deep CCD data asymptotically, reaching completeness at the noise level. The program has many adjustable parameters: detection threshold, sky histogram, detection filter, segmentation sensitivity, intensity moments weighting kernel, and objects classification parameters.

A subsequent streamlined software package based upon similar operating principles, \SExtractor \citep[][version 2.25.3 being used in this study]{Bertin1996}, was developed for faster automated photometry. While based partly on similar precepts to that of \textsc{FOCAS}, it was designed to have a more global and computationally efficient segmentation algorithm, necessarily trading some control over optimal filters in each phase that was present in \textsc{FOCAS}. The \SExtractor package is capable of covering a wider range of object sizes however, and also outputs a sky estimation map and a detected source catalogue. For sky estimation, the local background around objects is clipped iteratively at $\pm3\sigma$ around the sky median. Like a single-pass FOCAS, it produces a sky that is biased high. The \SExtractor software package has been instrumental in the development of several image processing pipelines in recent years, notably the \textsc{SIGMA} \citep{Kelvin2010,Kelvin2012} and \textsc{GALAPAGOS} \citep{Barden2012,Hiemer2014} packages, both of which have been used to rapidly process and analyse millions of galaxies across multiple wavelengths using data from a wide variety of surveys. The \SExtractor package itself has since been rewritten as \textsc{SourceExtractor++}\footnote{\url{https://sourcextractorplusplus.readthedocs.io}}, incorporating a number of novel developments made in recent years. Furthermore, building upon the successes of \SExtractor, a range of contemporary software packages which utilise similar operating processes have also become available, such as \textsc{ProFound} \citep{Robotham2018} and \textsc{MTObjects} \citep{Teeninga2015}.

More recently, the \textsc{Gnu Astronomy Utilities} \citep[\Gnuastro;][]{Akhlaghi2015,Akhlaghi2019a} software package (version 0.17 used here\footnote{\url{http://www.gnu.org/software/gnuastro}}) has been released, providing a suite of programs and functions for use in the manipulation of astronomical data. One such constituent program is \NoiseChisel, a program for the detection of faint, extended, nebulous astrophysical objects. \NoiseChisel relies on a non-parametric algorithm rather than any particular function or fitting technique to identify contiguous regions whose noise properties are altered by the underlying faint astrophysical objects buried beneath. Other signal processing algorithms, like those mentioned above, are biased towards detecting objects that are sufficiently similar to the detection kernel being used. As \NoiseChisel does not depend on knowing the shapes or profiles of sources beforehand, and does not require any a priori knowledge of the spatial properties of astrophysical objects, it is therefore unbiased in this regard. Counter to the top-down operating principles of many prior source detection software packages, \NoiseChisel operates on a bottom-up philosophy, attempting first to characterise the background noise in an image before classifying the residual signal flux. This approach has potential advantages in the detection and characterisation of LSB light, but also carries certain risks with regards the possible inclusion of noise in estimates of signal flux, as further discussed later in this study.

A modern detection algorithm is being developed for use with the \VRO (hereafter, \RO), formerly known as the Large Synoptic Survey Telescope, which will undertake a 10-year Wide-Fast Deep (WFD) survey of the southern sky called the \LSST \citep[LSST,][]{Ivezic2019}. A version of this software is currently used in the data reduction pipeline for the Hyper Suprime Cam Subaru Strategic Program \citep[HSC-SSP;][]{Aihara2018a,Aihara2018b,Bosch2018,Bosch2019}. This algorithm first identifies high signal-to-noise stars on the stellar locus, using these to model the spatially varying PSF over the entire image and subsequently facilitating characterisation of simply connected regions above a given detection threshold as detections (see Section \ref{sec:dmstack2018} for further details). Detected sources are then masked, so a polynomial model can be fit to the remaining pixels to evaluate a sky background model. This process is repeated iteratively at the CCD level until a final PSF model, detection catalogue, and sky background model are achieved.

When reducing and analysing astronomical imaging data, it is often best to separate the two tasks of sky estimation and catalogue generation \citep[e.g., object photometry; also see][]{Akhlaghi2019a}. A Bayesian statistical approach to sky estimation which uses the known faint galaxy counts has recently been pursued by \cite{Ji2018}. It was shown in simulations that robust sky precision of $\about4\,\mathrm{ppm}$ is in principle possible.

In this study, we explore a number of alternative background estimator configuration modes which have the potential to be more accurate and less impacted by the extended surface brightness profiles of galaxies and the density of the local image environment as compared to traditional techniques. The primary challenge to be addressed is the performance of these techniques over realistic large deep fields with a broad range of object densities and sizes. Building upon prior efforts in this field, we present a range of sky estimation configuration modes which are designed to accurately preserve sky flux over large fields in realistic deep imaging, paving the way for future LSB science cases. For a fully realistic approach, we apply these configuration modes to full image simulations based upon recent deep survey imaging, testing the robustness of sky estimation outputs and source photometry for each.

The remainder of this paper is structured as follows. Section \ref{sec:simdata} provides an overview of our simulated dataset using the \GalSim software package. Section \ref{sec:sextraction} discusses all sky estimation software packages and distinct software configuration modes explored within this study. Results and discussion are presented in Section \ref{sec:results}, and conclusions are summarised in Section \ref{sec:conclusions}. For file storage purposes, FITS imaging has been compressed/uncompressed using both the NASA HEASARC \textsc{fpack}/\textsc{funpack} compression format \citep{Pence2009,Pence2010} and the GNU Gzip\footnote{\url{https://www.gnu.org/software/gzip}} compression format\footnote{With the former preferred for noisy float-type imaging and the latter for integer-type or truncated floating point data.}. All data and associated code produced for use within this study has been archived on the \textsc{Zenodo} open-access repository\footnote{\url{https://doi.org/10.5281/zenodo.7067465}}.

\section{Simulated Data}
\label{sec:simdata}

\begin{figure*}
    \centering
    \includegraphics[width=1.0\columnwidth]{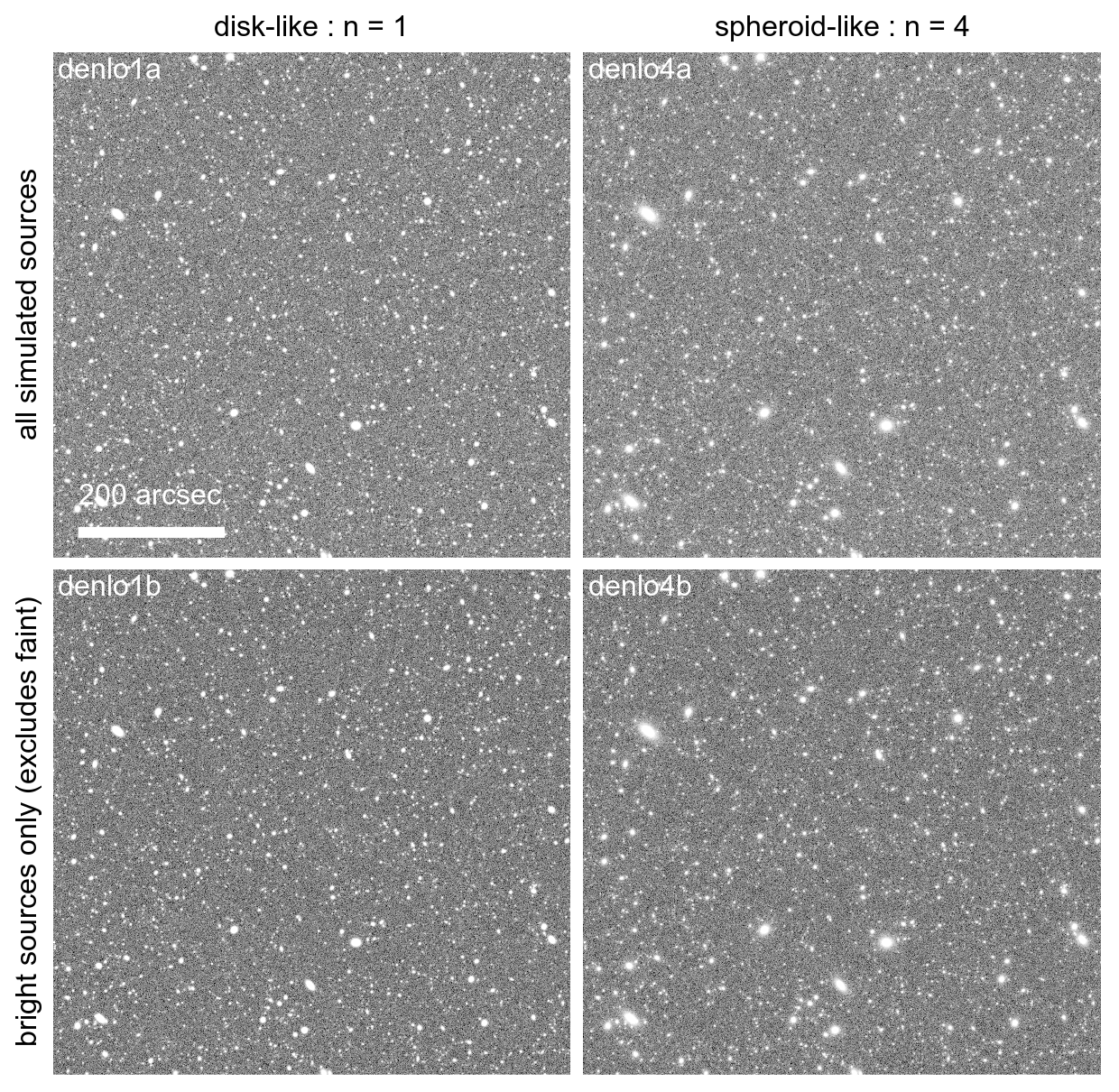}
    \includegraphics[width=1.0\columnwidth]{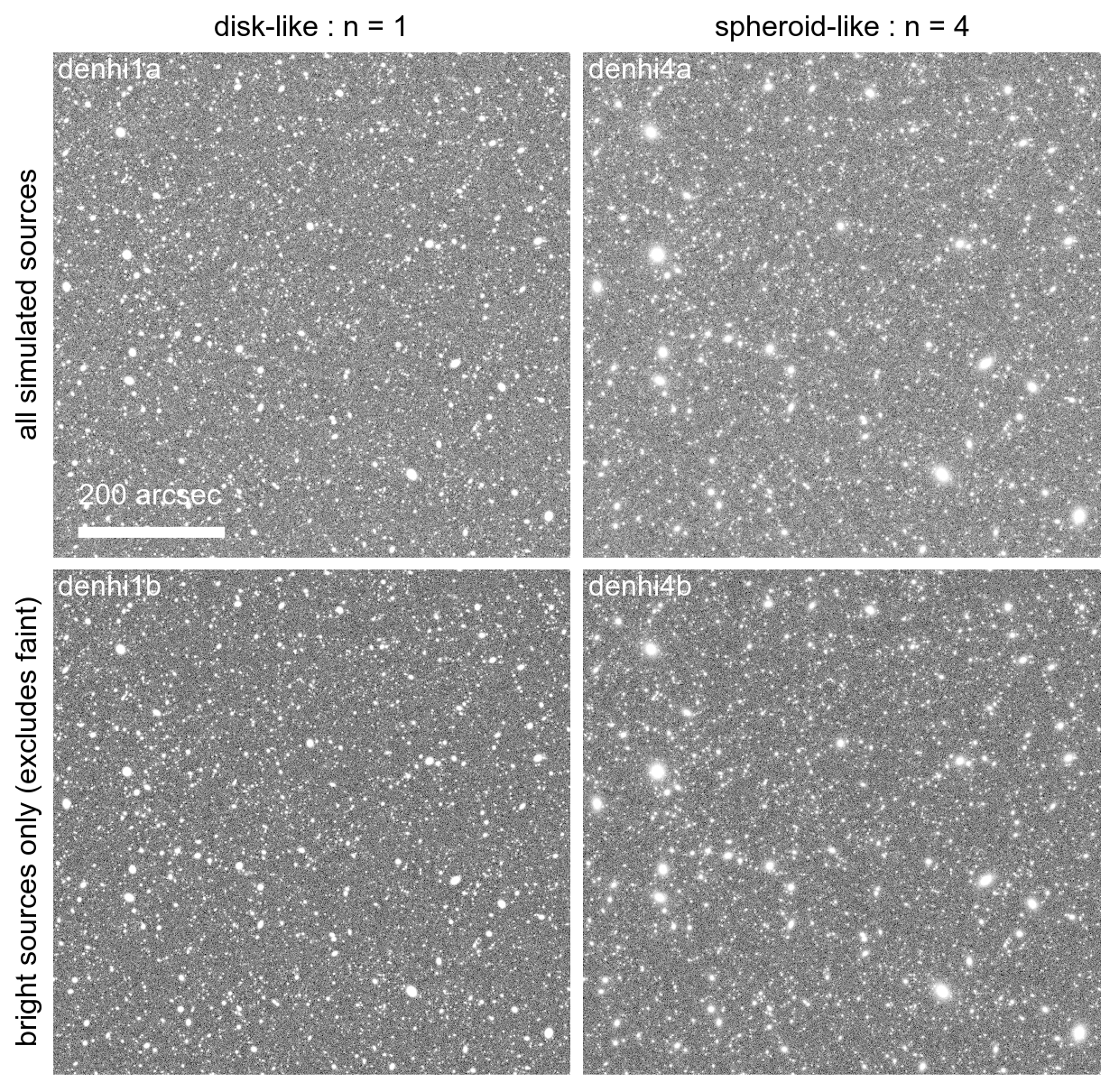}
    \caption{All eight simulated \GalSim fields for the low density `denlo' region (left four panels) and the high density `denhi' region (right four panels). Fields containing only disk-like exponential ($n=1$) or spheroid-like de Vaucouleurs ($n=4$) sources are identified by the labels at the top of the figure. Fields containing the full simulated sample or the bright-end sub-population only (i.e., excluding faint missing sources) are labelled along the left side of each density quartet. The scale inset into the top left panel of each density quartet applies equivalently to all panels. All sources are PSF-convolved, and Poisson noise equivalent to HSC-SSP data has been added. Images are arctan scaled. The impact of source population (i.e., contrasting the top row with the bottom) and source profile type (column-to-column) is visibly evident on the apparent background level.}
    \label{fig:simstampfull}
\end{figure*}

To test various sky estimation routines, we require an input dataset with known quantities such as sky level and source profile type. We therefore construct a series of simulated images using the \GalSim \citep{Rowe2015} software. All images consist of a flat sky level with a flux pedestal of zero counts. A series of simulated sources are injected into these data spanning three different binary configuration modes: source population, source density, and source profile type. For source population, the simulated field is generated in two modes: once containing all sources from the bright end down to the faintest limit explored in this study ($m_{r,\mathrm{total}}=30$ mag), and a second time excluding the typically undetected faint sources which largely constitute the extragalactic background light. For source density, both low- and high-density fields are simulated. Finally, for source profile type, both disk-like exponential and more extended de Vaucouleurs type sources are injected. For the sake of simplicity, we opt to populate these regions with extended-type \citet{Sersic1963,Sersic1968} sources alone, avoiding the additional complication of point sources which typically contribute less to sky estimation errors (excepting the extreme bright-end population of stars) than extended profiles. Taking all three binary-type configurations into account, our final simulated dataset consists of eight simulated fields. These simulated fields are shown in Figure \ref{fig:simstampfull}.

Each simulated region is designed to mimic a single Hyper Suprime-Cam \citep[HSC,][]{Miyazaki2012,Miyazaki2018} Subaru Strategic Program \citep[SSP,][]{Aihara2018a} Public Data Release \citep[PDR1\footnote{\url{https://hsc-release.mtk.nao.ac.jp}},][]{Aihara2018b} patch, spanning exactly $4200\times4100$ pixels in dimension (approximately
$706''\times689''$). Following a series of tests, tract-patch 8283-38 was selected as a basis for the low density region, whilst tract-patch 9592-20 was selected as a basis for the high-density region. The \SExtractor software package is used to generate initial source catalogues across these two fields. Further information on the selection procedure of these two fields and their analysis may be found in Section \ref{sec:hsc}.

\begin{figure*}
    \centering
    \includegraphics[width=0.9\textwidth]{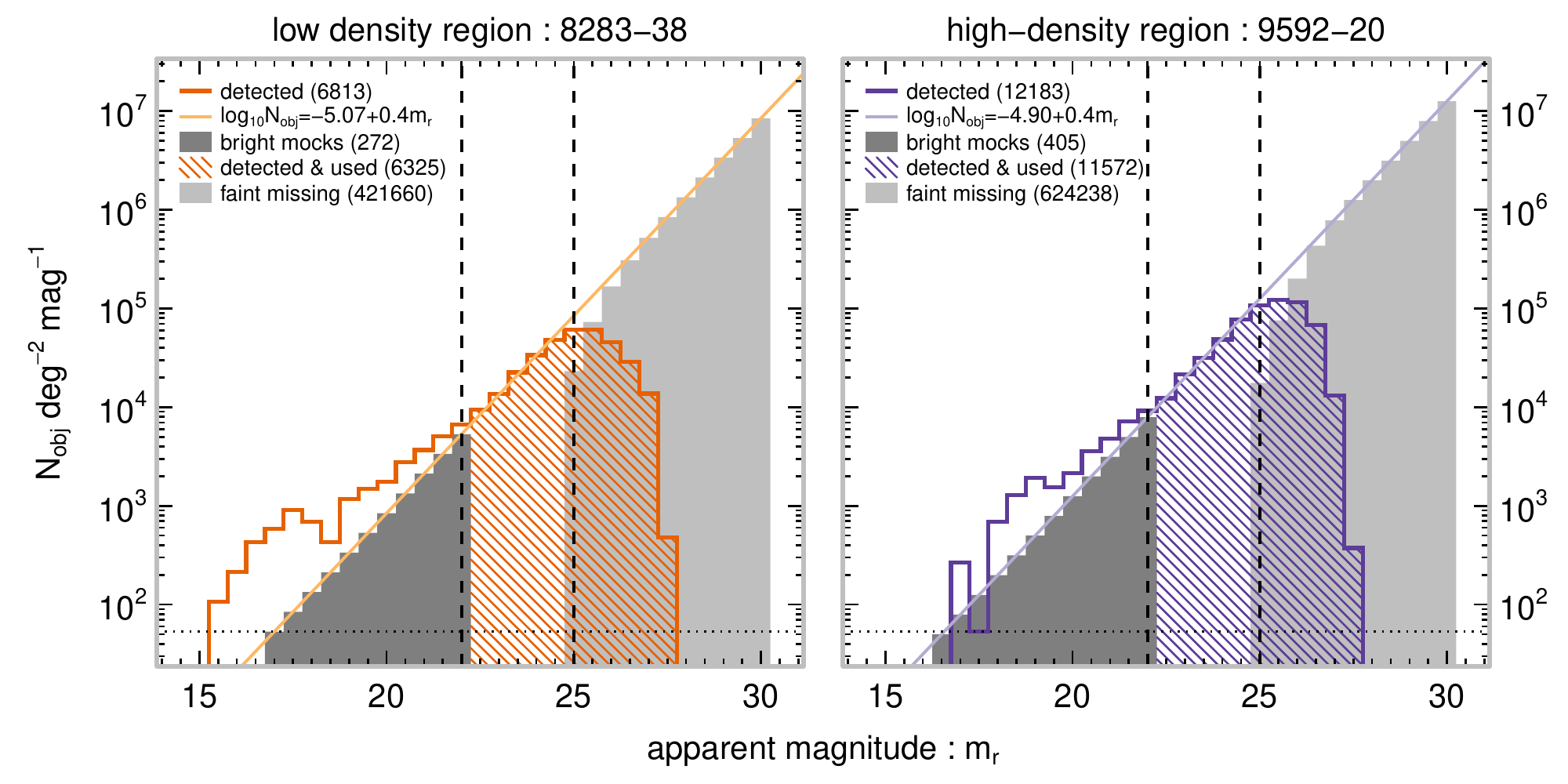}
    \caption{The number of sources per square degree per magnitude as a function of apparent $r$-band magnitude. The low density region (tract-patch 8283-38) is shown in the left panel, whilst the high density region (tract-patch 9592-20) is shown in the right panel, as indicated. Within each panel, the detected source number density profile is represented as a solid emboldened line. A log-linear best fit line to the detected source profile in the range $22<m_r<25$ mag (boundaries represented by vertical dashed lines) is shown using a solid thin line. The dotted horizontal line is equivalent to one object per the area of our sample region per magnitude. Detected sources which are directly used in the simulated input catalogue (detected \& used) are found within the hatched area. Simulated sources at the bright end ($m_r<22$ mag; bright mocks) and faint end ($m_r>25$ mag; faint missing) are shown using dark grey and light grey shaded regions, respectively. These sources are defined such that their summation in addition to any relevant detected \& used sources is equivalent to the fitted log-linear best fit line. Total object numbers for each regime in addition to best fit line parameters are shown in the inset legends. The three flux-associated populations defined here are used to define our input simulated number densities as a function of magnitude.}
    \label{fig:numcounts}
\end{figure*}

Figure \ref{fig:numcounts} shows the number of sources per square degree and per magnitude as a function of apparent $r$-band magnitude. The low density region is shown in the left panel, and the high density region in the right panel. The detected populations from our basis datasets are shown using bold solid outline histograms. Using these detections, simulated sources which map to the thin solid log-linear lines are constructed.

In both the low- and high-density region, the simulated population may be subdivided into three distinct regimes: bright mock sources ($m_r<22$ mag); detected \& used sources ($m_r>22$ mag), and; faint missing sources ($m_r>25$ mag). Further information on these simulated data may be found in Section \ref{sec:galsim}. In brief, detected sources fainter than $m_r=22$ mag are taken as-is, with their measured parameters used to construct simulated sources. A log-linear trend line is fitted to the detected population in the magnitude range $22<m_r<25$, with the faint end limit chosen as the magnitude at which the detected population begins to become incomplete and turn over. The bright end extrapolation of this trend line determines the bright mock sources injected into these simulated data. Bright detected sources are not directly used, owing to a greater uncertainty on their measured parameters. Finally, a faint-end extrapolation of the fitted trend line is used to generate faint sources down to $m_r=30$ mag.

\section{Source Extraction and Sky Estimation Techniques}
\label{sec:sextraction}

We make use of three distinct source extraction software packages: \SExtractor, \textsc{Gnu Astronomy Utilities} (\Gnuastro), and the \RO \LSSTSPs \citep[hereafter, \LSSTPs]{Bosch2018,Bosch2019}. These three packages are chosen due to their significant usage within the astronomy community at large, and (notably in the case of the \LSSTPs) due to their anticipated importance in the reduction of future datasets.

Both \SExtractor and \Gnuastro are initially operated in a default mode configuration to provide a baseline comparison as to its capabilities. As is common in the literature, each package is further run in various configuration modes with the aim of trying to optimally extract the background sky and generate high-quality source catalogues. As a reminder, simulated background sky maps are flat with a flux pedestal of zero counts. An optimal recovered sky map would therefore return a sky level around zero and relatively uniform in nature.

The \LSSTPs are operated in two default configurations to replicate the behaviour of the \LSSTPs at or around the time of the publication of HSC SSP PDR1 data (our basis dataset). As a caveat, we note that a number of significant improvements in the sky estimation capabilities of the \LSSTPs have taken place since HSC SSP PDR1 data was released, however, we stress that the precise form of the data used to build our simulated dataset is largely irrelevant here. As real HSC data is only being used to initialise a simulated dataset constructed by \GalSim, we would not expect any of our primary conclusions to change should we adopt more recent data released as part of the HSC SSP. We therefore provide these benchmark results by way of comparison, and to facilitate the discussion surrounding sky estimation and source characterisation.

The following sections present two novel techniques aimed at assisting in sky estimation: dilated masking and modelled masking. Subsequent sections provide specific details on each distinct configuration mode explored within this study.

\subsection{Magnitude Dependent Dilated Masking}
\label{sec:dilatedmasking}

As shown in \citet{Ji2018}, insufficient masking of bright extended sources can overestimate any estimation of the background sky. Flux in the wings of such sources can easily leak out beyond a segmentation mask, causing a sky estimation algorithm to interpret these pixels as sky and erroneously attempt to subtract the associated flux. This effect is catastrophic for those interested in low surface brightness science, as it removes much of this LSB signal before scientific analyses can occur. Here we present the dilated masking method, a means by which such flux can be accurately accounted for in order to mitigate the effect of sky over-subtraction.

Dilated masking refers to the process of enlarging the segmented mask map associated with a particular object in such a way that those pixels containing flux from the wings of the source are brought underneath the mask. It is often beneficial to vary the amount of such dilated masking according to the magnitude of the underlying source, as in \citet{Ji2018}, with brighter source masks dilated by a greater amount. Following experimentation with our input simulated dataset and software packages, we opt to expand the segmentation mask about a given detected object by a number of pixels\footnote{The pixel scale of HSC data is $0.168\;\mathrm{arcsec}/\mathrm{pixel}$} $d$, where the extent of $d$ is related to its detected $r$-band magnitude $m_r$ as given by the following equation:
\begin{equation}
    \label{eq:dilmask}
    d = \lfloor10^{-0.2\left(m_r-25\right)}\rfloor + 5
\end{equation}
Relative to a bright source, progressively fainter objects have their segmentation mask expanded by decreasing amounts, with the power law component of $d$ flattening out to zero pixels at magnitudes fainter than $m_r=25$ mag. Each masked region is additionally expanded by a fixed pixel amount of $5$ pixels. Segmentation masks are expanded using the \Gnuastro \textsc{arithmetic} software package. Masks are dilated one pixel at a time, alternating between 8-connected dilation and 4-connected dilation, in order to ultimately achieve a dilated mask which isn't biased toward either procedure. The implementation of this dilated masking prescription as applied to \SExtractor and \Gnuastro outputs is shown in Section \ref{sec:sextractordilated} and Section \ref{sec:gnuastrodilated}, respectively.

\subsection{Modelled Masking}
\label{sec:modelledmasking}

Modelled masking describes the process of constructing a parametric model for select sources, subtracting them from the science image, and reattempting background estimation on the residual image. The generation of a parametric model allows for the extrapolation of such a profile out to large radii and into the noise, rather than using a pixel mask which truncates at a hard isophotal value. Such a concept is not new, with a similar approach previously used in the SDSS \citep{Lupton2001,Aihara2011a,Blanton2011}. There, two-dimensional `cmodel' galaxy source models are subtracted from the image in order to improve their estimate of the sky, with the outer parts of galaxies considerably more extended than had been before. It's especially important for the chosen form of the model to sufficiently mimic the observed data in the wings of the source, as choosing an inappropriate model may degrade any estimate of the sky rather than improve it.

We opt to model all detected sources brighter than $m_r=25$ mag with a PSF-convolved single \Sersic profile. The \citet{Sersic1963,Sersic1968} profile has long been successfully used to describe galaxy surface brightness profiles for a wide range of galaxy types \citep[e.g.,][]{Caon1993,Trujillo2004,Graham2005a,Hill2011,Simard2011,Kelvin2012,Kelvin2014a,Kelvin2014b}. Furthermore, our input simulated data have been generated using a range of PSF-convolved \Sersic profiles as described in Section \ref{sec:galsim}. As a caveat, whilst fitting \Sersic models to sources with a known \Sersic profile is a best-case scenario (with true galaxies likely more complicated than this), adopting PSF-convolved \Sersic models as our base descriptor of source surface brightness profiles allows us to make a direct comparison of sky estimation software packages without needing to account for potential model inaccuracies. Our rationale for limiting model fitting to bright sources alone is twofold. Firstly, if extended sources do act to negatively impact sky estimation routines, then it's the brightest of these sources with their broad wings overlapping other sources that will have the largest per-object impact overall. Secondly, bright sources contain a larger number of pixels above the noise threshold. As a consequence, their resultant best-fit models are therefore of a higher certainty.

All detected sources brighter than $m_r=25$ mag are fit using the \SExtractor software package. Whilst \SExtractor is well known for its source extraction capacity, its model fitting capabilities are less familiar. In concert with \PSFEx, \SExtractor is able to accurately and rapidly fit multiple model types to either all detected sources or a subset of detected sources\footnote{Subsets of the total detection catalogue are defined via source matching to $x$/$y$ coordinates and magnitude as provided by an \texttt{ASSOC} catalogue.}. Available model types include a PSF-like point source (\texttt{POINTSOURCE}), an exponential disk-like profile (\texttt{DISK}) and a generic \Sersic-like spheroid profile (\texttt{SPHEROID}). Specifying any of the associated profile fitting parameter names in the input parameter configuration catalogue will enable source fitting. A catalogue generated from an initial default run of \SExtractor and our previously described \PSFEx PSF model (see Section \ref{sec:psf}) are used as an input. We opt to fit our magnitude-limited subset sources using the \texttt{SPHEROID} \SExtractor model type\footnote{Complete \SExtractor catalogue parameters used for fitting in this instance are: \texttt{NUMBER}, \texttt{X\_IMAGE}, \texttt{Y\_IMAGE}, \texttt{FLUX\_SPHEROID}, \texttt{MAG\_SPHEROID}, \texttt{SPHEROID\_REFF\_IMAGE}, \texttt{SPHEROID\_ASPECT\_IMAGE}, \texttt{SPHEROID\_THETA\_IMAGE}, \texttt{SPHEROID\_SERSICN}, \texttt{CHI2\_MODEL}, \texttt{FLAGS\_MODEL} and \texttt{VECTOR\_ASSOC}. Further information may be found in \citet{Bertin1996} and associated documentation.}.

Once a catalogue of best fit \Sersic parameters is known, we use the \GalSim software package to construct model imaging. Whilst \SExtractor is also capable of outputting model imaging (using the \texttt{MODELS} check image type), these images were found to contain several edge artefacts that are not easily removed\footnote{We direct the interested reader to Section \ref{sec:postagestampsizes} for further details.}. For each source, the \SExtractor best fit output provides a starting $x$/$y$ coordinate location, total luminosity, half light radius, axis ratio, ellipticity and \Sersic index. In addition, \GalSim requires the user to specify the postage stamp box size within which the source is generated. As discussed in Section \ref{sec:postagestamps}, this postage stamp size should ideally encompass the entirety of our simulated image, mitigating potential edge effects. Due to computational expense however, some lower limit must be specified. We once again define a postage stamp size for each source such that the source profile reaches a surface brightness limit of $\mu_r=35\;\mathrm{mag\;arcsec}^{-2}$ at the extreme outer edge of the box, with a minimum box size of $11\times11$ pixels. This ensures that any profile boundary falls well within the noise, effectively eliminating its impact on our subsequent analyses. The resultant model image is finally subtracted from the initial science image, providing a new dataset upon which to estimate the sky with most of the bright source contaminant flux removed. The implementation of this modelled masking procedure as applied to \SExtractor and \Gnuastro outputs is shown in Section \ref{sec:sextractormodelled} and Section \ref{sec:gnuastromodelled}, respectively.

\subsection{\SExtractor Configurations}
\label{sec:sextractor}

\subsubsection{Default \SExtractor Configuration}
\label{sec:sextractordefault}

The default setup of \SExtractor used at this stage of processing is identical to that outlined in Section \ref{sec:sample}, with the convolution kernel, stellar classification neural network file and configuration file all the same. The threshold for source detection is maintained at $1.5\sigma$ of the background estimate. Having excluded detected sources, background sky in \SExtractor is estimated across the image within a default of $64\times64$ pixel \texttt{BACK\_SIZE} mesh elements. A bicubic-spline interpolation is subsequently applied to all mesh elements to produce a background map, allowing for the exclusion of erroneous mesh values. In addition to the output catalogue parameters defined in Section \ref{sec:sample}, we also opt to output \texttt{ISOAREA\_IMAGE}, providing data on the total area above the analysis threshold for each detected object. The \SExtractor package is run upon each simulated image, producing an output catalogue and check images for the segmentation map (\texttt{SEGMENTATION}), background pedestal map (\texttt{BACKGROUND}) and background RMS map (\texttt{BACKGROUND\_RMS}). Further information on the default operation of \SExtractor may be found in \citet{Bertin1996}.

\subsubsection{Modified \SExtractor Configuration}
\label{sec:sextractormodified}

The \SExtractor package is often modified in order to better tailor its outputs to the specific scientific needs of the input dataset. A sky estimation and source detection setup which may prove optimal for the characterisation of point-like sources is often wholly inadequate for the purposes of extended source analysis. Modifications are also often made to address common complaints with the default outputs of \SExtractor \citep[see, e.g.,][]{Haussler2007, Haussler2013, Simard2011, Barden2012, Kelvin2012, Hiemer2014}. Principal amongst these are issues surrounding the derived sky level, an inadequate smoothing kernel, and inaccurate source segmentation. There is a long and inglorious history to such segmentation failures in top-down auto-detect algorithms. For a number of years in the field of Source Extraction \citep[see e.g. \textsc{FOCAS},][]{Jarvis1981} such segmentation failures have often been encountered when using low threshold runs of \SExtractor, largely in situations where there is a triangle of objects that fail proper segmentation and photometry. This geometry can create a situation where there is no 1-D saddle point in the surface brightness.

We adopt five key \SExtractor configuration modifications to address these issues. First, we increase the background mesh element size (\texttt{BACK\_SIZE}) to $128\times128$ pixels (from a default of $64\times64$ pixels). An increase of this type reduces the risk that any individual mesh element becomes severely compromised by source flux from a single bright object or cluster of bright objects. Second, we also increase the background mesh filter size (\texttt{BACK\_FILTERSIZE}) to $5$ (from a default of $3$). This filtering size determines the mesh super-pixel area over which mesh elements are filtered to detect and eliminate background estimate outliers. A larger filter size increases the identification of outliers, providing a more coherent global sky estimate. Third, to assist with initial source detection, we reduce the detection and analysis thresholds (\texttt{DETECT\_THRESH} and \texttt{ANALYSIS\_THRESH}) to $0.5\sigma$ (from a default of $1.5\sigma$), and fourth, we replace the default smoothing kernel with a Gaussian of FWHM $\Gamma=2$ pixels. A larger smoothing kernel reduces the impact of contaminant noise spikes present within the background and allows for a lower overall source detection threshold. Finally, we reduce the minimum segmentation contrast parameter (\texttt{DEBLEND\_MINCONT}) to $0.00005$ (from a default of $0.005$). Reducing this value increases the chance that faint local peaks will be included as separate objects. Whilst this does not directly impact the final sky estimate, it does produce superior segmentation results within contiguous regions falling above the detection threshold, necessarily improving source extraction catalogue data.

\subsubsection{\SExtractor with Dilated Masks}
\label{sec:sextractordilated}

It is not natively possible to apply the dilated masking process outlined in Section \ref{sec:dilatedmasking} within \SExtractor itself. Implementation of this procedure therefore requires modifying the data that is fed into \SExtractor. First, dilated masks are produced using outputs from a default \SExtractor run. Using the input science image, each dilated masked pixel is set equal to \texttt{NaN}. Such pixels will not be used by \SExtractor to construct a background map. This new manually masked image is processed using the same default \SExtractor setup as previously used.

\begin{figure*}
    \centering
    \includegraphics[width=1.0\textwidth]{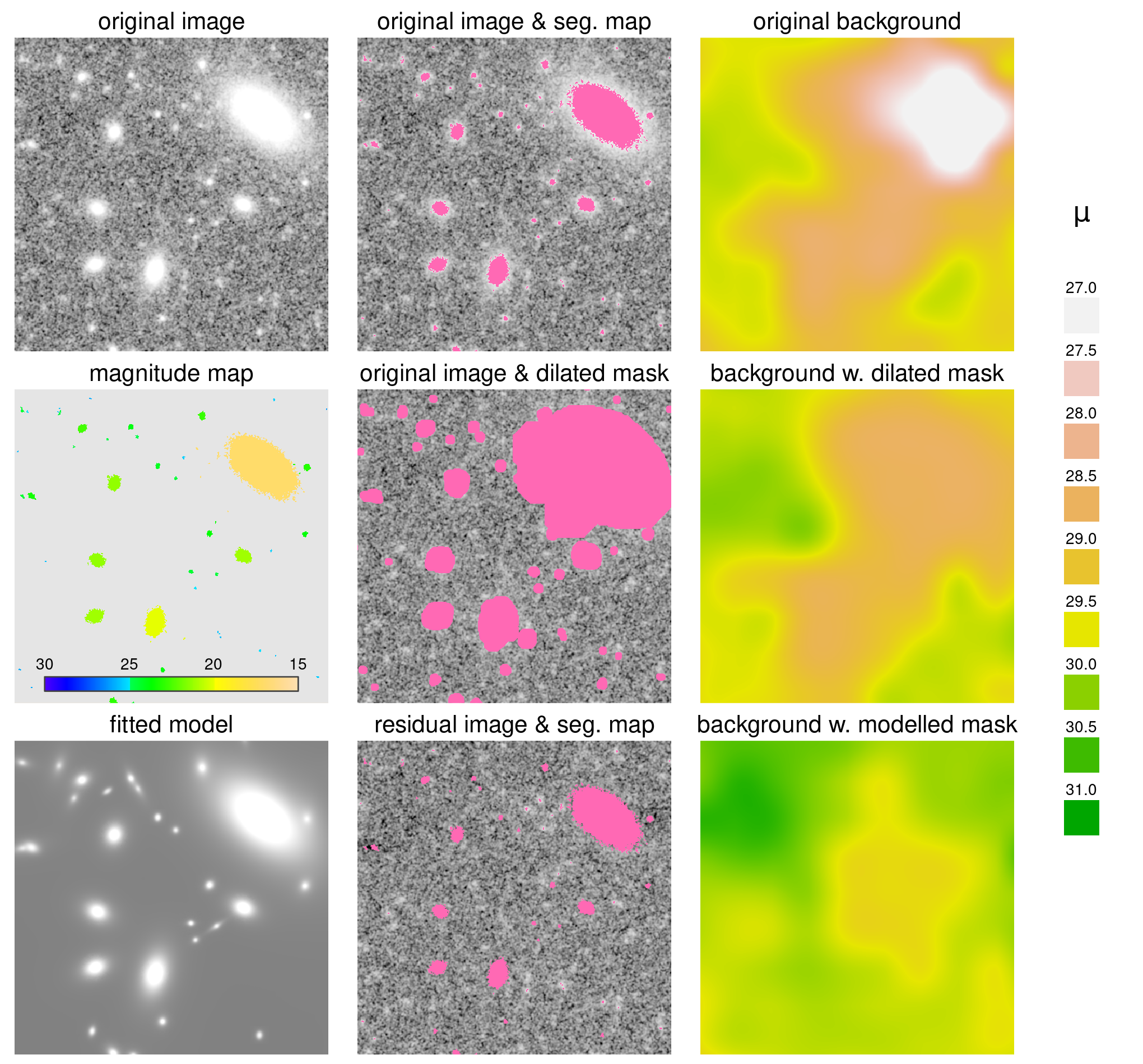}
    \caption{Examples of dilated masking and modelled masking sky estimation techniques as applied to a low density simulated field and processed using the \SExtractor software package. Top row, left: the original simulated dataset, corresponding to a zoomed in section of the low density field shown in Figure \ref{fig:simstampfull}. Top row, centre: the original segmentation map overlaid upon the original image. A significant amount of flux can be seen leaking out beyond the default segmented regions, thereby contaminating any estimate of the sky. Top row, right: the original estimate of the background sky as found when using default segmentation masks. A notable contamination due to the bright extended source within this field is evident. Middle row, left: the segmentation map for this region as given by a default run of \SExtractor. Each segmented region is colour coded according to its underlying detected source magnitude, as shown by the inset colour bar. Middle row, centre: the new dilated mask overlaid upon the original image. The extent of mask dilation is magnitude dependent, as given by Equation \ref{eq:dilmask}. These larger dilated masks are superior at masking signal around the wings of bright sources. Middle row, right: the new estimate of the background sky as found when using the dilated masks shown here. Much of the aforementioned background sky contamination is improved, resulting in a flatter and more accurate sky estimate. Bottom row, left: the successfully fitted single \Sersic models based upon detected sources brighter than $m_r=25$ mag. Bottom row, centre: the residual image generated after the fitted model is removed from the original image, overlaid with the default \SExtractor segmentation map. Note how much of the prior contaminant flux observed leaking from beyond the edge of the segmented regions has been removed. Bottom row, right: the new estimate of the background sky generated when using modelled masks. Colours correspond to surface brightness levels in units of $\mathrm{mag\;arcsec}^{-2}$ shown in the outlying legend. The sky estimate is significantly improved relative to both the default sky map and the dilated mask sky map, no longer exhibiting strong source-correlated sky biases. Images are arctan scaled, with noisy imaging smoothed by a Gaussian kernel of $\Gamma=3\;\mathrm{pix}$.}
    \label{fig:mask_sex}
\end{figure*}

An example of the dilated masking routine as applied to simulated data processed by \SExtractor is shown in Figure \ref{fig:mask_sex}. The top-row left panel shows a zoomed in $n=4$ low density simulated field (denlo4a), arctan scaled and smoothed with a Gaussian kernel of $\Gamma=3\;\mathrm{pix}$. The middle-row left panel shows the segmentation map for this region as produced using a default \SExtractor run. Each segmented region is colour coded according to the total detected $r$-band magnitude of the underlying source, as represented by the inset colour bar. The top-row central panel shows the default \SExtractor segmentation map overlaid upon the initial image, whilst the middle-row central panel shows our magnitude dependent dilated masks overlaid upon the initial image. Finally, the top-row and middle-row right panels show the output background sky maps as determined using the original default segmentation map and the new dilated mask map, respectively. Bottom-row panels are discussed in the next section.

As shown, a significant amount of flux is evidently leaking from the outskirts of the default segmentation map. Further, the amount of flux leakage appears linked to the brightness of the underlying source, with the majority of the contamination in this particular field linked to the large bright source in the upper right corner. As these initial simulated fields consist of a flat and equal to zero background map, recovery of a faint and relatively uniform background model is preferred. In this case, application of magnitude-dependent dilated masks significantly reduces the amount of contaminant flux present in the remaining background pixels, markedly improving the resultant background sky estimate.

\subsubsection{\SExtractor with Modelled Masks}
\label{sec:sextractormodelled}

Following an initial run of \SExtractor and \GalSim in order to generate free-\Sersic models for all sources brighter than $m_r=25$ mag, the model plane is subtracted from the science image to produce a residual image. The removal of bright source wings allows for a more accurate estimate of the sky in these regions, whilst simultaneously markedly improving source characterisation for small faint sources previously hidden beneath the wings of their bright neighbours. The centres of bright astrophysical sources are notoriously difficult to fit, with even a slight model inaccuracy resulting in a pixel value offset of many multiples of the background RMS. To prevent such regions having any impact upon our sky estimation routine, we manually apply the original default \SExtractor segmentation mask to the residual image, setting masked pixels equal to \texttt{NaN}. Consequently, centroid pixels are ignored by \SExtractor when generating sky estimates and whilst performing source extraction.

An example of the modelled masking routine as applied to our simulated data is shown in Figure \ref{fig:mask_sex}. The top and middle rows have already been discussed in the prior section, showing the original simulated image, the original background map, and the dilated masking results. The bottom row left panel shows the fitted PSF-convolved \Sersic model plane generated for this region. All detected sources brighter than $m_r=25$ mag are fitted. The bottom row centre panel shows the residual image generated once the fitted model is removed from the original image, also with the original mask map overlaid. As can be seen, much of the contaminant flux leaking out from the segmented regions has been successfully accounted for. The bottom right panel shows the background sky resulting from a model masked processed image. The difference between this background map and the original is obvious, with multiple orders of magnitude difference in the resultant background sky estimate. As these simulated data consist of a flat and equal to zero sky pedestal, the background map produced via the modelled masking procedure is clearly outperforming the other methods shown here. Notably, the impact of the singular bright source has been almost completely accounted for, leaving behind a much more uniform sky estimate in its place.

\subsection{\Gnuastro Configurations}
\label{sec:gnuastro}

\subsubsection{Default \Gnuastro Configuration}
\label{sec:gnuastrodefault}

The \Gnuastro software suite is highly modularised, necessitating the use of several sub-packages to fully process any given dataset. Each simulated image is initially processed by \NoiseChisel, the \Gnuastro source detection and sky estimation package designed to optimally extract sources with extended profiles embedded within noisy data. In brief, pixel data is first convolved with a Gaussian smoothing kernel and thresholded at some pixel value quantile (e.g., the $30^{\mathrm{th}}$ percentile) within mesh elements of default size $30\times30$ pixels, splitting the image into regions of noise and regions of signal. A series of erosions and dilations are subsequently performed on the regions of noise (i.e., the noise is `chiselled' away) and false detections are removed. Any remaining regions of signal undergo further pixel dilation. Once regions believed to contain true signal flux are determined, the \Segment package splits these regions into distinct sources, performing clump-based segmentation where required.

The \Segment package assigns pixels into one of three distinct types: noise, clump, and object. Noise regions are those where no significant signal from physical objects have been detected. Clumps are connected regions associated with local maxima on the image. Objects correspond to all pixels which are believed to contain physical object flux. Each object may contain zero, one, or many distinct clumps, and clump information is used to assist in object segmentation \citep[see][for further information]{Akhlaghi2015}. As the focus of this study is on the optimal extraction of the background sky, our primary interest is the distinction between signal and noise. We therefore opt here to utilise the object catalogues for further study throughout.

Once segmentation has been finalised, the \MakeCatalog package is used to convert processed pixel data into an output source catalogue\footnote{Output \Gnuastro \MakeCatalog columns are: \texttt{ids}, \texttt{x}, \texttt{y}, \texttt{brightness}, \texttt{magnitude}, \texttt{semimajor}, \texttt{semiminor}, \texttt{geosemimajor}, \texttt{geosemiminor}, \texttt{sn}, \texttt{axisratio}, \texttt{positionangle}, \texttt{sky}, \texttt{std}, \texttt{area} and \texttt{upperlimit}. A full description for each column output may be found in \citet{Akhlaghi2015} and \citet{Akhlaghi2019a}}. Further information on the operation of all \Gnuastro packages may be found in \citet{Akhlaghi2015}, with subsequent updates in \citet{Akhlaghi2019a}.

\subsubsection{Modified \Gnuastro Configuration}
\label{sec:gnuastromodified}

As with \SExtractor, a number of modifications are typically made to the operation of \Gnuastro to improve or tailor its output results. Owing to the modularised nature of \Gnuastro, some modifications are made to all utilised sub-packages, whilst others are specific to individual sub-packages. We detail all modifications here.

First, as with \SExtractor, we opt to increase the default mesh size of $30\times30$ pixels used during initial background estimation using the \texttt{tilesize} argument. The tile size determines the mesh size within which initial background thresholding occurs. In the presence of large sky gradients or large contaminant sources, an overly-large tile size may lead to a significant amount of data either being discarded or erroneously classified by \NoiseChisel. However, a larger mesh size will also reduce the chance that any single mesh element is entirely dominated by signal flux, and enforce the generation of a flatter and inherently less-responsive sky model. During testing, the recovered sky level appeared to be somewhat unresponsive to the choice of mesh size (within a factor of 2 of the default value), having a relatively larger impact only on the RMS of the background map. Therefore, in keeping with the logic previously used with \SExtractor, we opt to increase the tile size to $64\times64$ pixels\footnote{An attempt was also made to increase the \Gnuastro tile size to $128\times128$ pixels, in keeping with the modified \SExtractor runs discussed previously. Unfortunately however, these runs fail with some of our input datasets, with \Gnuastro reporting that not enough neighbours could be found for close neighbour interpolation.}. The tile size parameter is globally provided to \NoiseChisel, \Segment and \MakeCatalog.

Second, we opt to reduce the threshold down to which \NoiseChisel detection maps are expanded using the \texttt{detgrowquant} parameter. The detection growth quantile (default $0.9$) determines down to what quantile `true' detections are expanded out into the noise. This growth mechanism has potential advantages over a traditional dilation, as it allows the user to follow the shape of the profile out into the noise. Such an effect is crucial in accurately capturing more of the diffuse low surface brightness light typically located around the wings of bright extended sources. However, setting this value too low leaves open the possibility that spurious noise signals will end up being drawn into the final resultant detection maps, artificially inflating the flux associated with detections. We opt to lower the detection growth quantile parameter to $0.7$, somewhat expanding our detection footprints relative to the defaults to capture more of the potential low surface brightness flux.

We note that a number of additional parameters were explored and rejected here, owing to their failure to significantly improve sky estimation or source extraction results during testing. These include modifications in: the smoothing kernel (\textsc{kernel}), the maximum mean and median quantile difference per tile (\textsc{meanmedqdiff}), the quantile threshold on the convolved image (\textsc{qthresh}), the quantile threshold determining which pixels are excluded from erosion (\textsc{noerodequant}), the selection of sky tiles (\textsc{minskyfrac}), and quantile of the signal-to-noise ratio distribution of clumps in undetected regions (\textsc{snquant}).

\subsubsection{\Gnuastro with Dilated Masks}
\label{sec:gnuastrodilated}
\begin{figure*}
    \centering
    \includegraphics[width=1.0\textwidth]{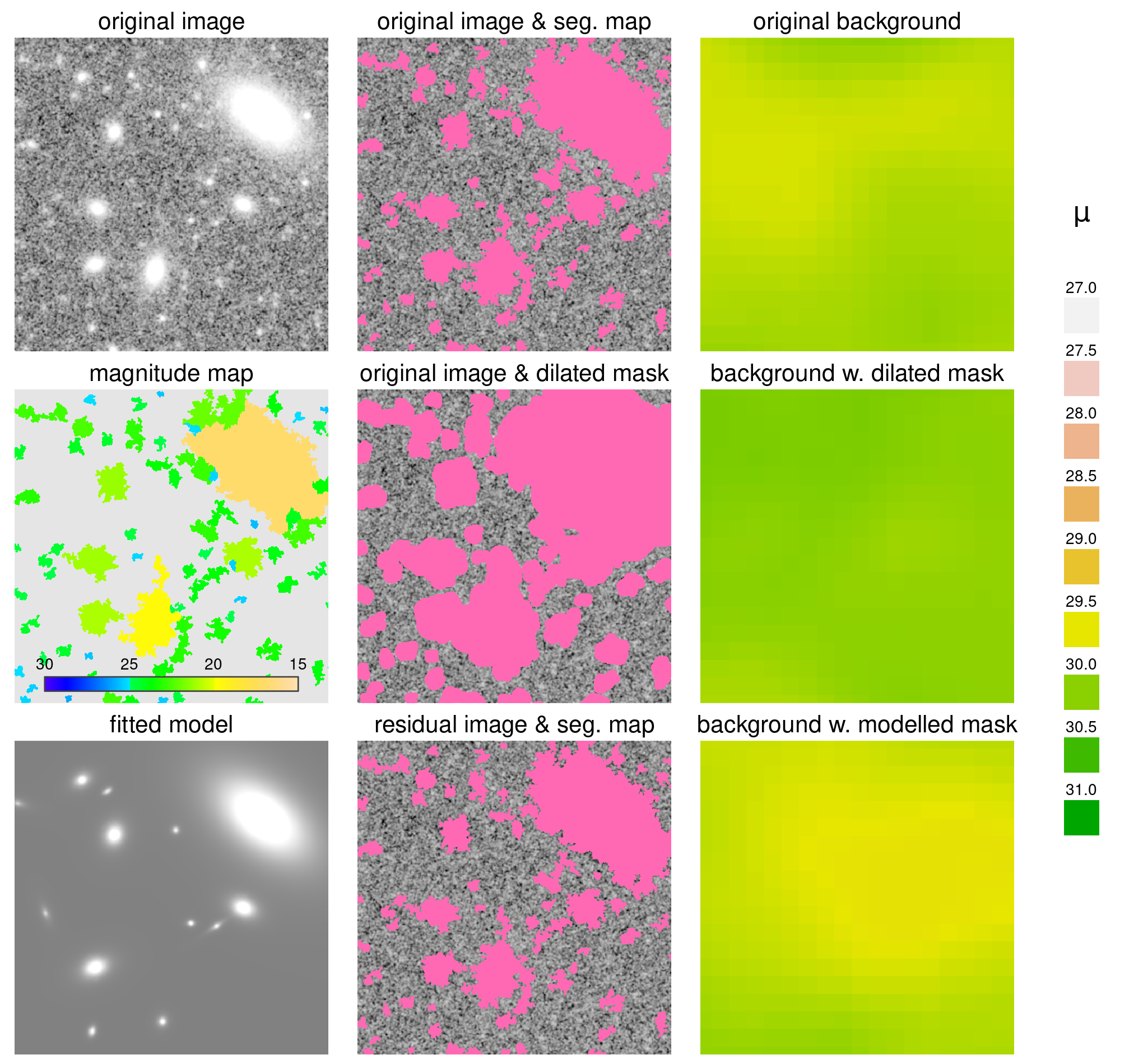}
    \caption{As Figure \ref{fig:mask_sex}, but for simulated data initially processed using the \Gnuastro software package.}
    \label{fig:mask_gnuastro}
\end{figure*}

Figure \ref{fig:mask_gnuastro} shows the dilated masking routine as applied to simulated data which has been processed by \Gnuastro. Each panel is equivalent to the panels previously shown in Figure \ref{fig:mask_sex}, excepting that the segmentation maps and subsequent dilated masks (and model masks, discussed in the next section) are based upon \Gnuastro source extraction processing. As above, the effect of the dilated masking procedure is to notably reduce the impact of flux leakage upon the final background sky estimate. As these simulated data have been constructed with a flat and equal to zero sky, a faint and uniform background model is optimal here. We also note however that the original segmented region produced by \Gnuastro covers a much larger fraction of the original image than was the case with \SExtractor. Consequently, the original default background sky produced by \Gnuastro is markedly less impacted by the bright source in this region than the default outputs from a \SExtractor run. This will be further discussed in Section \ref{sec:results}.

\subsubsection{\Gnuastro with Modelled Masks}
\label{sec:gnuastromodelled}

Figure \ref{fig:mask_gnuastro} (bottom row) shows the effect of modelled masking as applied to simulated data which has been processed by \Gnuastro. Simulated data have been constructed with a flat and equal to zero sky pedestal. As above, each panel is equivalent to the panels previously shown in Figure \ref{fig:mask_sex}, except that the segmentation maps and derivative modelled mask data are based upon \Gnuastro source extraction processing. The impact of modelled masks appears to have a negligible impact on the resultant background map. With the initial segmentation map already reasonably extensive, as noted above, the model masking procedure has little extra to contribute. A more complete discussion of the impact of modelled masking will be further discussed in Section \ref{sec:results}.

\subsection{Configurations of the \LSSTPs}
\label{sec:dmstack}

The \RO \LSSTPs were designed to optimally process HSC and LSST type imaging, providing a suite of complex image processing routines to ultimately produce high fidelity output catalogue and imaging data. This software stack has been working with HSC data since HSC's first light, making it an ideal candidate for use with the simulated HSC-type imaging produced for this study. We process our simulated data with two version 19.0.0 configurations of the \LSSTPs. First, these data are analysed with a specific configuration of the \LSSTPs designed to mimic an earlier state of the code base utilising a sixth-order polynomial fit to describe the background model. The code base at this time is largely equivalent to that used to initially produce HSC-SSP PDR1 data, and serves as a useful baseline. Second, we process our simulated datasets with an updated configuration of the \LSSTPs which employs a more local 128-pixel spline to describe the background model. Both configurations are further discussed in the sections below, with full details to be found in \citet{Bosch2018} and \citet{Bosch2019}.

\subsubsection{\LSSTPs Sixth Degree Polynomial Configuration (\DMA)}
\label{sec:dmstack2018}

The first configuration of the \LSSTPs explored here is one which generates a background map using a sixth degree polynomial fit to detected background pixels (hereafter referred to as `\DMA', for brevity). Simulated image data is initially convolved with a smoothing kernel, and contiguous regions above some threshold level are identified. In a standard run of the \LSSTPs (i.e., using true astrophysical data), high signal-to-noise stars are identified on the stellar locus using a placeholder Gaussian PSF to perform a first detection pass. These bright stars are then used to model the spatially varying PSF over the image using \PSFEx. A circular Gaussian whose RMS width is matched to the PSF is used as a detection kernel for performance reasons. However, as the simulated imaging constructed for use within this study contains no point sources (see Section \ref{sec:galsim}) we do not follow this procedure here. Instead, we construct a circular Gaussian whose RMS width is matched to the PSF described in section \ref{sec:psf}. This circularised Gaussian is subsequently used as the primary detection kernel.

Background estimation within the \LSSTPs is handled by the \textsc{SubtractBackgroundTask} method. The astrophysical background contains contributions from several sources: the night sky, optical ghosts and scattered light in the telescope, and undetected sources that make up the EBL amongst them. A parametric model based upon the sum total of these contributions to the background is constructed and subsequently subtracted out at the image level. The process consists of two high-level steps. First, the image is subdivided into $128\times128$ pixel bins, or \textit{super pixels}. Pixels which constitute each super pixel and do not correspond to any detected object are used to calculate the mean, variance, and pixel averaged centre of its host super pixel. Second, a 6th order two-dimensional Chebyshev polynomial model is fit to the mean values of the image, using the average pixel centres of each super pixel. Each bin is inverse weighted by its variance such that noisy super pixels with relatively few unmasked regions do not overly bias the fit.

\begin{figure*}
    \centering
    \includegraphics[width=1.0\textwidth]{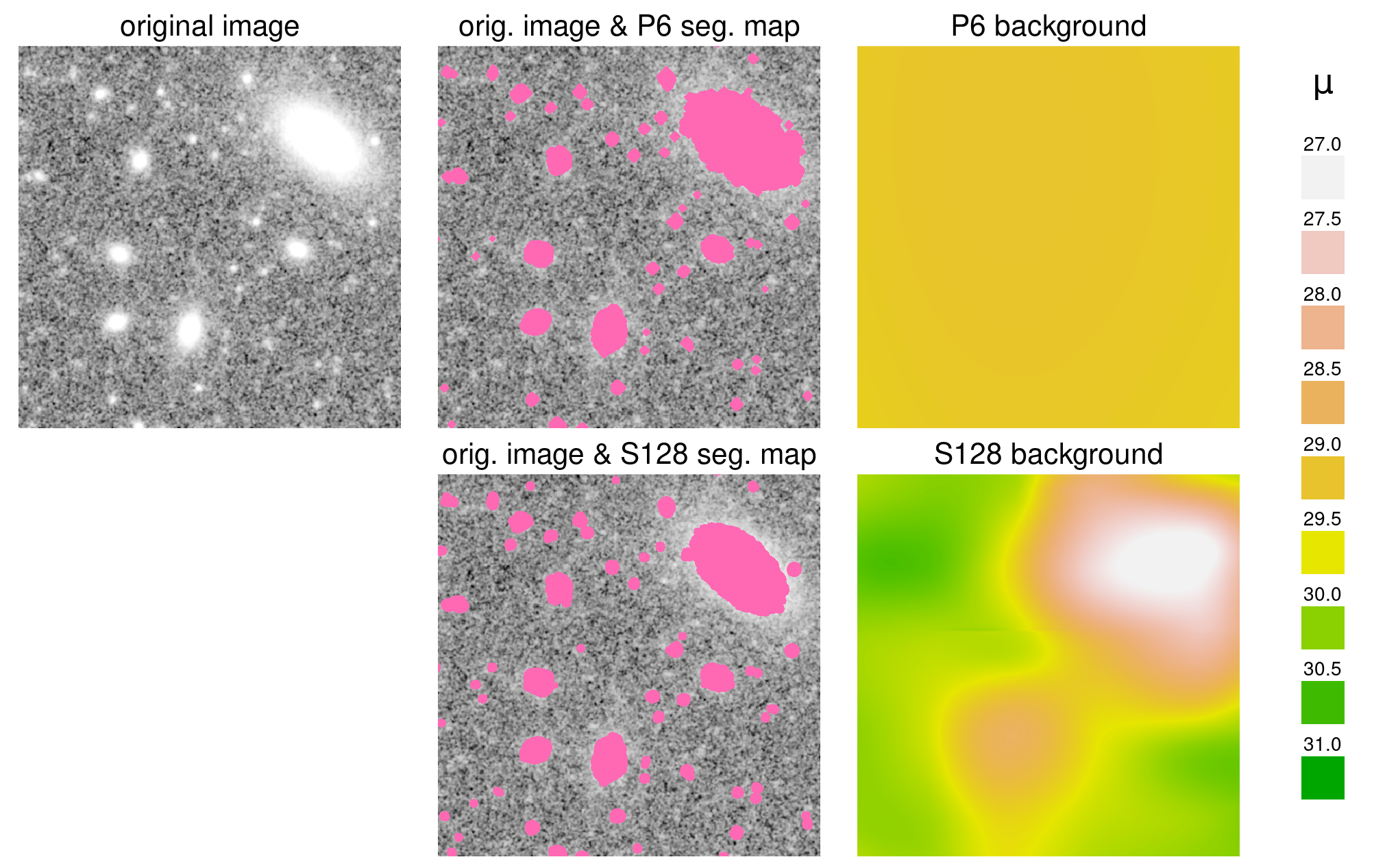}
    \caption{Similar to Figures \ref{fig:mask_sex} and \ref{fig:mask_gnuastro}, but for simulated data initially processed using the \LSSTSPs. A zoomed-in section of the original simulated image is shown on the left. The central column shows the original simulated image again in the background, with the segmentation maps returned by the \DMA (top) and \DMB (bottom) methods overlaid. The right column shows the resultant recovered background maps for this small zoomed in section of the sky, again for both the \DMA (top) and \DMB (bottom) methodologies. Whilst the \DMA configuration returns a flatter and more homogeneous sky pedestal, the \DMB configuration returns fainter sky levels (albeit with more fluctuation in the recovered sky map).}
    \label{fig:mask_dmstack}
\end{figure*}

Figure \ref{fig:mask_dmstack} shows the results of the \LSSTPs \DMA configuration as applied to these simulated data. The top row shows the original simulated image, the returned segmentation map, and the recovered sky map for a zoomed in section. As shown, the sky map is relatively homogeneous, albeit with a brighter sky pedestal than that recorded by some other techniques.

\subsubsection{\LSSTPs 128-Pixel Spline Configuration (\DMB)}
\label{sec:dmstack2020}

The second configuration of the \LSSTPs explored here generates a background map using a $128$ pixel spline fit to all identified background pixels (hereafter referred to as `\DMB', for brevity). This configuration makes use of an alternate detection algorithm which is better suited towards detecting EBL galaxies and LSB flux. In the \LSSTPs \DMA configurations, the detection threshold is fixed at $5\sigma$ across the whole field, where $\sigma^2$ is the variance. However, the variance estimation may be incorrect after convolution owing to the fact that convolution correlates the noise. This leads to an under-performance in the detection algorithm, where recognisable-by-eye sources are not detected \citep[see][Figure 5]{Aihara2019}. To mitigate these effects, we run the high level process \textsc{DetectCoaddSourcesTask} in the \DMB configuration, which adopts the \textsc{DynamicDetectionTask} in the \LSSTPs in lieu of the default detection package used in the \DMA configuration. In the \textsc{DynamicDetectionTask}, mock PSFs are injected into the image in regions of mostly empty sky, ensuring that detected objects are avoided. The ratio between the standard deviation of PSF fluxes of the points to the variance over the effective area of the PSF provides a correction factor to the detection threshold, dynamically adjusting it across the image.

To implement this optimised setup, image data is initially processed in a manner consistent with that described above. Sources are detected, masked, and the remaining simulated imaging is subdivided into $128\times128$ super pixel bins. A $3\sigma$ clipped mean is computed among all non-masked pixels inside each super-pixel, and a two-dimensional 6th order bilinear spline is fit to these super pixel means. This creates a full resolution image of the sky background model. This procedure is designed to be consistent with the operations performed to produce HSC-SSP PDR1 data.

The bottom row of Figure \ref{fig:mask_dmstack} shows the returned segmentation map and recovered background sky for a zoomed in segment of low density simulated data. Whilst fainter sources appear to exhibit segmentation footprints similar in scale to those from the prior \DMA configuration, the segmentation footprint for the brightest source in this region is notably smaller in area. This has a significant impact on the recovered background map, as shown. However, excepting such bright source contamination, the overall recovered sky is significantly fainter here than with the \DMA configuration.

\section{Results}
\label{sec:results}

The procedures above provide background sky estimates and source detection results for a number of simulated datasets. A total of eight HSC-SSP PDR1 fields consisting of a flat and equal to zero sky level are simulated in three binary flavours: fields with faint EBL sources either included or excluded; fields of either a high or a low source density; and fields populated exclusively with either exponential ($n=1$) or de Vaucouleurs ($n=4$) radial profile source types. These data are processed in $10$ software configuration modes: four configurations of the \SExtractor software package, four configurations of the \Gnuastro software suite, and two configurations of the \LSSTPs.

Our full sky estimation results are shown in Table \ref{tab:results}. Each cell in this table shows the estimated mean and standard deviation (in units of nJy per square arcsec\footnote{measured counts on an image are converted to the calibrated unit nanoJanskys using the conversion factor $1\,\mathrm{count}\sim 57.5\,\mathrm{nJy}$. The pixel scale is $0.168\;\mathrm{arcsec}/\mathrm{pixel}$. Results are presented in nanoJanskys to facilitate comparison with other datasets and to allow for negative flux values to be presented in the case of background subtracted data.}) of the recovered background sky model for each flavour of simulated data (per column) and software configuration mode (per row). As a reminder, all simulated datasets are background subtracted (i.e., the input sky was set to zero), so recovered sky values closer to zero are more desirable here. Common trends become apparent, and are explored below. We begin by discussing global sky estimation trends visible between all flavours of binary data type regardless of any specific software configuration mode in Section \ref{sec:flavour}. Following this, we then compare each software configuration mode against one another in the context of output sky levels in Section \ref{sec:mode} to discover desirable sky estimation and source extraction techniques. Finally, we move beyond analysis of the output sky levels and sky maps to take a closer look at the fidelity of output source catalogues in Section \ref{sec:fidelity}.

\begin{table*}
    \setlength{\tabcolsep}{5pt}
    \begin{tabular}{ l | r | r | r | r | r | r | r | r }
    \multicolumn{1}{r}{source population :} & \multicolumn{4}{c}{including all sources} & \multicolumn{4}{c}{excluding faint sources}\\
    \multicolumn{1}{r}{source density :} & \multicolumn{2}{c}{low density} & \multicolumn{2}{c}{high density} & \multicolumn{2}{c}{low density} & \multicolumn{2}{c}{high density}\\
    \multicolumn{1}{r}{source type :} & \multicolumn{1}{c}{1} & \multicolumn{1}{c}{4} & \multicolumn{1}{c}{1} & \multicolumn{1}{c}{4} & \multicolumn{1}{c}{1} & \multicolumn{1}{c}{4} & \multicolumn{1}{c}{1} & \multicolumn{1}{c}{4}\\
    \multicolumn{1}{r}{label :} & \multicolumn{1}{c}{(denlo1a)} & \multicolumn{1}{c}{(denlo4a)} & \multicolumn{1}{c}{(denhi1a)} & \multicolumn{1}{c}{(denhi4a)} & \multicolumn{1}{c}{(denlo1b)} & \multicolumn{1}{c}{(denlo4b)} & \multicolumn{1}{c}{(denhi1b)} & \multicolumn{1}{c}{(denhi4b)}\\
    \hline
    \SExtractor & & & & & & & & \\
    \hspace{25pt}default & $13.9\pm3.1$ & $16.7\pm5.7$ & $20.7\pm3.3$ & $25.4\pm6.8$ & $3.6\pm2.8$ & $5.7\pm5.6$ & $6.0\pm3.1$ & $9.9\pm6.9$\\
    \hspace{25pt}modified & $14.0\pm1.7$ & $16.5\pm2.2$ & $20.9\pm1.4$ & $24.9\pm2.5$ & $3.6\pm1.1$ & $5.6\pm1.8$ & $6.1\pm1.4$ & $9.6\pm2.7$\\
    \hspace{25pt}w. dilated masks & $10.5\pm2.1$ & $13.8\pm3.2$ & $16.0\pm2.3$ & $20.8\pm3.3$ & $0.9\pm1.8$ & $3.2\pm2.8$ & $1.7\pm1.8$ & $5.4\pm3.3$\\
    \hspace{25pt}w. modelled masks & $8.4\pm2.6$ & $8.7\pm3.5$ & $12.4\pm3.3$ & $12.5\pm4.5$ & $-1.0\pm2.3$ & $-1.8\pm3.3$ & $-1.2\pm2.6$ & $-2.5\pm4.3$\\
    \hline
    \Gnuastro & & & & & & & & \\
    \hspace{25pt}default & $9.3\pm1.3$ & $11.6\pm1.5$ & $13.1\pm1.6$ & $16.5\pm1.7$ & $0.4\pm0.9$ & $1.9\pm1.1$ & $0.5\pm0.9$ & $2.6\pm1.3$\\
    \hspace{25pt}modified & $7.8\pm0.8$ & $9.8\pm0.8$ & $11.1\pm1.1$ & $14.5\pm0.8$ & $-0.5\pm0.7$ & $0.4\pm0.7$ & $-0.7\pm0.8$ & $0.7\pm0.6$\\
    \hspace{25pt}w. dilated masks & $8.9\pm1.5$ & $10.7\pm1.5$ & $12.6\pm0.8$ & $15.7\pm1.6$ & $0.1\pm1.2$ & $1.1\pm1.2$ & $0.7\pm1.1$ & $1.2\pm1.3$\\
    \hspace{25pt}w. modelled masks & $9.2\pm1.4$ & $11.2\pm1.5$ & $13.2\pm1.6$ & $15.9\pm1.7$ & $0.2\pm0.9$ & $1.1\pm1.3$ & $0.6\pm1.1$ & $1.9\pm1.3$\\
    \hline
    \LSSTPs & & & & & & & & \\
    \hspace{25pt}\DMA & $12.6\pm0.4$ & $16.1\pm1.4$ & $18.3\pm0.5$ & $23.9\pm1.5$ & $2.2\pm0.3$ & $5.1\pm1.3$ & $3.4\pm0.4$ & $8.1\pm1.4$\\
    \hspace{25pt}\DMB & $10.0\pm1.5$ & $11.9\pm4.9$ & $14.2\pm1.8$ & $19.8\pm5.9$ & $3.2\pm1.3$ & $3.3\pm4.9$ & $0.2\pm1.5$ & $5.3\pm6.1$\\
    \end{tabular}
    \caption{Estimated background statistics as determined by various image characterisation software packages. Each column represents a unique flavour of simulated field as defined in Table \ref{tab:simtab}. Each row represents a different sky estimation software package configuration mode as detailed in Section \ref{sec:sextraction}. Values shown here represent the output mean and standard deviation in units of $\mathrm{nJy\;arcsec}^{-2}$. More accurate (i.e., closer to zero) sky estimation results are consistently recovered in: fields with faint sources accounted for and excluded; fields of relatively low source density, and; in fields comprised of relatively more compact $n=1$ type sources.}
    \label{tab:results}
\end{table*}

\subsubsection{Trends by Flavour: Source Population, Density and Type}
\label{sec:flavour}

Simulated fields with faint EBL sources included always return brighter estimated sky levels than fields with such sources excluded, as might be expected. This is true for all software configuration modes and for all flavours of source population, density and profile type explored here. Many of these faint sources are exceedingly difficult to disentangle from background noise, leading to their being missed from initial source detection and consequently having their flux entered into background estimation routines. Mischaracterisation of such sources is the single largest contributor to sky estimation offsets, as evidenced here. The effect of including faint background sources causes recovered mean background sky levels to be anywhere from $6.8$ to $15.8$ $\mathrm{nJy\;arcsec}^{-2}$ brighter. This equates to recovered mean background sky levels being brighter when faint EBL sources are included by anywhere in the range $1.8\le\sigma\le39.5$. On average, recovered sky levels are $9.4\sigma$ brighter when faint EBL sources are included. Such large offsets evidence the importance in accounting for extremely faint sources in the first instance, and also underlines the significant role they play in modifying output background models. This effect was also found in \citet{Ji2018}, where the sky misestimation produced by a default \SExtractor run was reduced by $8\sigma$ in the case of a random galaxy distribution when using an improved sky background estimator.

The impact of field density on the resultant background model is also significant, with $36$ out of the total $40$ combinations of software configuration modes and flavours explored here returning a brighter mean background sky in the high density regime than in its equivalent low density counterpart. Notably, all simulated fields with faint EBL sources included return a brighter sky in the high density regime. The effect of switching from a low density field to a high density field returns mean background sky values anywhere in the range $2.9$ $\mathrm{nJy\;arcsec}^{-2}$ fainter to $8.6$ $\mathrm{nJy\;arcsec}^{-2}$ brighter. This equates to recovered mean background sky levels in high density regimes falling in the range $2.3\sigma$ fainter to $15.2\sigma$ brighter, with an average of $2.2\sigma$ brighter. When considering only the more realistic `a'-type simulated datasets (i.e., only considering those that also include faint EBL sources), recovered mean background sky levels are brighter in the high density field relative to their low density equivalents anywhere in the range $1.1\le\sigma\le15.2$, with an average value of $3.6\sigma$ brighter. These data show that high density fields act to notably impact the sky estimation routines explored here. A higher source density results in a greater number of undetected galaxies that ultimately contribute to contamination of the sky estimate. There are of the order $200000$ more galaxies in the high density regime as compared to the equivalent low density regime. As a consequence, not only are more galaxies prone to being lost in the EBL, but a higher number of the brightest of galaxies will also throw extra flux into the background in their bright extended wings.

Finally, source profile type also has a notable impact on resultant background map levels. Of the $40$ software configuration modes and flavours compared here, $38$ return a brighter sky in those fields populated with $n=4$ de Vaucouleurs type profiles than in their equivalent $n=1$ exponential counterparts. As with source density, the subset of fields that include faint EBL sources always return a brighter background sky in the $n=4$ scenario relative to the $n=1$ case. Switching from $n=1$ to $n=4$ profiles results in estimated background sky levels returned in the range $1.2$ $\mathrm{nJy\;arcsec}^{-2}$ fainter to $5.6$ $\mathrm{nJy\;arcsec}^{-2}$ brighter. This equates to $n=4$ simulated fields being anywhere from $0.5\sigma$ fainter to $12.4\sigma$ brighter than their $n=1$ equivalents, with an average significance of $2.5\sigma$ brighter. When considering only the more realistic `a'-type simulated datasets with faint EBL sources included, recovered background sky maps are brighter in the $n=4$ case by anywhere in the range $0.0\le\sigma\le11.8$, with an average significance of $2.7\sigma$ brighter. These data show that fields populated with more extended sources often act to increase the resultant estimated background pedestal. Extended de Vaucouleurs type sources contain significantly more flux in their wings than their equivalent exponential-type profiles, with the latter concentrating more of their flux in the relatively high surface brightness core regions. Once a source has been detected, an exponential-type source profile will necessarily leak less flux into the background map (beyond its masked detection footprint region) than an equivalent de Vaucouleurs type source, and therefore will contribute less toward background contamination.

\begin{figure}
    \centering
    \includegraphics[width=\columnwidth]{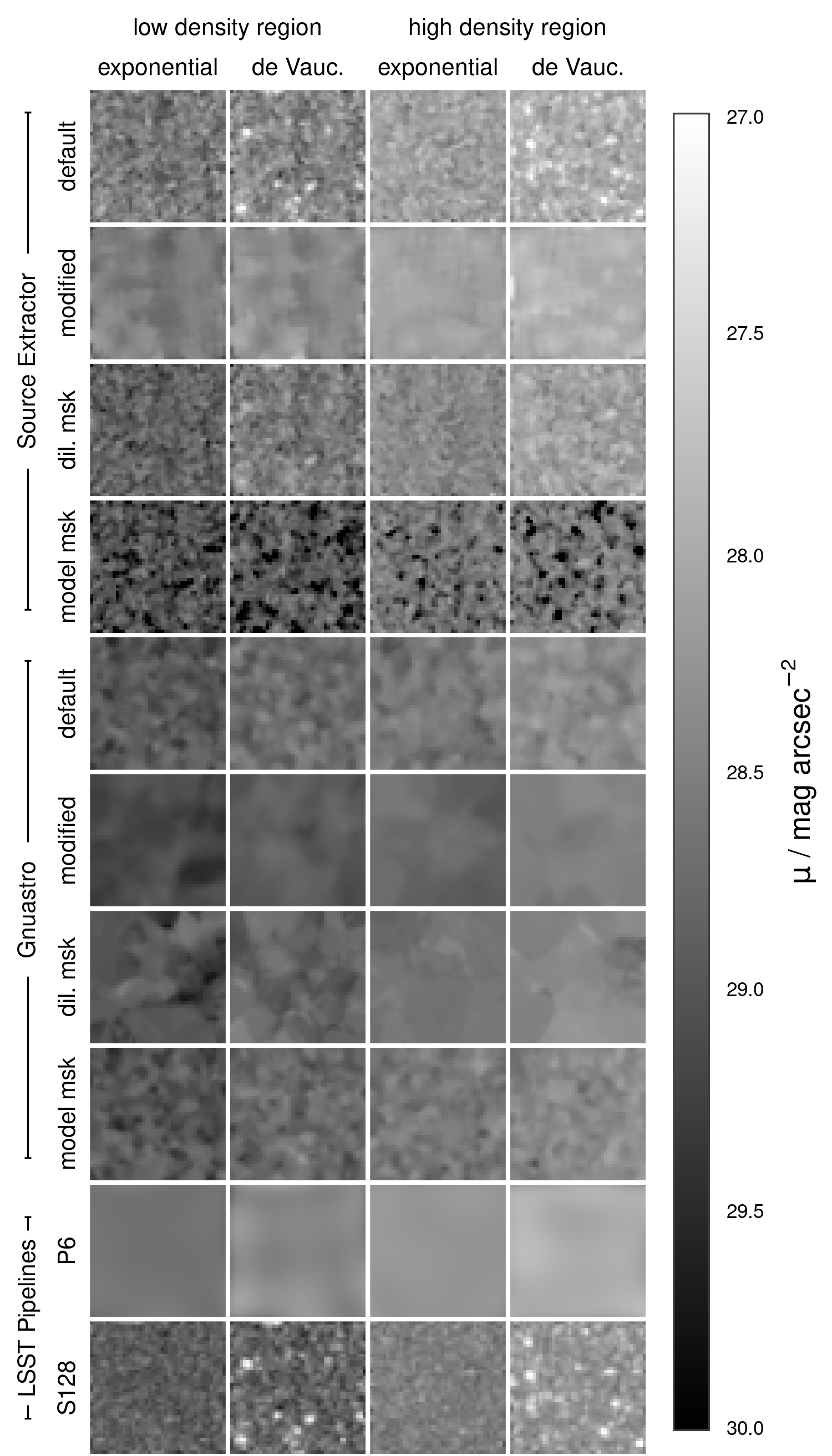}
    \caption{Recovered background sky maps generated by various sky estimation and source extraction configurations. Rows depict different software configuration modes. Columns represent the recovered background sky maps generated from different input simulated datasets. Each original $4200\times4100$ pixel image has been mean-binned into $100\times100$ pixel super-pixels and colour-coded according to its mean surface brightness value, as shown by the outlying colour bar. High density simulated regions are brigher than their low density counterparts by $\Delta\mu_{\mathrm{hi}-\mathrm{lo}}=-0.43\;\mathrm{mag\;arcsec}^{-2}$ on average, owing to the greater number of contaminant sources littering the field. Similarly, simulated regions populated with extended de Vaucouleurs sources are brighter than their exponential source equivalents by $\Delta\mu_{\mathrm{4}-\mathrm{1}}=-0.21\;\mathrm{mag\;arcsec}^{-2}$ on average. Note how some sky estimation software configuration modes also recover a background sky with relatively fewer spatial fluctuations, an indicator that those particular configurations are less impacted by singular bright sources.}
    \label{fig:skymaps}
\end{figure}

We show full recovered background sky maps in Figure \ref{fig:skymaps}, visualising each output background model across the entire field. Each panel shows a different recovered background model as a function of software configuration mode (per-row) and simulated data type flavour (per-column). Each pixel in the original $4200\times4100$ pixel background map has been mean-binned into super-pixels of $100\times100$ pixels each. Super-pixels are colour-coded according to their equivalent surface brightness value, as shown by the associated colour bar. As such, and for clarity, we show here only the four `a'-type full simulated datasets with faint EBL sources included (denlo1a, denlo4a, denhi1a and denhi4a). The spatial trends seen here sync with the global average trends presented above. Background maps derived from high density simulated regions are brighter than their low density counterparts by $\Delta\mu_{\mathrm{hi}-\mathrm{lo}}=-0.43\;\mathrm{mag\;arcsec}^{-2}$ on average. Similarly, background maps derived from fields containing highly extended de Vaucouleurs type ($n=4$) sources are brighter than their exponential ($n=1$) counterparts by $\Delta\mu_{\mathrm{4}-\mathrm{1}}=-0.21\;\mathrm{mag\;arcsec}^{-2}$ on average. These global mean offsets again underline the impact of source density and profile type on background modelling, with denser regions and regions comprised of more extended sources notably more affected by sky estimation contaminant flux. Global mean offsets do not tell the whole story however, with some background models containing a large number of high frequency (small spatial scale) background fluctuations relative to other background models, which remain relatively flat across the field. Such high frequency fluctuations, smaller than the noise level on occasion, are potentially linked to the adopted background mesh size. This is one potential indicator that some configurations are more impacted by singular bright sources than others. In such cases, the detection and measurement of sources may not reach as deep.

\begin{figure*}
    \centering
    \includegraphics[width=0.9\textwidth]{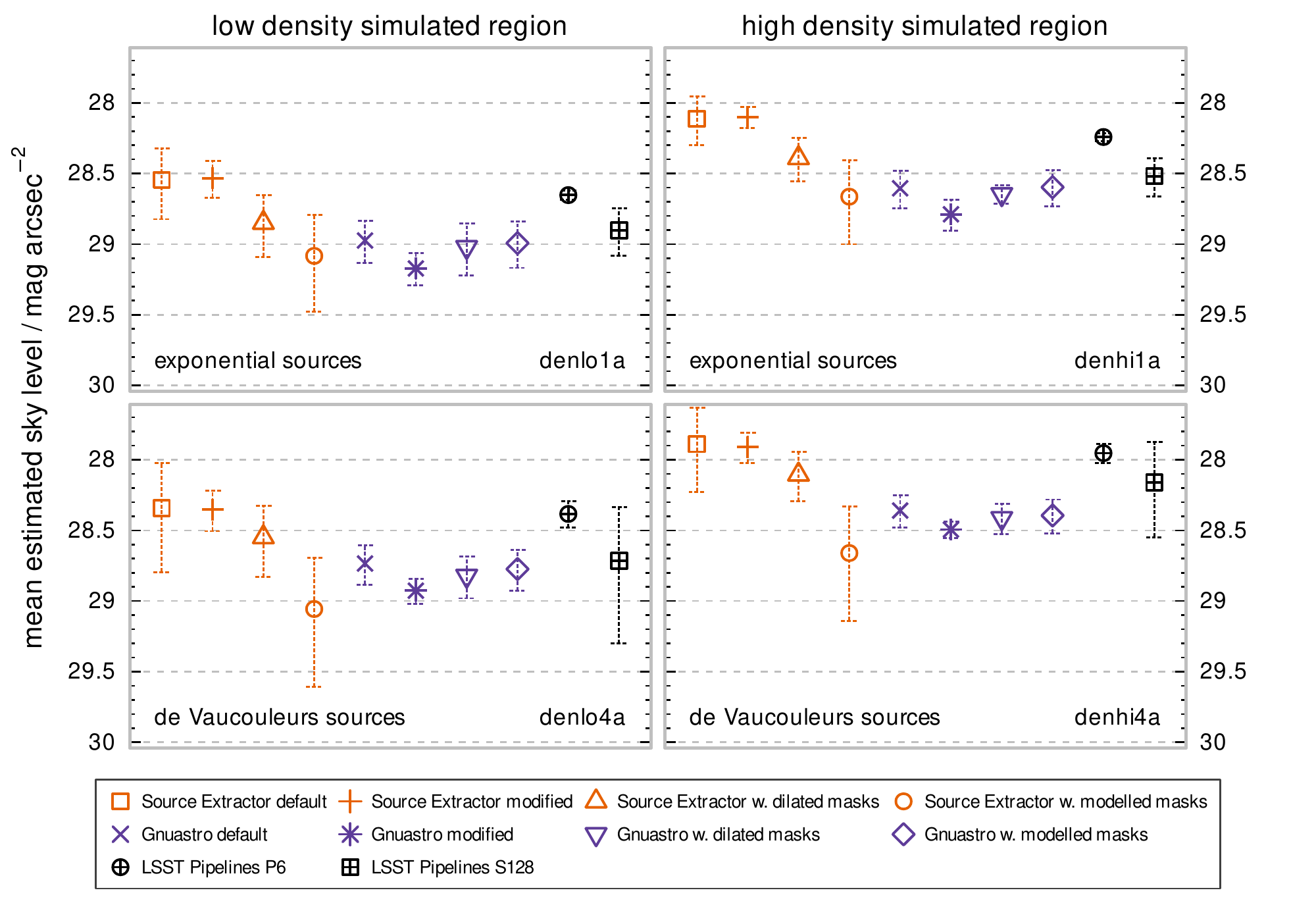}
    \caption{Mean estimated sky levels recovered for our complete simulated data as generated by \SExtractor (orange), \Gnuastro (purple) and the \LSSTPs (black) operating in various configurations (shown within the legend). Distinct software configuration modes are equally spaced along the x-axis within each panel, for clarity. The uppermost two panels represent simulated regions consisting of exponential ($n=1$) sources alone. The lower two panels represent simulated data consisting of de Vaucouleurs ($n=4$) sources alone. The left side panels represent low density simulated regions, whilst the right side panels represent high density simulated regions. Specific simulation data IDs are inset within each panel (see Table \ref{tab:simtab}). Error bars represent $\pm1\sigma$ about the mean. The impact of source density and source profile shape on the resultant sky level estimates shown here is evident, with higher density and extended source types (i.e., denhi4a) consistently brighter by up to an order of magnitude than that of their lower density or less extended simulated imaging counterparts. Various software configurations appear to consistently outperform others, with the discrepancy between the best and worst performing configurations similarly spanning half a magnitude. The shape of sky level offsets between configurations appears relatively static as a function of field type, only varying in the overall amplitude.}
    \label{fig:skyvala}
\end{figure*}

Figure \ref{fig:skyvala} shows mean recovered sky levels in units of $\mathrm{mag\;arcsec}^{-2}$ returned by all software configuration modes (see legend) and full (`a'-type) simulated data flavours (per panel). Software packages are uniquely colour coded, with \SExtractor results in orange, \Gnuastro results in purple, and \LSSTPs results in black. Background levels estimated from low density simulated data are shown in the two panels on the left, and their equivalents from high density regions are shown in the two panels on the right. The units along the $x$ axis are arbitrary, to separate out the points. Data points relating to modified, dilated masking and modelled masking are all based upon a default configuration (e.g., the dilated masking procedure is not based upon a modified configuration). Fields populated with exponential ($n=1$) sources are shown in the uppermost two panels, whilst fields populated with de Vaucouleurs ($n=4$) type sources are shown in the lower two panels. The global trends discussed above are also evident here, with the field density and profile type acting to modulate the resultant background level by up to half a magnitude. As also shown in Table \ref{tab:results} and Figure \ref{fig:skymaps}, Figure \ref{fig:skyvala} additionally acts to highlight the variation in recovered background level as a function of specific software configuration.

\subsubsection{Trends by Software Configuration Mode}
\label{sec:mode}

The faintest sky maps tend to be returned by the modelled masking variant of \SExtractor, all variants of \Gnuastro, and the \DMB variant of the \LSSTPs. The most significant improvement over a default configuration mode is the application of modelled masking to \SExtractor. When this technique is applied, resultant sky maps are fainter than default \SExtractor by up to an order of magnitude. The model masked \SExtractor sky map outputs are typically fainter than all other configuration modes in the extended de Vaucouleurs datasets, albeit with a larger scatter, and fainter than all configuration modes other than modified \Gnuastro in the case of exponential datasets. The modelled masking technique is ideally suited to the relatively smaller segmentation masks produced by \SExtractor. Unlike the dilated masking approach, modelled masking does not reduce the total number of pixels used to estimate the background; a significant benefit in crowded fields. Conversely, out of the box, default \SExtractor returns some of the brightest sky maps; a well known issue, often encountered in the literature. The modifications made to \SExtractor do not appear to significantly alter its resultant output sky level relative to its default configuration, although they do act to reduce the scatter. Our primary \SExtractor modifications are an increase in the background mesh element size (and associated mesh filtering) within which sky is initially estimated, a reduction in the source detection threshold (i.e., allowing more flux to be classified as signal rather than noise), and changes to the smoothing kernel and segmentation parameters. Many of these changes are often adopted in the literature to improve derived catalogue parameters, yet they appear to have little to no impact on the resultant average sky level. The dilated masking \SExtractor technique also performs well, notably improving on the average sky level relative to the default results by up to a quarter of a magnitude. This procedure has significant benefits, as explored in \citet{Ji2018}, helping reduce contamination at the background estimation step by masking more of the potentially contaminant flux in the wings of bright sources.

Excepting modelled masking, sky recovered by \Gnuastro are of the order half a magnitude fainter than \SExtractor when comparing like-for-like configuration setups. The consistently faint sky levels returned by \Gnuastro, regardless of configuration mode, are likely attributed to its bottom-up approach towards source identification and characterisation (in contrast to the top-down flux thresholding techniques employed by \SExtractor and the \LSSTPs) and the resultant extensive segmentation maps which leave little room for further modification. The modifications made to \Gnuastro here (an increase in the background mesh size and a reduction in the growth quantile threshold) act to improve the recovered sky map by a quarter of a magnitude, and simultaneously reducing the scatter (i.e., the background maps are less undulating than their default equivalents). The impact of dilated masking and modelled masking relative to default \Gnuastro appears slight. This is likely due to the extensive pixel masks removing additional information from what was, already, a sparse dataset.

The \DMB variant of the \LSSTPs performs comparably to default \Gnuastro, albeit exhibiting a larger scatter. Conversely, the \DMA variant of the \LSSTPs typically returns the smallest recovered sky level scatter relative to other software modes explored here, with a typical resultant sky level intermediate to the default outputs given by \SExtractor and \Gnuastro, and of the order $\about0.25$ mag brighter than \DMB.

\subsubsection{Trends in Output Source Catalogue Fidelity}
\label{sec:fidelity}

\begin{figure*}
    \centering
    \includegraphics[width=0.9\textwidth]{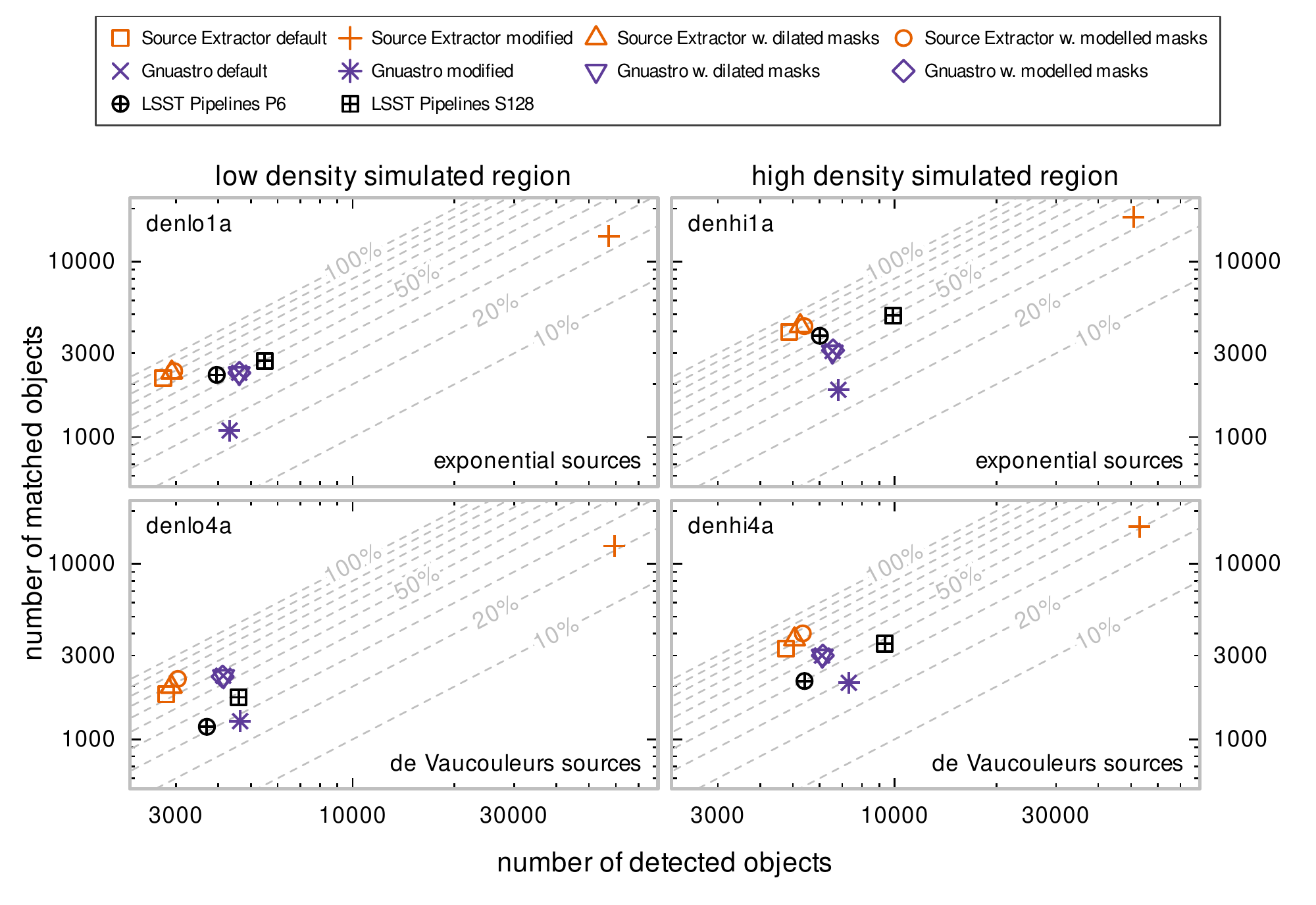}
    \caption{Number of true matched objects as a function of the total number of detected objects given by the various sky estimation and source extraction software configuration modes listed in the legend. Output sources are matched to their corresponding simulated input counterpart if they satisfy three criteria: the $x$ and $y$ output pixel positions fall within $5$ pixels of the input source, and the recovered magnitude is within $\pm0.5$ mag of its input total magnitude. Dashed grey lines represent matched-to-total ratios in $10\%$ increments. The \SExtractor outputs typically return the highest fidelity ratios, with \Gnuastro and the \LSSTPs instead leaning towards recovery of more objects.}
    \label{fig:ndetnmatcha}
\end{figure*}

Whilst the previous sections focus on analysis of the output sky models, the fidelity of the resultant source catalogues is also of interest. To facilitate this analysis we generate a series of matched catalogues for each data output produced here, matching the output catalogue to the known simulated input catalogue using the \texttt{STILTS}\footnote{\url{http://www.star.bris.ac.uk/~mbt/stilts}} software package. We perform a 3-dimensional match between the output and input catalogues based upon $x$ and $y$ pixel coordinates and total magnitude. To match, the output pixel coordinate must be within $5$ pixels of its input location (independently on both axes), and the output magnitude must be within $0.5$ magnitudes of the input magnitude. Magnitude estimates are derived from the Kron-like $\mathrm{FLUX\_AUTO}$ value with \SExtractor outputs, the $\mathrm{BRIGHTNESS}$ value with \Gnuastro outputs, and the $\mathrm{base\_SdssShape\_instFlux}$ value with \LSSTPs outputs. Figure \ref{fig:ndetnmatcha} shows the number of true matched objects ($y$-axis) as a function of the total number of detected objects ($x$-axis) for various software configuration modes listed in the legend and simulated data type flavours. Data point type and colour coding is identical to that previously used. Slanted dashed grey lines represent different fidelity fractions in increments of $10\%$.

We find that \SExtractor default-based modes of operation (i.e., default, dilated masking and modelled masking) produce the highest fidelity catalogues, with approximately three quarters of output sources finding a corresponding true match in their equivalent original input catalogue. These catalogues also return the fewest total number of objects however, indicating that default \SExtractor is well optimised towards returning the largest of sources in any given field of view. Modified \SExtractor operations drastically increase both the total number of output sources (by a factor of $\about20$ on average) and the number of matched sources (by a factor of $\about6$ on average), with a corresponding reduced fidelity ratio of around a quarter.

Default-based \Gnuastro operational modes (i.e., default, dilated masking and modelled masking) similarly cluster together, with a matched fidelity ratio of $\about50\%$. These outputs do find a larger number of total sources than their equivalent \SExtractor runs, however, fewer of these are able to be matched to a known simulated input. Modifications made to \Gnuastro act to somewhat increase the total number of detected objects (by $\about10\%$ on average, except in the denlo1a scenario where a slight decrease is recorded), however, any increase is at the expense of source fidelity, with a drastically reduced recovered fidelity ratio of $\about30\%$ or lower.

The two \LSSTPs modes of operation perform fairly consistently and comparably to the default \Gnuastro outputs, returning fidelity ratios in the range $\about30\%$ to $\about60\%$. Simulated regions populated with exponential sources tend to lead to higher fidelity output \LSSTPs catalogues on average, with a larger number of spurious sources for no discernible increase in total numbers found in the equivalent de Vaucouleurs populated fields. Progressing from the \DMA mode of operation to the \DMB variant approximately maintains a similar fidelity level, improving in both total number of detected objects and the equivalent number of true matched sources.

\begin{figure*}
    \centering
    \includegraphics[width=1\textwidth]{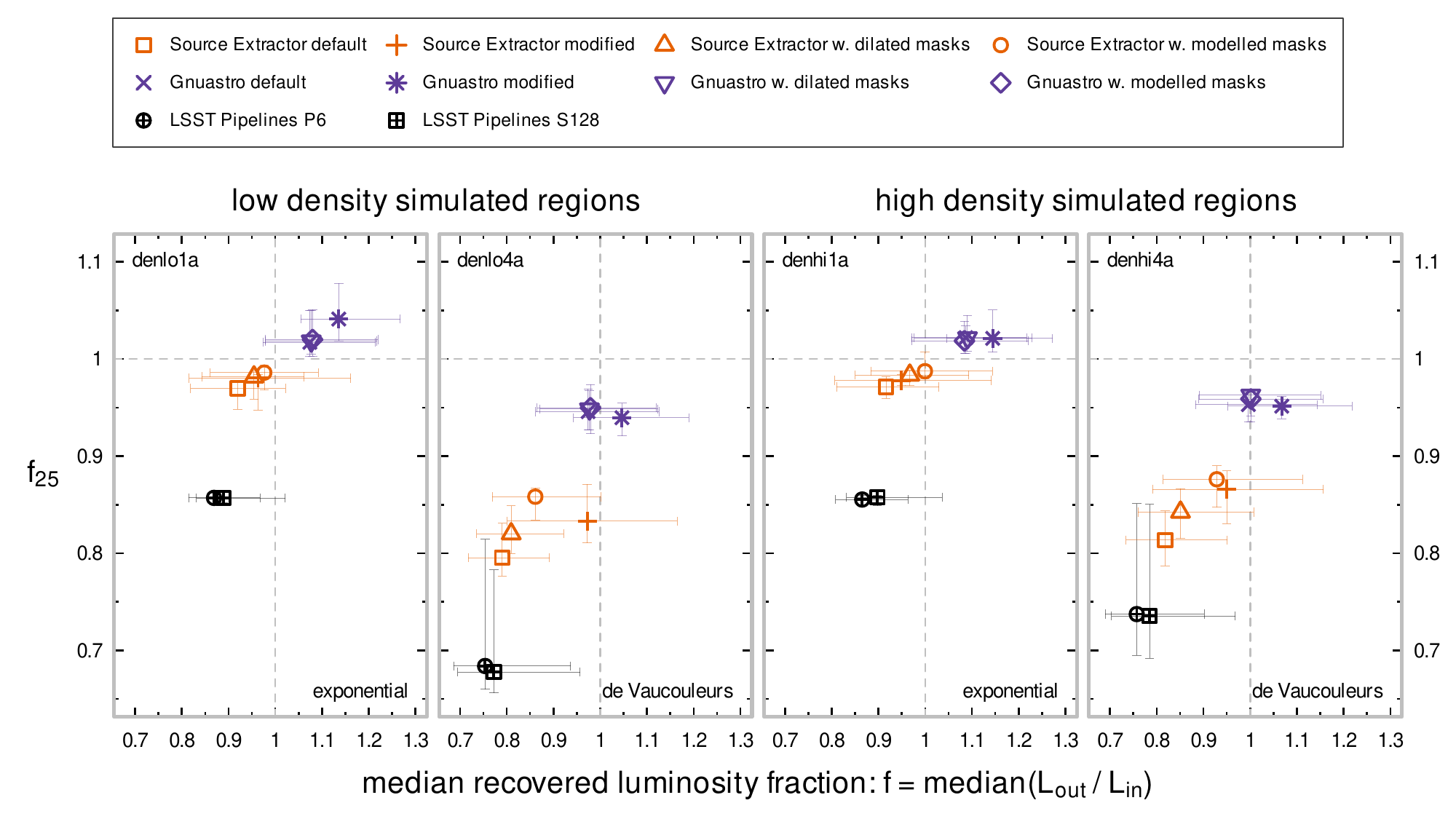}
    \caption{Median recovered luminosity fractions (output/input) for matched sources as determined by various source extraction techniques listed in the accompanying legend. Each panel represents a different simulated input dataset flavour, as indicated by the supporting label text. Within each panel, the $x$-axis represents the median recovered luminosity fraction $f$ for all matched sources, where $f=\mathrm{median}(\mathrm{L}_{\mathrm{out}}/\mathrm{L}_{\mathrm{in}})$, whilst the $y$-axis represents the median recovered luminosity fraction for the 25 largest sources alone ($f_{25}$, i.e., those sources that have the largest impact on background sky estimation and are equivalently most severely impacted by it). Error bars represent $25^{\mathrm{th}}$ and $75^{\mathrm{th}}$ percentile ranges. These results show that \SExtractor performs well with exponential type sources, accurately recovering close to $100\%$ of the simulated input flux, whilst under-estimating total flux for the more extended de Vaucouleurs type sources by $\about15\%$, as expected. Conversely, \Gnuastro performs well for the majority of de Vaucouleurs types, missing $\about\mathrm{few}\%$ of the flux for the largest de Vaucouleurs type sources, however, incorrectly over-estimating flux for exponential type sources by $\about10\%$, and always over-estimating flux when run its modified configuration mode.}
    \label{fig:lumfrac}
\end{figure*}

In addition to overall numbers, we also want to explore whether the output magnitudes being returned by our software package of choice accurately portray the true input magnitudes in our simulated datasets. Figure \ref{fig:lumfrac} shows median recovered luminosity fractions for the matched source subsets described above. The $x$-axis shows the median recovered luminosity fraction $f$ for all matched sources, where $f=\mathrm{median}(\mathrm{L}_{\mathrm{out}}/\mathrm{L}_{\mathrm{in}})$. As $\mathrm{L}_{\mathrm{in}}$ is a pure simulated input quantity, whereas $\mathrm{L}_{\mathrm{out}}$ is a measured output quantity, ideal values of $f$ should ideally fall in the range $0<f<1$, with higher values preferred. The $y$-axis represents the median recovered luminosity fraction for the 25 largest matched sources alone ($f_{25}$). The latter of these two parameters is of particular interest, as these sources have the largest individual impact on background sky estimation results. We find a strong bimodality in $f$ as a function of source profile type, as might naively be expected, with similar overall trends observed in both regardless of field density.

Here it becomes apparent that \SExtractor appears optimally configured for the accurate characterisation of exponential type sources, with all \SExtractor software variants clustering just below $f=1$ and $f_{25}=1$, i.e., their median output luminosities accurately represent their inputs. Recovered \SExtractor results for de Vaucouleurs populated fields are less consistent, with typical median recovered luminosity fractions for de Vaucouleurs type sources in both field density regimes returning $f\sim0.85$ (and equivalently for the largest 25 sources), i.e., $\about15\%$ of their input flux is missing in its output flux estimate. Some flux loss for more extended sources is to be expected, as discussed above, so these results are not too surprising. The value of $f$ varies significantly with \SExtractor software configuration, with the default mode of operation performing worst, and the modified and modelled masking procedures performing significantly better on average. Several such configuration modes push the \SExtractor results up to $f\sim0.95$, however, all continue to struggle for the largest 25 sources.

Conversely, the \Gnuastro package appears designed for the optimal extraction of de Vaucouleurs type sources, more accurately picking up the LSB flux in the wings of these objects. All default-based configuration variants of \Gnuastro return $f\sim1$ in the case of de Vaucouleurs type simulated datasets. Modified \Gnuastro tends to report an \textit{excess} of flux in de Vaucouleurs type objects by $\about5\%$. This is likely due to our choice to lower the detection growth quantile parameter to $0.7$ from its default value of $0.9$, which improves the sky estimate at the expense of object fidelity. The recovered flux for the 25 largest de Vaucouleurs sources falls somewhat lower, at $f_{25}\sim0.95$ (i.e., the brightest 25 sources are reported to be $\about5\%$ fainter than their input \Sersic magnitude). This in-built facility to identify LSB flux around de Vaucouleurs sources does not similarly lead to a successful outcome in the case of exponential-type populated fields. Recovered exponential source flux appears to be over-estimated by $\about10\%$ (and $\about15\%$ in the case of a modified \Gnuastro configuration). Unfortunately, it appears that \Gnuastro tends towards inaccurately incorporating flux from surrounding faint EBL sources and noise spikes into these exponential source footprints, falsely inflating their resultant flux estimates above and beyond their known simulated input.

Of all three software packages explored here, results from the \LSSTPs consistently recover the lowest flux fractions. In the exponential scenario, recovered fluxes are $\about10\%$ lower then their known inputs. When considering the largest 25 sources alone, recovered fluxes are $\about15\%$ lower than their known inputs. The situation is notably more extreme in the case of fields populated with de Vaucouleurs sources, with equivalent recovered values of $f\sim0.8$ and $f_{25}\sim0.7$, respectively. This potentially indicates that a significant amount of flux from these sources is leaking into the resultant sky estimate, severely contaminating resultant sky maps.

\section{Conclusions}
\label{sec:conclusions}

Using a series of simulated images designed to mimic a wide range of Hyper Suprime-Cam Subaru Strategic Program survey data, we examine how varying the galaxy source population, density, and surface brightness profile type influences the background measured across a typical patch of sky. In an era of increasingly deep, wide-area survey science, the potential for LSB discovery is greater than ever before. The potential pitfalls of background over-subtraction on LSB science however is significant \citep[see for example][Figure 5]{Aihara2019}, with much of this faint flux at severe risk of contamination or destruction during image processing.

This study makes use of three contemporary image processing software packages: \SExtractor, \Gnuastro, and the \LSSTSPs. Each package is operated in multiple configuration modes to explore optimal strategies towards recovering a fainter overall background sky, and ones in which low surface brightness features are maximally preserved. Both \SExtractor and \Gnuastro are operated in four distinct modes: default, modified, with dilated masking, and with modelled masking. The default mode of operation explores how to software performs `out-of-the-box', whilst the modified mode of operation makes a number of changes commonly recommended in the literature which are designed to improve sky estimates and source catalogue fidelity. The dilated masking procedure follows the methodology initially outlined in \citet{Ji2018}, whereby the source flux masks are artificially expanded around sources to mask signal flux leakage which would otherwise contaminate the background model. The amount of mask expansion is related to the magnitude of the underlying source. Finally, we present here the modelled masking approach, whereby single-\Sersic parametric models are fit to a subset of detected sources and subtracted. This approach may be considered analogous to a weighted pixel mask technique, whereby an attempt is made to mask only the signal flux out into the noise, rather than using a fixed isophotal threshold mask as with a traditional segmentation map. Following subtraction of the model plane, the sky level is then estimated from the residual image. Finally, the \LSSTPs are operated in two distinct modes of operation: a sixth-degree polynomial background mapping approach (\DMA) and an updated modified 128-pixel spline fitting configuration (\DMB).

We summarise here our principal conclusions on the nature of background sky estimation as a function of source population parameters and data processing methodology:

1. The faintest and most accurate sky maps are returned when using the modelled masked variant of \SExtractor, all variants of \Gnuastro, and the \DMB variant of the \LSSTPs. These software configurations produce sky maps of the order half a magnitude fainter than a default run of \SExtractor on the same data. The dilated masking variant of \SExtractor also performs well, returning sky maps of the order a quarter of a magnitude fainter than a default run of \SExtractor. The apparent success in estimating the sky for techniques which artificially inflate the pixel masks around bright sources is evidence for the significant issue of flux leakage in contaminating background estimation efforts. Should such LSB flux be considered noise, rather than signal, its chances of being subtracted away in the background model are significant.

2. The `top-down' source extraction software package \SExtractor identifies a far smaller fraction of pixels in any given field as signal than do the `bottom-up' routines utilised by \Gnuastro. Top-down tools attempt to identify bright source peaks above a limiting isophote, with any flux not successfully detected ultimately entering into the estimate of the background sky. Conversely, the bottom-up routines of \Gnuastro instead attempt to identify contiguous regions of noise, assigning everything else to a detection. In densely packed environments, or fields comprised of more extended sources, bottom-up techniques such as those utilised within \Gnuastro result in most sources being linked by `flux-bridges'. Whilst this may have the potential to accurately capture most instances of flux leakage into the background model, this also leaves behind a relatively smaller area of non-signal sky upon which to base a reliable sky model. A key strength of the masked modelling approach when applied to top-down sky estimation techniques such as \SExtractor is that no additional pixels are required to be masked than would otherwise be by default. This is a crucial consideration for future deep surveys, where the source density will only increase.

3. Adopting optimal source detection configurations for the three software packages given above produces multiple outcomes. Modifications explored here include varying the mesh grid size to be used for determination of the sky level on an image, changing the source detection and analysis thresholds, and modifying the smoothing kernels used to associate regions of noise, amongst others (see text for further details). A negligible change in the mean sky level was found when comparing modified \SExtractor against its default configuration. Both the \Gnuastro and \LSSTPs however recover sky levels up to a quarter of a magnitude fainter than that found previously. We note however that the background map returned by modified \SExtractor and modified \Gnuastro are notably flatter than that returned by the \DMB variant of the \LSSTPs, with the latter of these showing distinct background peaks at the position of bright sources. The ideal background map should not display any such peaks, with the presence of these instead indicating that a significant amount of light from singular bright sources has inadvertently leaked through and contaminated the background map estimation.

4. Output source catalogues provided by \SExtractor tend to recover the fewest total number of sources relative to the other software packages explored here. Of these catalogues however, $\about75\%$ of sources can be matched to a known simulated input source. A notable exception is in the case of the modified \SExtractor configuration, whereby the total number of detections are vastly increased by a factor of $\about20$ yet the number of matches by a factor of only $\about6$. Output \Gnuastro source catalogues tend to recover slightly more sources than provided by \SExtractor, yet at the expense of match-to-total ratio\footnote{As noted in Section \ref{sec:gnuastrodefault}, here we opt to make use of the \textit{object} catalogues provided by default by the \Gnuastro \Segment software. For the data explored here, the associated \textit{clump} catalogue (the high signal-to-noise cores of the brightest sources) tends to return far fewer sources. For example, for the \textit{denhi4b} simulated dataset, the default output \Gnuastro catalogue contains 3720 clumps and 5261 objects. This is caused by some `signal' pixel regions not containing any significant flux peak to constitute a clump. Furthermore, some clumps are merged into a single object because the maximum value of the signal-to-noise in the `river' pixels between two clumps falls below the default value of $1$ (a value which may be over-ridden by the \Segment parameter \textit{objbordersn}). Utilising a clump catalogue instead of an object catalogue may well return a higher match-to-total ratio, however, the clump catalogue by definition does not encompass all non-noise pixels detected in an image. Conversely, the object catalogue \textit{is} directly related to those pixels which were not found to be noise (information which is used to assist in sky estimation), therefore making it the more appropriate choice for our further analysis in this paper. It is tempting to consider making use of a modified object catalogue, setting \textit{objbordersn} to some very high value (i.e., never merging multiple clumps together into one object). This would have the advantage of returning the maximum number of objects at the expense of never merging together clumps which are sub-components of a single object, as is the case with, for example, a star-forming knot within an extended source. As some form of clump merging will always be required in real data analysis, we do not consider this a viable approach to further explore in this study.}, with a reduced fidelity factor of $\about50\%$. As with \SExtractor, modified \Gnuastro catalogues recover a higher total number of sources than by default (excepting in a low density exponential environment) and also exhibit significantly worse fidelity ratios of $\about30\%$. Both iterations of the \LSSTPs explored here return fidelity ratios of $\about50\%$, performing better in high-density exponential type simulated regions, and worse in low-density de Vaucouleurs type fields. Switching from the \DMA variant to the \DMB variant approximately maintains the fidelity ratio whilst increasing the total number of detections. These results suggest that top-down tools such as \SExtractor have been designed with the detection of the brightest of sources in mind, resulting in a reduction in the number of false positives than would otherwise be returned.

5. The \SExtractor routines successfully capture almost all of the light associated with exponential type simulated sources, in both high and low density environments. Exponential source fluxes output by default \SExtractor modes of operation return $\about95\%$ of the known input flux. This rises up to $\about100\%$ of the known input flux being correctly identified in the case of \SExtractor with modelled masking. For de Vaucouleurs type galaxies, default \SExtractor returns a lower recovered luminosity fraction of $\about80\%$ (i.e., $\about20\%$ of the known input light is missing), rising to $\about90\%$ in the case of \SExtractor with modelled masks and higher still when utilising a modified configuration. From this we deduce that \SExtractor appears optimally configured for the detection of more compact sources, with modifications or the application of modelled masking required to perform equivalently with more extended sources.

6. The \Gnuastro routines successfully capture almost all of the light associated with extended de Vaucouleurs type sources, in both high and low density environments. Output fluxes for de Vaucouleurs sources are within a few percent of their known input flux, with the exception of \Gnuastro operating in a modified configuration which results in an overestimation of the source flux by $\about5\%$. In the case of exponential simulated inputs, \Gnuastro tends to overestimate the measure of flux by $\about10\%$, excepting \Gnuastro operating in a modified configuration whereby flux estimates are $\about15\%$ brighter than their known inputs. This tendency for recovered exponential flux estimates to be $\about10\%$ brighter than their known simulated input is a serious cause for concern, implying that a significant amount of contaminant flux is being included at the source detection phase. As \Gnuastro has a wide range of configuration parameters, further such modifications in addition to those tested here should be sought out to attempt to mitigate this undesirable effect.

7. Both variants of the \LSSTPs tested here recover somewhat lower flux fractions than had been initially input. In the case of exponential type sources, the \LSSTPs recover $\about90\%$ of the input light. This reduces to $\about75\%$ of the input light in the case of de Vaucouleurs type sources. Slight $\about\mathrm{few}\%$ improvements are noted when switching from the \DMA to the \DMB variant, in both high and low density environments. As the top-down sky estimation and source extraction philosophies behind \SExtractor and the \LSSTPs are similar in nature, it is likely that application of either the dilated masking procedure or the modelled masking approach would have a similarly beneficial effect.

8. The faint extragalactic background light (EBL) component, largely associated with undetectable sources, plays a significant role in modifying the overall recovered sky level pedestal. Simulated fields containing an EBL component are brighter in the range $1.8\le\sigma\le39.5$ than an equivalent field with the EBL component removed, with an average of $8.3\sigma$ brighter. As such sources are ubiquitous, their impact on local background estimation is simpler to predict at present. Future surveys probing to deeper depths may ultimately require a more considered approach towards characterising such a component.

9. The recovered background sky is considerably biased bright in regions of high source density relative to less densely packed fields. Simulated high density environments return a background sky brighter than their low density equivalents in the range $1.1\le\sigma\le15.2$; $3.3\sigma$ on average. The additional flux thrown out into the background map by these extra sources acts to notably contaminate sky estimation routines, with denser regions more severely impacted. This is an important consideration for analysis of clustered environments, and an increasingly important consideration for future deep surveys such as the LSST.

10. Regions of sky containing extended de Vaucouleurs type galaxies are more difficult to derive an accurate background estimate for than regions populated with less extensive exponential type sources. Of these two types, fields containing de Vaucouleurs sources return sky estimates that are brighter than those equivalents from fields containing exponential sources in the range $0.0\le\sigma\le11.8$, with an average significance of $2.4\sigma$ brighter. The de Vaucouleurs surface brightness profile, often associated with spheroidal elliptical galaxies, throws out significantly more flux into its wings than does an exponential profile. Other sources with bright extended wings, such as the extended halo of a PSF, also have the potential for a significant fraction of their light to be left undetected and unaccounted for beneath the limiting isophote, further contributing to background sky estimate contamination. These results highlight the importance for future optimal LSB detection strategies to be sensitive to the form of the bright sources they neighbour.

The choice of software package and its configuration is based on the specific science question one wishes to answer. In the case of LSB science, the ability to recover a faint and flat sky is paramount. Of the software types and configurations tested here, the results of this study suggest that the faintest and most reliable sky maps are recovered when the modelled masking technique is applied to data processed by \SExtractor, producing the largest average sky level improvements of any technique tested herein. The distinct advantage of this approach is that it requires significantly smaller detection footprints to operate successfully, with the dual bonus advantages of a lower false-positive detection rate and the tendency not to over-estimate source flux in output source catalogues. The top-down approaches of \SExtractor and the \LSSTPs seemingly lend themselves to a greater range of configurational possibility. Conversely, bottom-up source detection routines as utilised by \Gnuastro also return faint sky maps, however, at the expense of masking a much larger fraction of the field of view. Such a philosophy may not be sustainable as we enter an era of increasingly deep, wide-area survey science, wherein rising source densities will make it harder to find `empty' patches of sky from which to derive a background model. The accuracy of output background sky models are found to vary as a function of source population, density, and type, with different considerations to be made in the case of each. Subsequent modifications to contemporary image processing pipelines such as the the \RO \LSSTSPs which address these concerns would therefore be of significant benefit, not only to the LSB science community, but to the entire scientific user base as a whole.

\newpage
\section*{Acknowledgements}
\label{sec:acknowledgements}

We thank Robert Armstrong, Eli Rykoff and Jeff Carlin for their significant and helpful comments on earlier drafts of this paper. This material is based upon work supported in part by the National Science Foundation (NSF) through Cooperative Agreement 1258333 managed by the Association of Universities for Research in Astronomy (AURA), and the Department of Energy under Contract No. DE-AC02-76SF00515 with the SLAC National Accelerator Laboratory. Additional funding for Rubin Observatory comes from private donations, grants to universities, and in-kind support from LSSTC Institutional Members. This research was also supported by Department of Energy grant DE-SC0009999 and NSF/AURA grant N56981C.

The Hyper Suprime-Cam (HSC) collaboration includes the astronomical communities of Japan and Taiwan, and Princeton University. The HSC instrumentation and software were developed by the National Astronomical Observatory of Japan (NAOJ), the Kavli Institute for the Physics and Mathematics of the Universe (Kavli IPMU), the University of Tokyo, the High Energy Accelerator Research Organization (KEK), the Academia Sinica Institute for Astronomy and Astrophysics in Taiwan (ASIAA), and Princeton University. Funding was contributed by the FIRST program from Japanese Cabinet Office, the Ministry of Education, Culture, Sports, Science and Technology (MEXT), the Japan Society for the Promotion of Science (JSPS), Japan Science and Technology Agency (JST), the Toray Science Foundation, NAOJ, Kavli IPMU, KEK, ASIAA, and Princeton University.

This paper makes use of software developed for the Large Synoptic Survey Telescope. We thank the LSST Project for making their code available as free software at http://dm.lsst.org.

The Pan-STARRS1 Surveys (PS1) have been made possible through contributions of the Institute for Astronomy, the University of Hawaii, the Pan-STARRS Project Office, the Max-Planck Society and its participating institutes, the Max Planck Institute for Astronomy, Heidelberg and the Max Planck Institute for Extraterrestrial Physics, Garching, The Johns Hopkins University, Durham University, the University of Edinburgh, Queen’s University Belfast, the Harvard-Smithsonian Center for Astrophysics, the Las Cumbres Observatory Global Telescope Network Incorporated, the National Central University of Taiwan, the Space Telescope Science Institute, the National Aeronautics and Space Administration under Grant No. NNX08AR22G issued through the Planetary Science Division of the NASA Science Mission Directorate, the National Science Foundation under Grant No. AST-1238877, the University of Maryland, and Eotvos Lorand University (ELTE) and the Los Alamos National Laboratory.

This work was partly done using GNU Astronomy Utilities (Gnuastro, ascl.net/1801.009) version 0.17. Work on Gnuastro has been funded by the Japanese Ministry of Education, Culture, Sports, Science, and Technology (MEXT) scholarship and its Grant-in-Aid for Scientific Research (21244012, 24253003), the European Research Council (ERC) advanced grant 339659-MUSICOS, European Union’s Horizon 2020 research and innovation programme under Marie Sklodowska-Curie grant agreement No 721463 to the SUNDIAL ITN, and from the Spanish Ministry of Economy and Competitiveness (MINECO) under grant number AYA2016-76219-P.



\section*{Data Availability}
\label{sec:data}

All data and associated code produced for use within this study has been archived on the \textsc{Zenodo} open-access repository at \url{https://doi.org/10.5281/zenodo.7067465}.



\bibliographystyle{mnras}
\bibliography{bibtex}



\appendix

\section{The Hyper Suprime-Cam Subaru Strategic Program}
\label{sec:hsc}

To ascertain the veracity of any given sky subtraction methodology, a real-world observed dataset is required upon which to base a subsequent series of complex simulations. To fill this role, we opt for data taken from the Hyper Suprime-Cam \citep[HSC,][]{Miyazaki2012,Miyazaki2018} Subaru Strategic Program \citep[SSP,][]{Aihara2018a}. The HSC is an $870$ megapixel prime focus camera consisting of $116$ 2k $\times$ 4k $0.168\;\mathrm{arcsec}/\mathrm{pixel}$ CCDs mounted on the $8.2\mathrm{m}$ Subaru telescope at Mauna Kea in Hawaii. HSC-SSP survey data is segregated into three distinct layers: Wide ($1400\;\deg^2$), Deep ($27\;\deg^2$) and UltraDeep ($3.5\;\deg^2$), the former of which is utilised here. Wide-layer optical imaging is provided in the $g$, $r$, $i$, $z$ and $y$ broad-band filters, with a typical $5\sigma$ point source depth of $r\about26\;\mathrm{mag}$. Unless otherwise stated, primary analyses throughout the remainder of this study are based upon HSC-SSP $r$-band data.

The HSC-SSP Public Data Release 1 \citep[PDR1\footnote{\url{https://hsc-release.mtk.nao.ac.jp}},][]{Aihara2018b} dataset used in this study spans $\about100\;\deg^2$ from the initial 12 month observing period, providing full-depth, reduced, calibrated imaging and relevant associated catalogue data. PDR1 full-depth co-added imaging is stored in a hierarchical format. Top level $2.82\;\deg^2$ `tracts' are organised to efficiently tile across the entire survey region, with each tract further subdivided into $9\times9$ slightly overlapping `patches' of $0.038\;\deg^2$. HSC-SSP PDR1 imaging data is processed using the \textsc{hscPipe~4} pipeline \citep{Bosch2018,Huang2018}; an HSC-specific customisation of the \RO \citep[formerly Large Synoptic Survey Telescope,][]{Ivezic2008} \LSSTSPs Data Management (DM) software stack \citep{Juric2017}, which itself is based upon concepts initially developed for use within the Sloan Digital Sky Survey (SDSS) Photo pipeline \citep{Lupton2001}. The \textsc{hscPipe~4} pipeline is tasked with a number of processing steps, summarised briefly here: initial raw data acquisition; single-visit CCD image calibration (e.g., bad pixel mask, bias correction, flat-fielding); sky estimation; PSF generation; source extraction, segmentation and deblending; pixel resampling, and; image co-addition. Full details of this processing pipeline, notably the sky estimation methodology, can be found in \citet{Bosch2018} and references therein. The significant depth, multi-wavelength aspect, wide area coverage, and data-processing synergies with prior and future imaging surveys make the HSC-SSP an ideal candidate for further exploration of contemporary sky-subtraction algorithms. We note that a subsequent data release has occurred since the primary analyses within this paper were conducted. Whilst improvements have been made \citep[see][Figure 5]{Aihara2019}, the underlying effects explored in this paper remain valid here and for other comparable surveys.

\subsection{High/Low Density Sample Regions}
\label{sec:sample}

It is fair to assume that sky estimation routines perform differently in regions of differing source density. To that end, we require two sample regions upon which to base our subsequent simulations: one of relatively low source density and one of relatively high source density. A pseudo-random list of $57$ unique tract-patch pairs\footnote{All $57$ tract-patch pairs: 8279-07, 8280-18, 8281-28, 8282-37, 8283-38, 8284-47, 8285-27, 8520-03, 8521-46, 8522-50, 8523-51, 8524-52, 8525-53, 8526-21, 8762-00, 8763-14, 8764-15, 8765-22, 8766-23, 8767-24, 8768-61, 9314-08, 9315-48, 9316-57, 9317-58, 9318-17, 9346-16, 9347-41, 9348-42, 9349-26, 9370-02, 9371-36, 9372-43, 9373-44, 9374-45, 9375-31, 9557-06, 9558-55, 9559-56, 9560-62, 9561-40, 9562-54, 9589-05, 9590-32, 9591-33, 9592-20, 9613-01, 9614-30, 9615-34, 9616-35, 9617-25, 9618-60, 9800-04, 9801-11, 9802-12, 9803-13, 9804-10.} is generated from the HSC-SSP PDR1 dataset, providing an unbiased sample of HSC imaging data across a wide area. The $r$-band coadded calibrated exposure (\texttt{deepCoadd\_calexp}) data products containing the \textsc{hscPipe~4} background-subtracted science images, pixel masks and variance maps are downloaded for each of these tract-patch pairs.

In order to provide an initial estimate of field density for each tract-patch region, the \SExtractor \citep{Bertin1996} source detection software package (version 2.25.3) is applied to each science image. The \SExtractor package is operated using mostly the default configuration setup\footnote{Default \SExtractor configuration file and parameter lists are generated on the command line using the arguments \texttt{-dd} and \texttt{-dp}, respectively. Default \SExtractor neural network and convolution files are provided with the software.}, with the exception of detection threshold which is fixed to an absolute value of $0.07$ counts\footnote{Using the standard HSC-SSP flux to nanoJansky conversion factor, $0.07$ counts corresponds to $\about4\;\mathrm{nJy}$.}. The default operation of \SExtractor defines sources as those with significant flux above a $1.5\sigma$ background threshold. Maintaining such a relative threshold across each tract-patch pair would introduce a sky estimation dependency on source density, and unfairly bias against regions where the initial estimation of the background is inaccurate. Preliminary testing of our dataset found that an absolute threshold of $0.07$ counts is sufficient to provide an unbiased estimate of source density for each tract-patch pair in these data. See Section \ref{sec:thresholds} for more information on the derivation of this absolute detection threshold level. Using this absolute detection threshold, \SExtractor is once again ran on all $57$ tract-patch regions, generating an estimate of field density (number of objects per square arcminute) in addition to a host of per-object parameters particular to each detected object\footnote{Our output \SExtractor catalogue parameters are: \texttt{NUMBER}, \texttt{FLUX\_AUTO}, \texttt{MAG\_AUTO}, \texttt{KRON\_RADIUS}, \texttt{PETRO\_RADIUS}, \texttt{BACKGROUND}, \texttt{THRESHOLD}, \texttt{X\_IMAGE}, \texttt{Y\_IMAGE}, \texttt{A\_IMAGE}, \texttt{B\_IMAGE}, \texttt{THETA\_IMAGE}, \texttt{ELLIPTICITY}, \texttt{CLASS\_STAR} and \texttt{FLUX\_RADIUS}. A full description of each parameter may be found in \citet{Bertin1996}.}.

\begin{figure}
    \centering
    \includegraphics[width=\columnwidth]{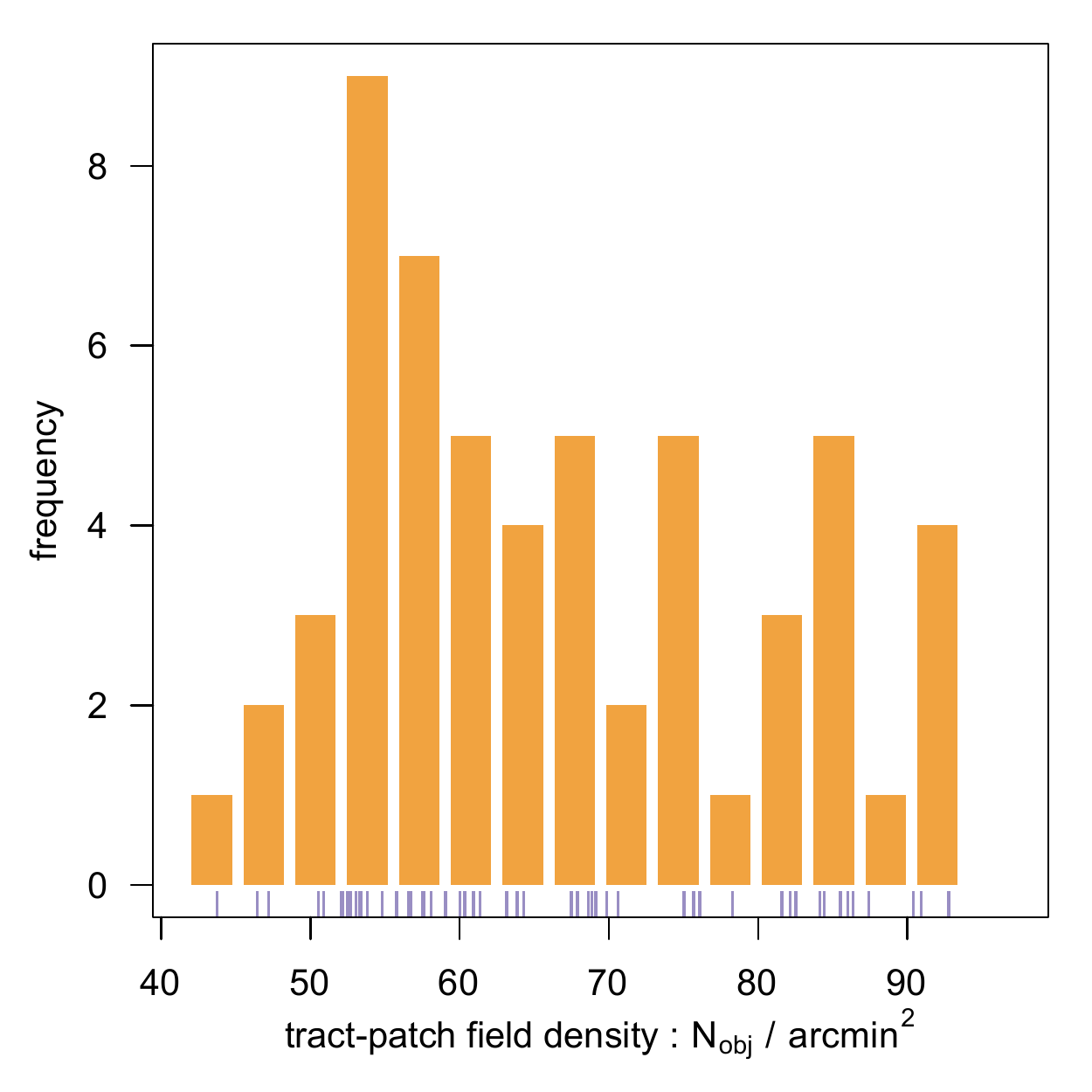}
    \caption{Initial estimate of field density for all $57$ tract-patch regions using a constant absolute detection threshold of $0.07$ counts. Each patch is approximately $0.196\;\deg$ on a side. Using these data, we select the tract-patch fields located at the 5th and 95th percentiles of recovered field density to respectively represent our low-density and high-density input observed datasets.}
    \label{fig:patchden}
\end{figure}

Recovered field densities for all tract-patch regions are shown in Figure \ref{fig:patchden}. Values range from $\about40$ to $\about100$ objects per square arcminute ($\about150000$ to $\about350000$ objects per square degree). To select two sample fields, we exclude outliers at both the extreme low density and high density terminus of this range, and select the tract-patch regions located at the 5th and 95th percentiles of recovered field density. The fields found at these percentiles are used as the bases for our low and high density regions, respectively. Tract-patch 8283-38 represents our low density region, containing $6813$ detected objects ($\about45\,\mathrm{N_{obj}\,/\,arcmin^2}$ or $\about181991\,\mathrm{N_{obj}\,/\,sq.\,deg}$), whilst tract-patch 9592-20 represents our high density region, containing $12183$ detected objects ($\about90\,\mathrm{N_{obj}\,/\,arcmin^2}$ or $\about325435\,\mathrm{N_{obj}\,/\,sq.\,deg}$).

\begin{figure*}
\begin{tabular}{rl}
\includegraphics[width=\columnwidth]{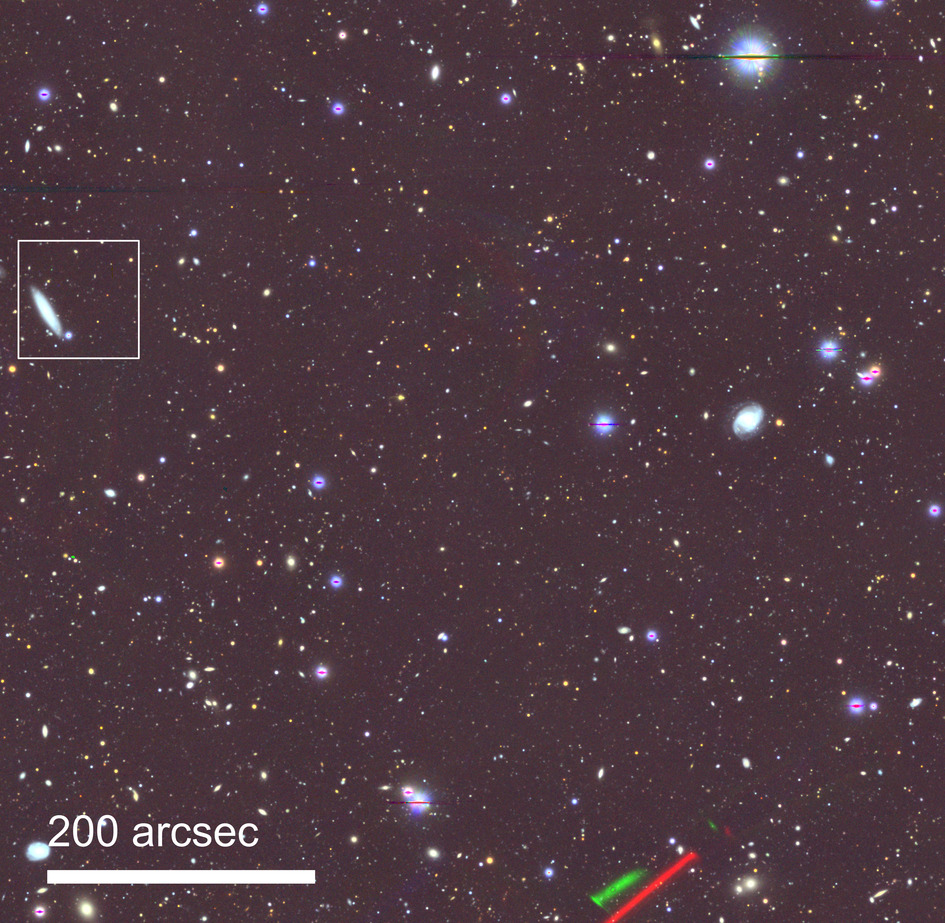}
&
\includegraphics[width=\columnwidth]{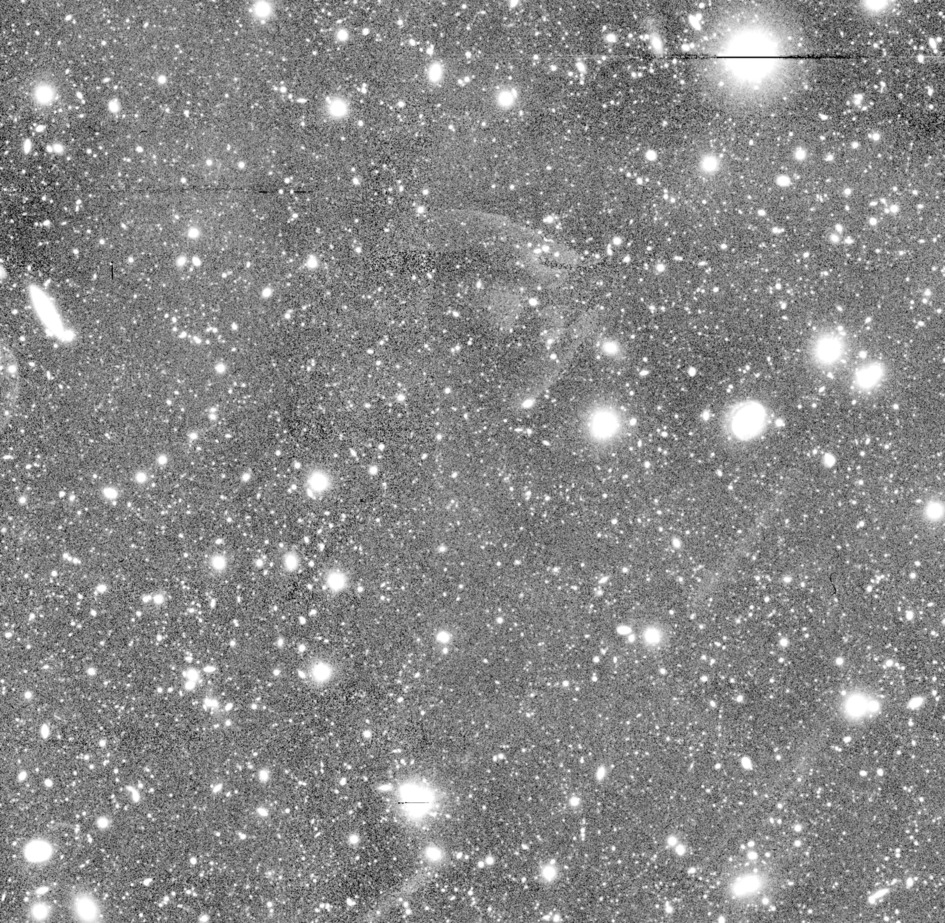}
\\\\
\includegraphics[width=\columnwidth]{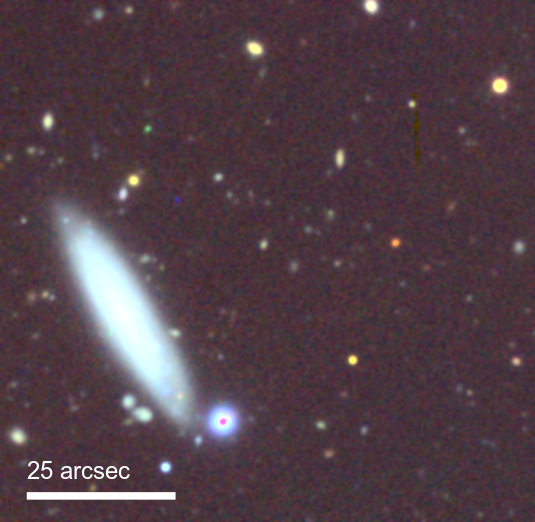}
&
\frame{\includegraphics[width=\columnwidth]{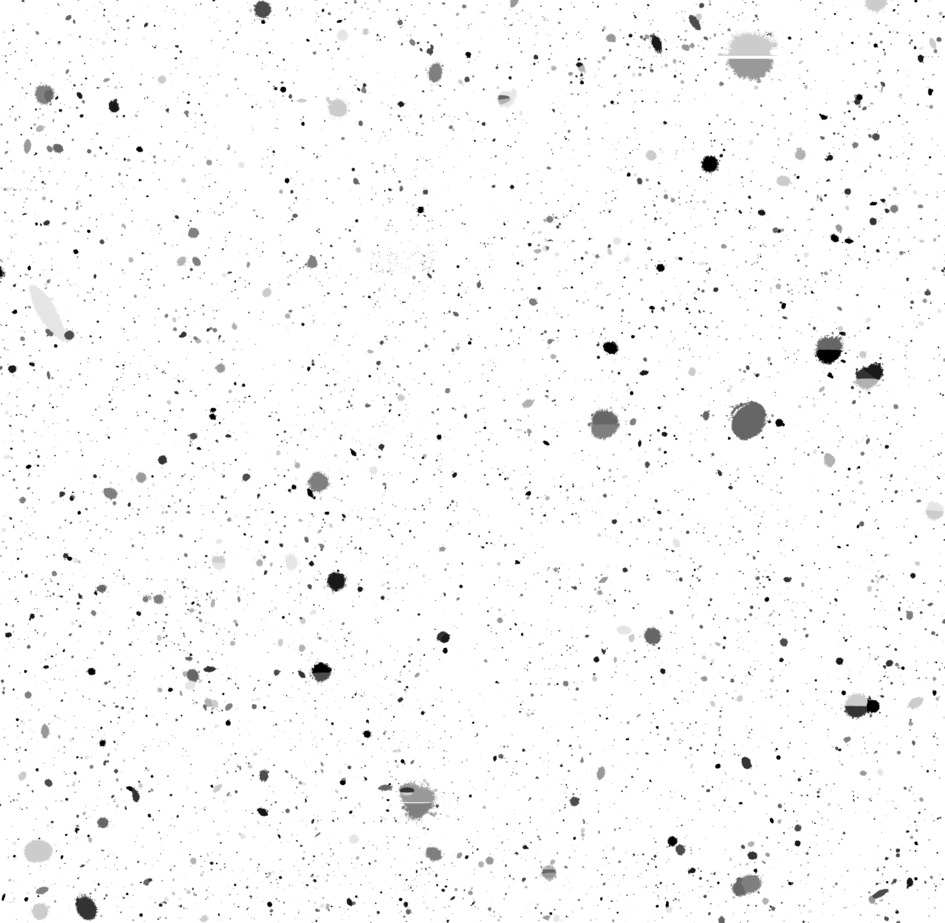}}
\end{tabular}
\caption{HSC-SSP PDR1 low density region, tract-patch 8283-38. Clockwise from top left: RGB postage stamp of the entire $706\times689\;\mathrm{arcsec}$ region; $r$-band postage stamp; segmentation map; and RGB postage stamp for a $90\times88\;\mathrm{arcsec}$ zoomed cutout. All postage stamps are arctan scaled and smoothed with a Gaussian kernel of $\Gamma=3\;\mathrm{pix}$. RGB images are generated using HSC-SSP $z$, $i$ and $r$ passbands. The white box in the upper-left panel shows the zoomed in region displayed in the lower-left panel. Shades of grey within the segmentation map are randomly assigned. This low-density region will be used as a basis for the generation of a low-density simulated field.}
\label{fig:stamplo}
\end{figure*}

\begin{figure*}
\begin{tabular}{rl}
\includegraphics[width=\columnwidth]{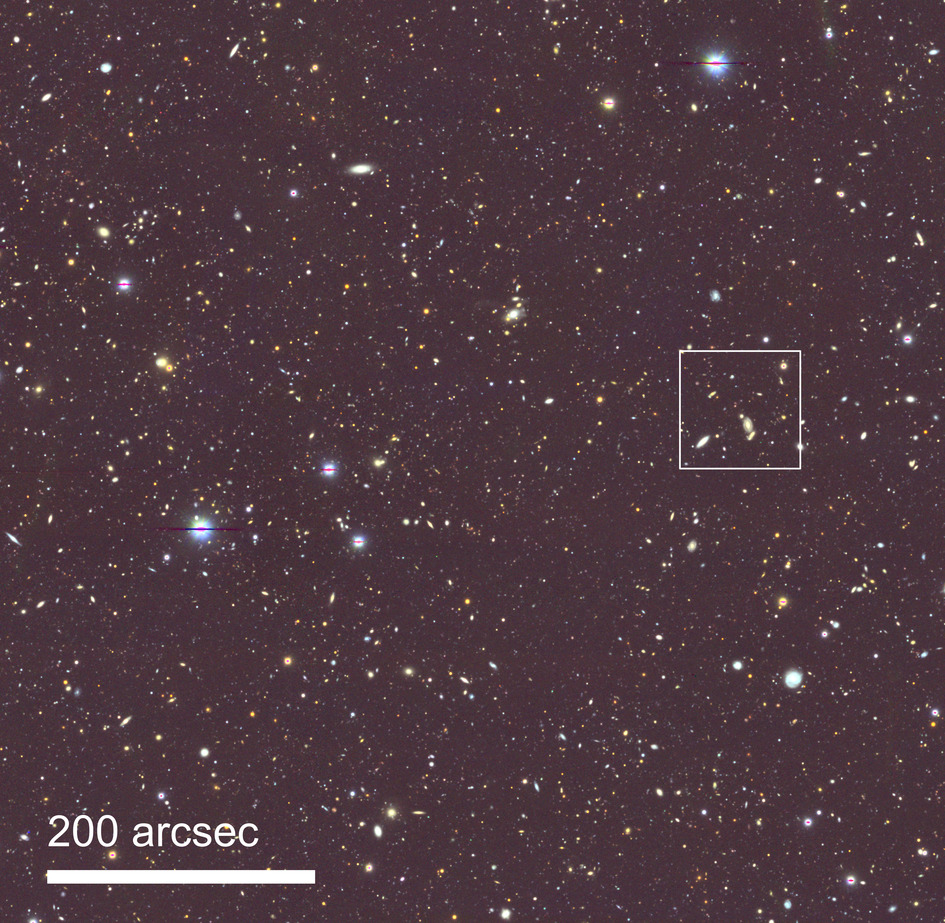}
&
\includegraphics[width=\columnwidth]{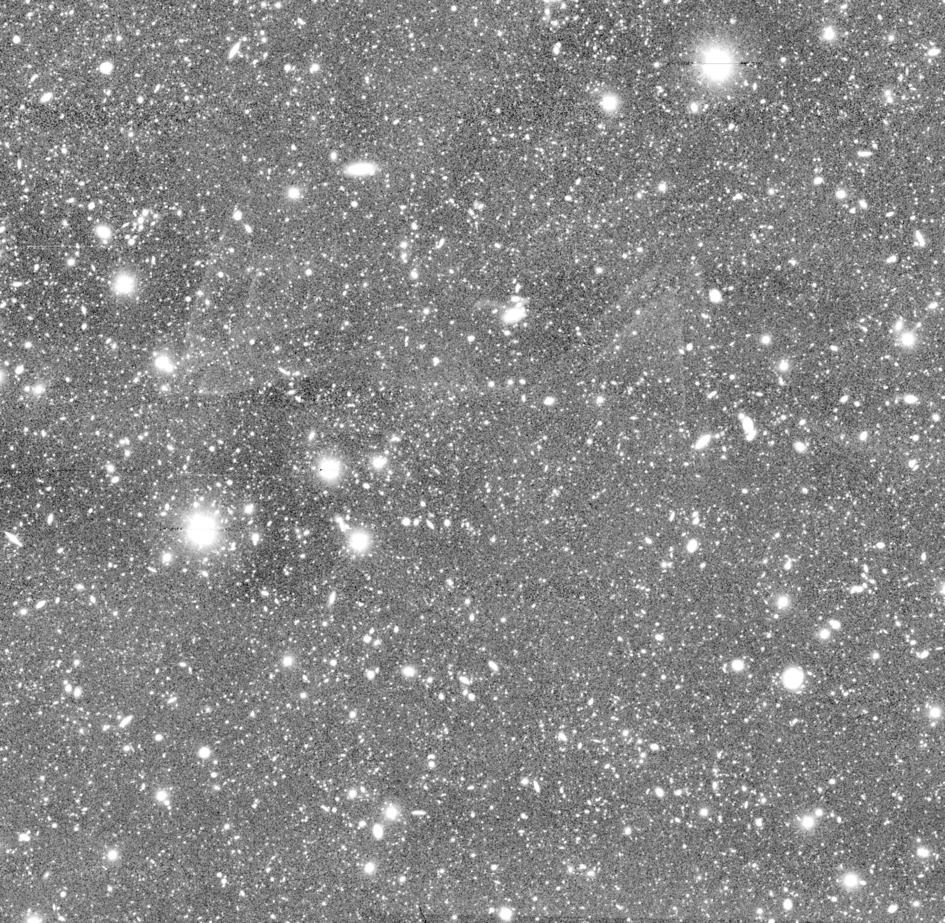}
\\\\
\includegraphics[width=\columnwidth]{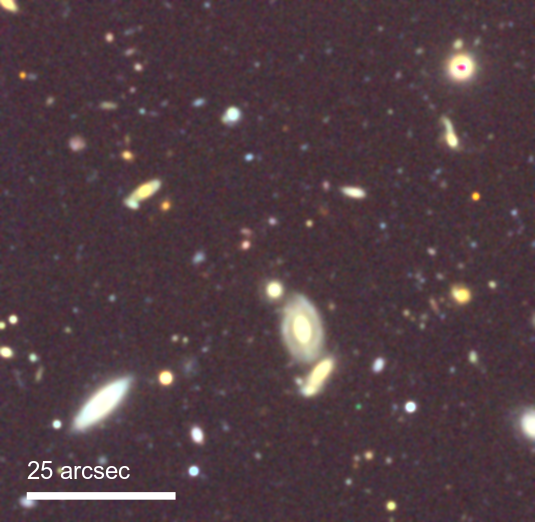}
&
\frame{\includegraphics[width=\columnwidth]{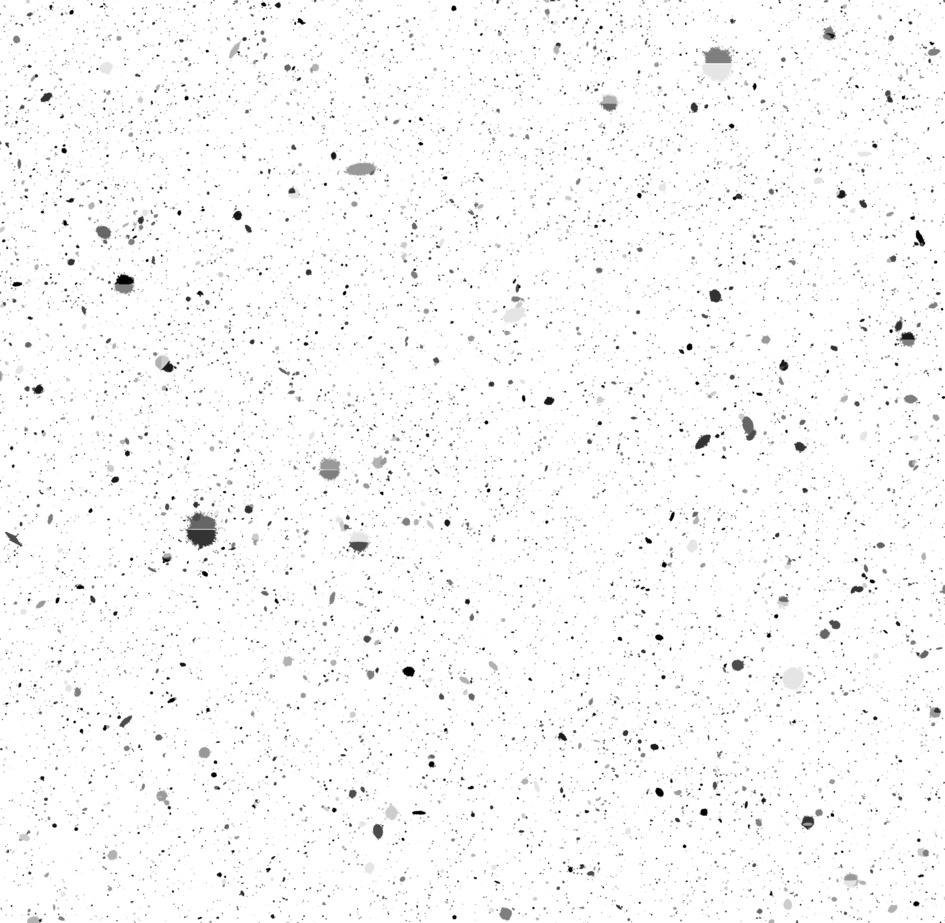}}
\end{tabular}
\caption{As Figure \ref{fig:stamplo}, but for the HSC-SSP PDR1 high density region, tract-patch 9592-20.}
\label{fig:stamphi}
\end{figure*}

Postage stamps and associated segmentation maps for the low density (tract-patch 8283-38) and high density (tract-patch 9592-20) regions can be found in Figures \ref{fig:stamplo} and \ref{fig:stamphi}, respectively (see captions for further details). As can be seen, the relatively high resolution and depth of the HSC dataset make these data an ideal candidate for further exploration of sky-subtraction methodologies. Several imaging artefacts remain in these data in addition to the obvious sky over-subtraction impact, notably, central saturation and diffraction spikes associated with bright point sources, optical ghosts, and shredding of bright objects. Whilst these artefacts do not severely impact our subsequent analyses, we highlight them here for reference.

\subsection{Point Spread Function}
\label{sec:psf}

It is necessary to convolve our simulated images with a realistic point spread function (PSF) in order to fully account for the impact of the imaging system upon the data. We opt to generate a model of the PSF in the low density region (tract-patch 8283-38) and utilise the resultant PSF model for all subsequent simulated imaging. We use the \PSFEx \citep[][]{Bertin2011} PSF extraction software package (version 3.17.1) to generate an empirical PSF from an input \SExtractor catalogue. The \SExtractor software is operated in a similar manner to that outlined in Section \ref{sec:sample}, with the exception that the output catalogue type must be in the FITS Leiden Data Analysis Center (LDAC) file format. This file format allows for more complex data and imaging to be stored in a single entity, facilitating later PSF extraction. The columns output in this catalogue must contain: VIGNET(101,101); X\_IMAGE; Y\_IMAGE; FLUX\_RADIUS; SNR\_WIN; FLUX\_APER(1); FLUXERR\_APER(1); ELONGATION; and FLAGS. The numbers inside parenthesis following VIGNET specify the postage stamp pixel cutout size around each input point source which will ultimately be used to help define the PSF. A full description of the remaining parameters can be found in \citet{Bertin2011}, and references therein. The resultant catalogue is used as an input to \PSFEx. As with \SExtractor, \PSFEx is largely operated using its in-built defaults. Sources identified as point-source-like are added to a library. From our $101\times101$ pixel vignettes, we request a single output snapshot PSF of size $55\times55$ pixels in dimension.

\begin{figure*}
    \centering
    \includegraphics[width=0.9\textwidth]{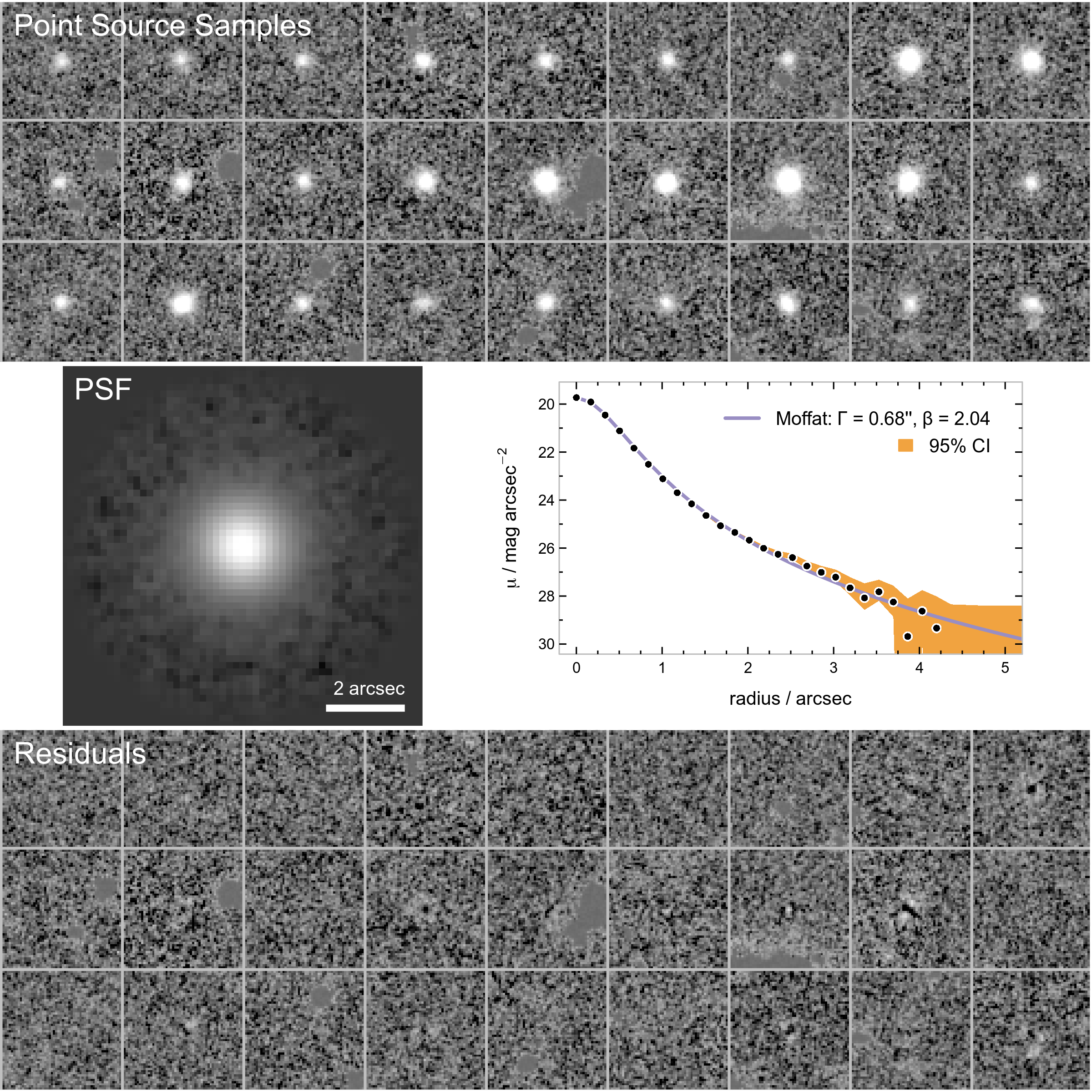}
    \caption{Visualisation of key outputs from the \PSFEx PSF modelling software, as applied to our HSC-SSP PDR1 low-density dataset (tract-patch 8283-38). Top: $27$ sample point sources, $\arcsinh$ scaled, from a total of $424$ used as an input to the PSF generation software. Middle-left: empirical PSF, log scaled, generated as the output of an $\mathrm{2}^{nd}$-order polynomial fit to the shape and position of all sample point sources. Middle-right: 1D surface brightness profile of the PSF, showing the PSF data (black points), $95\%$ confidence interval (shaded orange region), and a Moffat profile fit to the observed data (purple line). Fitted Moffat parameters are shown in the inset legend. Bottom: PSF residual images, $\arcsinh$ scaled, representing the results of a flux-scaled PSF subtraction from each point source in the top panel. Much of the point source flux has been successfully removed in the residual frames, evidencing the high-fidelity nature of the output PSF model.}
    \label{fig:psf}
\end{figure*}

The \PSFEx outputs are shown in Figure \ref{fig:psf}. The top panel shows a random selection of $27$ out of a total of $424$ point sources used to construct this PSF. Nearby neighbours which may act as a flux contaminant are automatically masked prior to use. The shape and position of these point sources on the original frame are used to construct a $\mathrm{2}^{nd}$-order polynomial model of the PSF, which is shown in the middle-left panel. This model exhibits good circular symmetry, with a measured ellipticity of $e=1-b/a\approx0.04$. Contaminant noise does become increasingly evident in the wings, however, the goodness of fit $\chi^2/\nu=0.99$ indicates a good match between the model and the input data. The empirical full-width at half-maximum is found to be $\Gamma=0.72\;\mathrm{arcsec}$. As shown in the middle-right panel, the PSF closely approximates a Moffat function down to very faint magnitudes, with fitted parameters and confidence intervals given in the inset legend. The result of subtracting a scaled version of the PSF to each of the input point source samples is shown in the lower panel. The majority of point source flux is removed in the residual frame, again evidencing the high-fidelity nature of our output PSF. This PSF is used for all subsequent relevant data analysis.

\section{\GalSim Simulations}
\label{sec:galsim}

\GalSim \citep{Rowe2015} is an open source software package designed to produce highly accurate simulated astronomical imaging. It has a number of powerful capabilities, including the ability to generate a broad range of galaxy and stellar profile types, perform rapid and accurate PSF convolution, and apply a physically motivated noise model. For these reasons, in addition to the speed and accuracy of the underlying base code, we opt to use the \GalSim package (version 2.2.6) to generate our simulated imaging.

Prior to running the \GalSim software, input catalogues and associated \texttt{YAML} configuration files must be generated to specify all necessary simulation parameters. Here we generate realistic simulated regions with noise approximating that found in HSC-SSP PDR1 data and populated with a range of objects exhibiting features similar to those found in our chosen low-density and high-density HSC-SSP PDR1 regions. Each region is exactly $4200\times4100$ pixels in dimension (approximately $706''\times689''$). As a caveat, and for the sake of simplicity, we opt to populate these regions with extended-type \citet{Sersic1963,Sersic1968} profiles alone, avoiding the additional complication of point sources which typically contribute less to sky estimation errors (excepting the extreme bright-end population of stars) than extended profiles. Each \Sersic profile requires a total of eight input parameters: pixel position in both $x$ and $y$; total luminosity in counts; half-light radii in pixels; an axis ratio (i.e., $q = 1-e = b/a$); a position angle in degrees; a \Sersic index $n$, and; a postage stamp size in pixels. The latter of these parameters provides the edge size of a square stamp on which to draw the objects within \GalSim\footnote{Postage stamp size is not strictly required in order to generate a \Sersic profile when using \GalSim, with a default option available. However, for reasons further described below, we require that this parameter be specified here.}.

Input catalogues must mimic the distribution of features in the sample regions as closely as possible. A number of objects will however be missing from any detection catalogue, owing to either bright-end confusion or faint-end incompleteness. Some such parameters for these extrapolated objects require a meaningful estimation, whilst others require explicit definition. In the following sections, we provide an overview of the derivation of each required simulation parameter, concluding with a summary of our simulated field data generation.

\subsection{Number Counts \& Magnitudes}
\label{sec:numcounts}

Perhaps the most fundamental parameter required to populate any simulated input catalogue is how many objects overall one wishes to simulate as a function of magnitude. As an initial guide, we have input catalogues for both sample regions from our initial \SExtractor runs. The solid emboldened lines in Figure \ref{fig:numcounts} show the detected number of sources per square degree and per magnitude as a function of apparent $r$-band magnitude, using a histogram bin width of $\Delta m_r=0.5$ mag. The low density region shown in the left panel, tract-patch 8283-38, contains a total of $6813$ detected sources. The high density region shown in the right panel, tract-patch 9592-20, contains a total of $12183$ detected sources. The dotted horizontal line in both panels represents the number density corresponding to one object per the area of a sample field ($\about0.038\;\deg^2$) per magnitude.

The brightest detected sources are found at $m_r\sim15.5$ mag in the low density region and $m_r\sim17$ mag in the high-density region. Number densities then rise in a somewhat stochastic nature down to $m_r\sim22$ mag in both fields. The fluctuations in source number density as a function of magnitude are as expected, owing to the aforementioned bright-end source confusion. In this regime, it is common for bright sources to become shredded, or for the flux in their wings to become erroneously attributed to or separated from a neighbouring source.

The source density profile continues to rise in the magnitude range $22<m_r<25$ mag (represented by two vertical dashed lines) in a log-linear fashion. A constrained log-linear fit to the detected source density profile in this regime is represented as a solid thin line in both panels. Following \citet{Ji2018}, the slope of this best fit line is kept fixed, with only the intercept parameter fitted. We find the low density regime source density profile goes as $\log_{10}N_{\mathrm{obj}}=-5.07+0.4m_r$. The high density regime source density profile goes as $\log_{10}N_{\mathrm{obj}}=-4.90+0.4m_r$.

At magnitudes fainter than $m_r\sim25$ mag, detected source densities begin to turn down, dropping off completely by $m_r\sim28$ mag. Such faint-end incompleteness is inherent in any source detection catalogue, as sources gradually become indistinguishable from the background noise and instead contribute to the faint undetected EBL.

Detected sources brighter than $m_r=22$ mag are significantly more numerous than that predicted by our best fit log-linear trend line, occasionally by almost an order of magnitude. It would be unwise to use these catalogue entries when defining a simulated input catalogue if indeed these sources are erroneously shredded sub-components of larger singular sources. For the purposes of defining a simulated catalogue, we opt to discard all detected sources brighter than $m_r=22$ mag and use the best fit trend line to define a replacement bright end dataset in its place. New bright mock sources are distributed uniformly within each $\Delta m_r=0.5$ mag bin. This procedure produces a total of $272$ new sources in the low density region and $405$ new sources in the high density region.

Remaining detected sources at intermediate and faint magnitudes are relatively free from such source processing issues. A single object detected in the low density region exhibited an erroneous flux value, and is consequently removed from all subsequent analyses. We therefore opt to propagate $6325$ such objects through to our simulated input catalogue from the low density region, and $11572$ from the high density region.

To account for known faint end incompleteness, we additionally derive a population of faint missing sources in the range $25<m_r<30$ mag. These sources are defined as the population required in addition to any detected and used sources which would satisfy our best-fit log-linear source density trend line. As with the bright mock sources, these new faint sources are distributed uniformly in magnitude space within each $\Delta m_r=0.5$ mag bin. As such, this provides an additional $421660$ faint end sources in the low density region, and $624238$ faint end sources in the high density region.

In summary, we define three distinct flux-associated source populations: bright mock sources, intermediate-brightness detected and used sources, and faint missing sources. These three populations are propagated through to our simulated catalogue and are represented in Figure \ref{fig:numcounts} as dark grey, hatched, and light grey shaded regions, respectively. In total, we simulate $428257$ sources in the low density regime and $636215$ sources in the high density regime. Magnitudes are converted to flux counts for injection using the standard HSC-SSP \texttt{deepCoadd\_calexp} zero point magnitude of $27.0\;\mathrm{mag}\;\mathrm{ADU}^{-1}$ \citep{Aihara2018b}.

\subsection{Half Light Radii}
\label{sec:magradii}

It's necessary and desirable for our simulated source sizes to mimic as closely as possible the distribution of size as a function of magnitude found in our observed data. The preliminary source detection catalogues provide half light radii (\texttt{FLUX\_RADIUS}) for our chosen intermediate brightness detected \& used sources; a starting point upon which to generate an appropriate size-magnitude distribution. Some form of extrapolation is however required to provide radii for the remaining brighter/fainter simulated source populations as outlined in Section \ref{sec:numcounts}.

\begin{figure}
    \centering
    \includegraphics[width=\columnwidth]{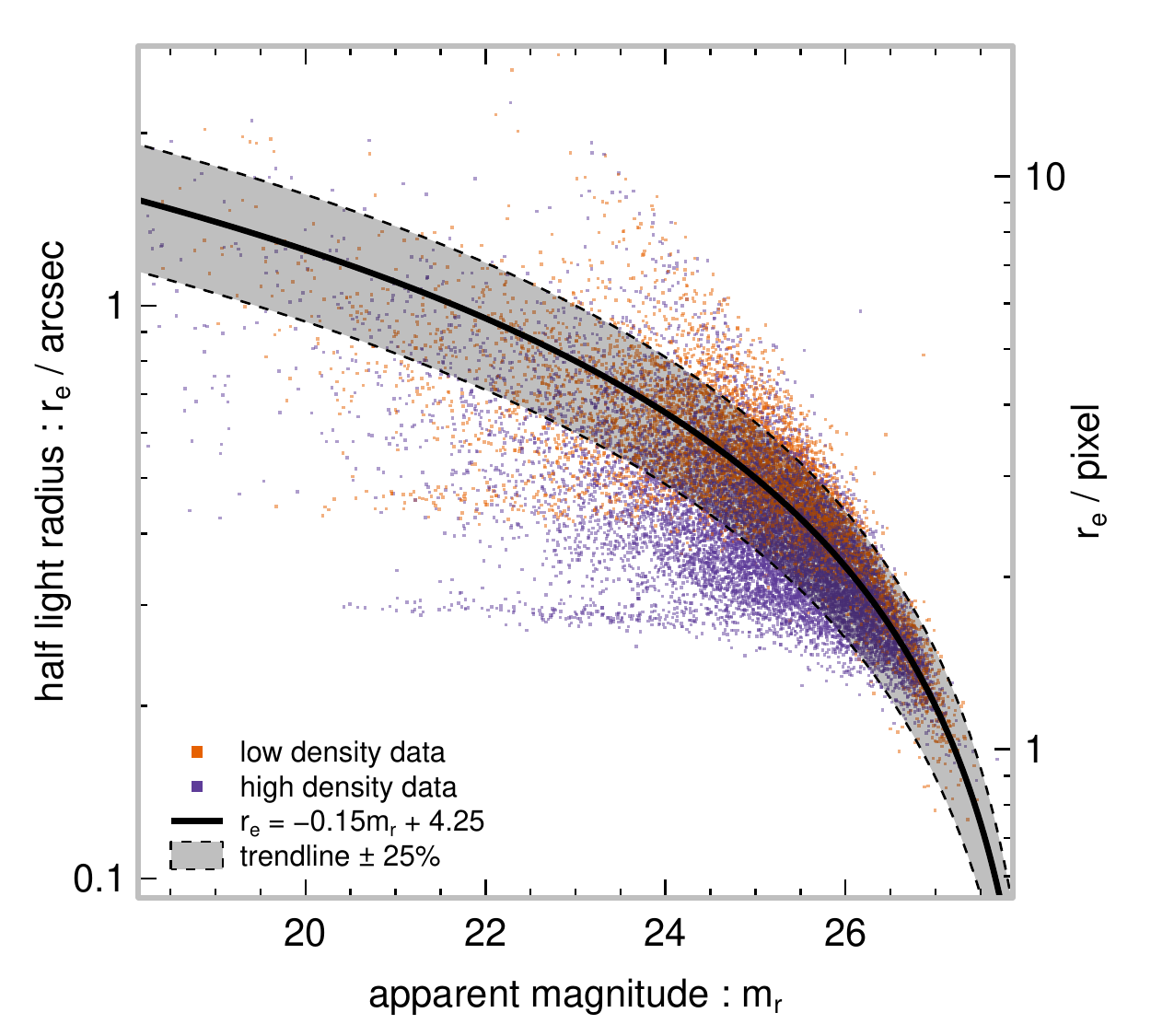}
    \caption{Recovered half-light radii as a function of apparent $r$-band magnitude for all extended-type ($\mathrm{CLASS\_STAR}<0.05$) detected sources. Sources from the low density region are shown in orange, whilst sources from the high density region are shown in purple. The solid bold black line represents a best fit trendline to those extended-type sources brighter than $m_r=22.25$ mag. Simulated bright mock and faint missing sources are randomly distributed within the $\pm25\%$ boundaries about this trendline down to an absolute minimum of $1$ pixel.}
    \label{fig:magradii}
\end{figure}

Figure \ref{fig:magradii} shows recovered half light radii as a function of apparent $r$-band magnitude for all extended-type detected sources in both the low-density region (orange data points) and high-density region (purple data points). We define extended-type sources as those with a \SExtractor stellar classification parameter in the range $\mathrm{CLASS\_STAR}<0.05$, removing obvious point sources. The application of this cut removes $386$ of $18995$ potential sources. Despite making this cut, two remnant point source spurs are evident at $r_{e,\mathrm{low}}\approx0.5$ and $r_{e,\mathrm{high}}\approx0.3$, with the low density region PSF evidently more extended than its high density region equivalent. The apparent bias in half light radius at any given magnitude as a function of field density is due to this effect, with sources in the low density regime (orange data points) appearing consistently larger on average by a factor of $\about2$ than their equivalent-brightness sources in the high density regime (purple data points).

We fit a rough linear trendline to detected extended-type sources brighter than $m_r=22.25$ mag using the \texttt{lm} linear minimisation routine \citep{Chambers1992} within the \textsc{R} programming language \citep{RCoreTeam}. The fit is constrained in order to accurately model the bulk population trend. We find a best fit line given by the equation $r_\mathrm{e}=-0.15m_r+4.25$, represented in Figure \ref{fig:magradii} as a solid bold line. Dashed lines bounding the shaded grey region represent the trend radius $\pm25\%$. We use this relation, alongside simulated bright mock and faint missing $r$-band magnitudes, to estimate simulated half light radii. Simulated radii are randomly uniformly distributed within the $\pm25\%$ boundary at a given magnitude down to an absolute minimum of $1$ pixel $\pm25\%$ (i.e., the faintest simulated source half light radii are randomly uniformly distributed in the range $[0.75,1.25]$ pixels).

\subsection{Ellipticities}
\label{sec:magellip}

\begin{figure*}
    \centering
    \includegraphics[width=0.9\textwidth]{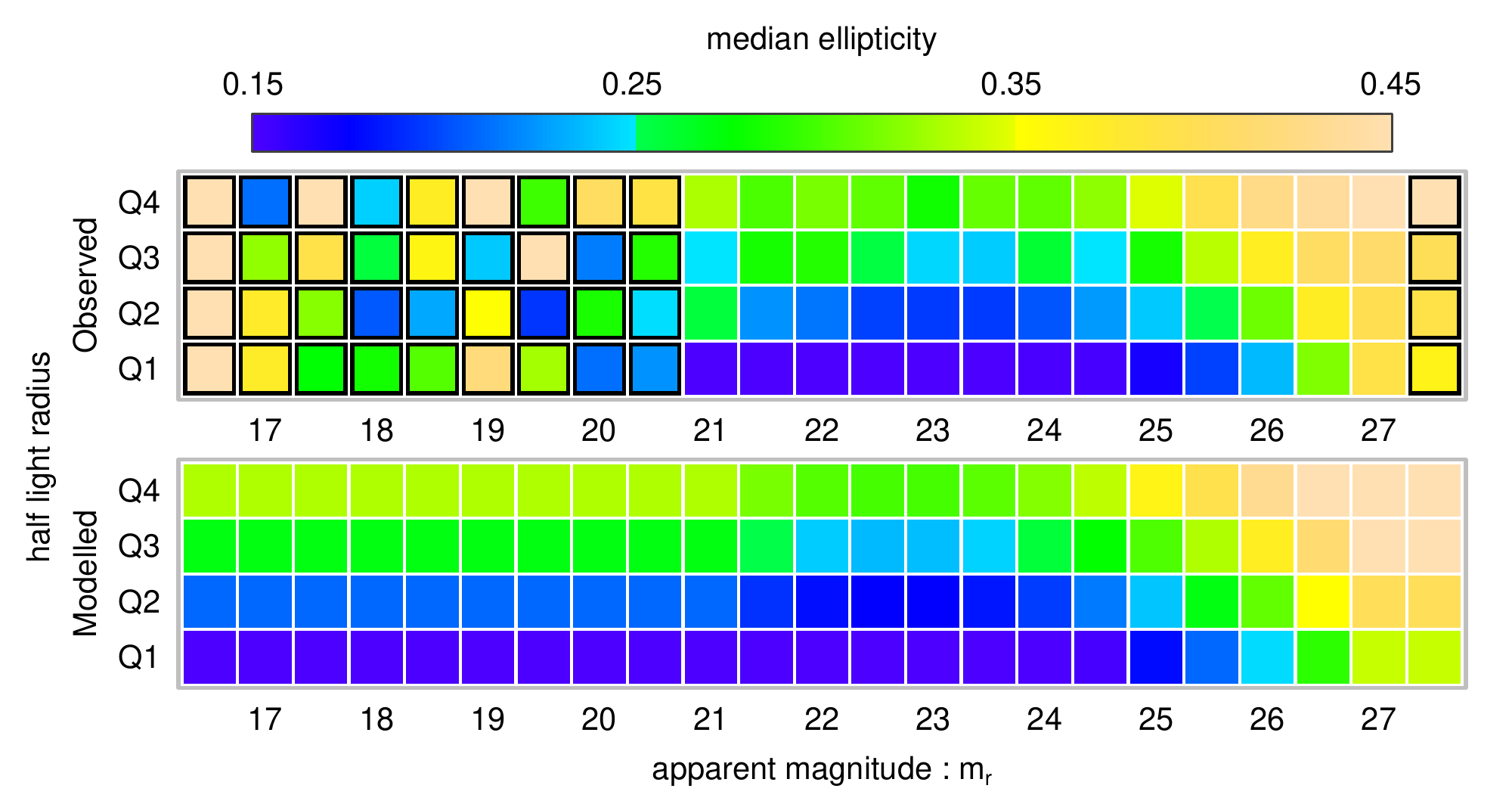}
    \caption{Median ellipticities in bins of half light radius and apparent $r$-band magnitude. Observed (detected) extended-type data is shown in the top panel, modelled (simulated) data is shown in the bottom panel. Observed bins enclosed within a solid black border represent minimally occupied bins containing $25$ or fewer sources. A fit to all remaining bins with $25$ or more sources each is used to construct the model in the lower panel (see text for more details).}
    \label{fig:magellip}
\end{figure*}

As with radii in Section \ref{sec:magradii}, simulated ellipticities (where $e=1-b/a$) are likewise based upon our combined low/high density preliminary detected source catalogues. Figure \ref{fig:magellip} shows median ellipticities according to the inset colour legend in bins of half light radius along the $y$ axis and apparent $r$-band magnitude along the $x$ axis. The top panel shows our observed (detected) extended-type dataset, whilst the bottom panel shows our resultant extrapolated modelled (simulated) ellipticity dataset. Apparent magnitude data along the $x$ axis is binned into $0.5$ mag bins (see Figure \ref{fig:numcounts}). Half light radii are subdivided into quartiles, thereby covering the full gamut of radii within each magnitude bin. For example, median half light radii at $m_r=21$ mag are $0.46''$ at $Q1$, $0.69''$ at $Q2$, $0.85''$ at $Q3$ and $1.04''$ at $Q4$, whilst median half light radii at the fainter $m_r=27$ mag are $0.19''$ at $Q1$, $0.21''$ at $Q2$, $0.23''$ at $Q3$ and $0.25''$ at $Q4$. Boxes bordered by a solid black outline in the observed (detected) top panel show those bins containing $25$ or fewer sources (equating to a total of $270$ sources from a potential $18609$), and therefore with a likely large error on the measure of ellipticity. As shown, the general trend is for larger objects to be more elliptical (less circular) at all magnitudes. Ignoring minimally occupied bins, we find a trend for detected sources becoming more circular as one progresses from bright magnitudes down to $m_r\approx23$ mag. Continuing on to fainter magnitudes reverses this trend, with median source ellipticities becoming more elliptical once more, culminating with one of the highest overall median ellipticity bins at $m_r=27$ mag. Following \citet{Binggeli1988}, these fainter sources are increasingly likely to be morphologically irregular, in agreement with the higher ellipticities recovered here.

All observed (detected) bins containing $25$ sources or more per bin (those without a surrounding solid black outline in Figure \ref{fig:magellip} in the magnitude range $21<m_r<27$ mag) are used to define a model fit to this feature space. Those bins containing $25$ or fewer sources exhibit a relatively large scatter on recovered median ellipticity, necessitating their exclusion here. The \texttt{lm} minimisation routine is used to fit a multi-dimensional model, with a best fit output given by:
\begin{equation}
    \label{eq:magellip}
    \tilde{e} = 6.31564 + 0.06165\,r_{e,Q} - 0.55233\,m_r + 0.01217\,m_r^2
\end{equation}
where $\tilde{e}$ is the median ellipticity for a given bin, $r_{e,Q}$ is the half light radius quartile bin number (namely: $1$, $2$, $3$, or $4$) and $m_r$ is the bin apparent $r$-band magnitude. This model is applied directly in the magnitude range $21<m_r<27$ mag, i.e., those magnitudes containing well-populated bins in Figure \ref{fig:magellip}. To avoid extrapolation error to brighter/fainter magnitudes (e.g., Runge's phenomenon), we opt to directly map model median ellipticities from the $m_r=21$ mag bin on to each bin in the range $m_r\leq20.5$ mag. Similarly, we opt to directly map model median ellipticities from the $m_r=27$ mag bin on to each bin in the range $m_r\geq27.5$ mag. This procedure defines the model shown in the lower panel of Figure \ref{fig:magellip}. Bright mock sources are assigned a simulated ellipticity given their simulated input half light radius and magnitude according to this model, with ellipticities randomly uniformly assigned in the range $\pm0.2$ about their modelled median ellipticity. Intermediate-brightness sources which are detected and used propagate their detected ellipticities directly. Following the trend in our faintest magnitude bin, faint missing sources are randomly uniformly assigned an ellipticity in the range $0.4\pm0.2$.

\subsection{Position Angles}
\label{sec:positionangle}

Associated with ellipticity is the position angle along which to align the source major axis. We opt to propagate position angles for those intermediate brightness detected and used sources directly into the simulated input catalogue. All remaining bright mock sources and faint missing sources are randomly uniformly assigned a position angle in the range $-90<\mathrm{P.A.}<90$ degrees.

\subsection{Locations within the Field}
\label{sec:pixelposition}

One of the most fundamental parameters required as an input to \GalSim is at what position along $x$ and $y$ should the simulated source be placed. We have three varying brightness datasets: bright mock; intermediate-brightness detected and used, and; faint missing, with each requiring a different positional prescription. The simplest is for intermediate brightness detected and used source positions, which are propagated directly into the final simulated catalogue. Faint missing sources are extremely numerous, with around half a million simulated sources injected into both sample regions. Owing to their significant abundance, we opt to randomly uniformly assign faint missing source position centroids along the $x$ axis in the range $0.5<x<4200.5$ pixels. Half-pixel start and end points reflect the \texttt{FITS} convention that the centre of the lower left corner pixel has the coordinate $[1,1]$ and a width of $1$ pixel. As with the $x$ axis, pixel position centroids along the $y$ axis are similarly randomly uniformly assigned in the range $0.5<y<4100.5$ pixels.

\begin{figure*}
    \centering
    \includegraphics[width=0.9\textwidth]{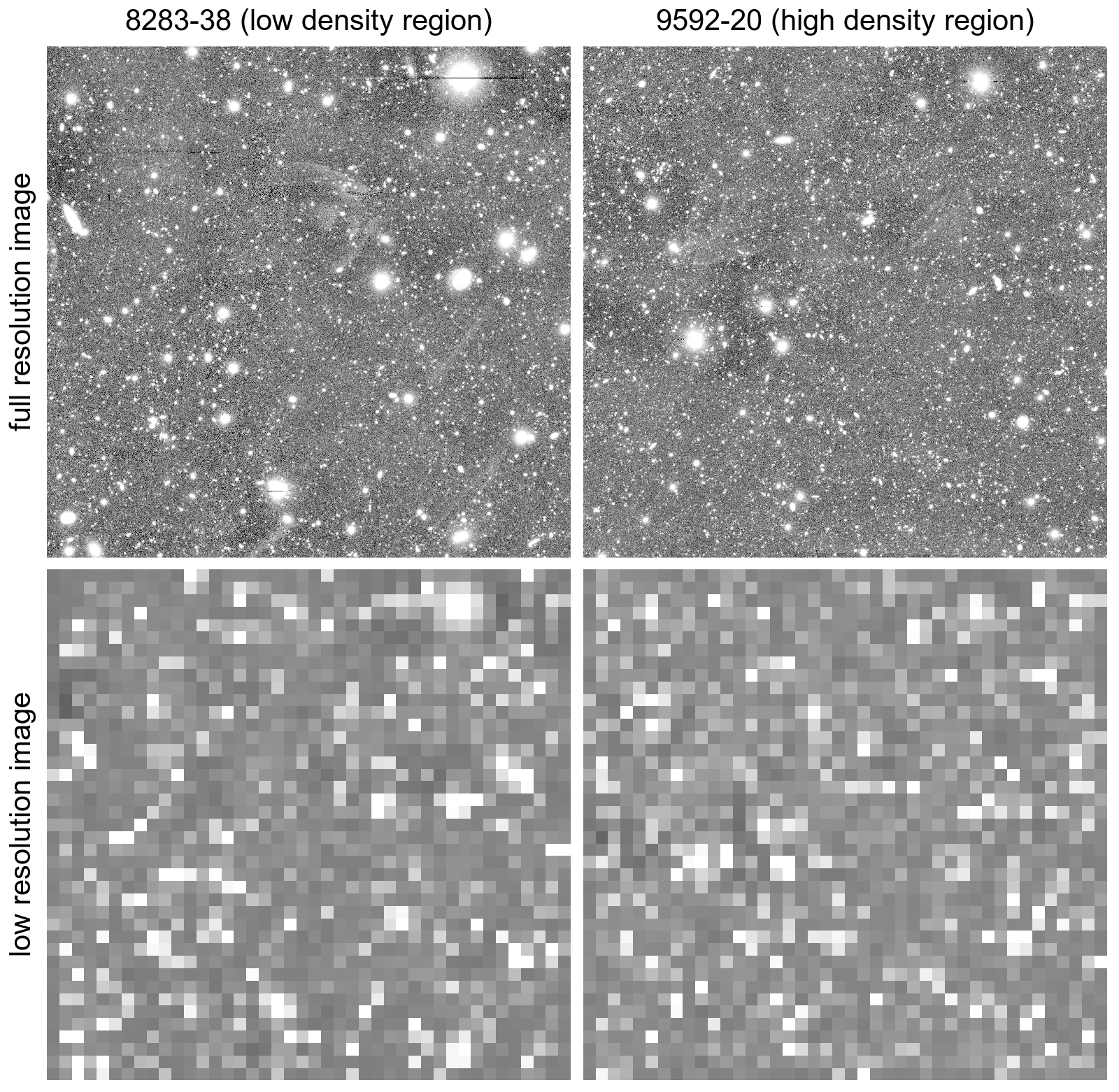}
    \caption{HSC-SSP PDR1 full-resolution (top) and artificially degraded low-resolution (bottom) imaging for both sample regions: low density on the left and high density on the right. Images are arctan scaled. Full resolution images are smoothed with a Gaussian kernel of $\Gamma=3\;\mathrm{pix}$ prior to plotting. Each super pixel in the low resolution image is a mean of $42\times41=1722$ ordinary pixels in the original data. Low resolution images are used to define a weight function when determining bright mock source centroid placement.}
    \label{fig:hireslores}
\end{figure*}

Determining the distribution of bright sources requires a different approach. The top two panels of Figure \ref{fig:hireslores} show arctan-scaled full-resolution images for our low density region (left) and high density region (right). As is evident, several bright sources may on occasion cluster together to form a loose group. Whether a `true' group in three-dimensional space or a mere projection effect, the resulting flux bridges that link these sources on the sky leave that region prone to sky over-subtraction effects. Any investigation of this effect must therefore place bright simulated sources in such a way so as to preserve any clustering signature. We use these input images as a basis for determining such signatures. The lower two panels in Figure \ref{fig:hireslores} show artificially degraded low resolution versions of their full resolution counterparts, with the low density region on the left and the high density region on the right. Each axis in the full resolution image is split into $100$ `super pixels', resulting in $4200/100=42$ super pixels along the $x$ axis and $4100/100=41$ super pixels along the $y$ axis. The mean flux value for each super pixel (equating to $42\times41=1722$ ordinary pixels) is calculated, resulting in the low resolution images shown here.

To preserve any clustering signature in our sample regions, low resolution imaging is used to define a positional weight function. A weight map is constructed by rescaling the low resolution images on to the range [$0.0\rightarrow1.0$]. Bright mock sources are iteratively assigned a randomly weighted position using the \texttt{sample} routine within \textsc{R} \citep{Ripley1987}.

\subsection{\Sersic Indices}
\label{sec:sersicindex}

Owing to the well known bimodality in galaxy radial profiles \citep{Baldry2004,Baldry2006,Driver2006,Kelvin2012,Kelvin2014a,Kelvin2014b,Taylor2015}, we opt to simulate \Sersic-like galaxies with \Sersic indices of $n=1$ (i.e., exponential disk-like) and $n=4$ (i.e., de Vaucouleurs type spheroid-like). Each simulated region is simulated twice, with each simulated source given a \Sersic index of $n=1$ and, separately, a \Sersic index of $n=4$. Simulating disk-like and spheroid-like sources allows us to explore the accuracy of our sky-estimation routine as a function of source profile type, with de Vaucouleurs sources exhibiting significantly more flux in their outer wings than their exponential counterparts.

\subsection{Postage Stamp Cutout Sizes}
\label{sec:postagestamps}

Postage stamp size is the final required per-object \GalSim input parameter. Each simulated source generated by \GalSim is extrapolated out to the limits of a user-provided box, known as the postage stamp. This limitation exists for reasons of computational efficiency, notably in reducing the number of pixels requiring PSF convolution. A-priori knowledge of all relevant \Sersic parameters (magnitude, half-light radius, \Sersic index and ellipticity) allows us to determine the surface brightness drop-off as a function of semi-major axis radius for each simulated source. We opt to define postage stamp boxes for each source such that a source aligned parallel to the $x$-axis would reach a surface brightness limit of $\mu_r=35\,\mathrm{mag}/\mathrm{arcsec}^{-2}$ at the boundary of the postage stamp. This limit is sufficiently faint to facilitate exploration of the background sky, yet not too deep to preclude the construction of these simulated frames. A minimum postage stamp size of $11\times11$ pixels is enforced\footnote{This postage stamp size limit is reached only in the exponential $n=1$ case for $\about12\%$ of sources.}. See Section \ref{sec:postagestampsizes} for further discussion on the importance of choosing an appropriate postage stamp cutout size.

\subsection{RMS and Sky Level}
\label{sec:rmsskylevel}

To ensure that output simulated imaging exhibits Poisson noise which mimics that found in coadded sky-subtracted HSC-SSP PDR1 data, \GalSim requires as an input an estimate of the original average sky level pedestal. Unfortunately, accessing these data for this PDR1 dataset is non-trivial, with subtracted background sky maps not provided nor any easily accessible record kept of subtracted sky levels. Consequently, we calculate a basic estimate of the initial sky level based upon the root mean square (RMS) of the flux in those pixels identified as background. Given typical averaged equivalent detector gain $g$ taken from the HSC-SSP PDR1 database and a measure of background RMS flux from our initial \SExtractor runs across both density regimes, we estimate the original sky level $f_\mathrm{sky}$ in $\mathrm{counts}$ using the relation $f_\mathrm{sky}=g\cdot\mathrm{RMS}^2$. Using this relation, in the low density regime (8283-38) we find $\mathrm{RMS}=5.43\times10^{-2}\,\mathrm{counts}$ and $f_\mathrm{sky}=1.645\,\mathrm{counts}$, whilst in the high density regime (9592-20) we find $\mathrm{RMS}=4.97\times10^{-2}\,\mathrm{counts}$ and $f_\mathrm{sky}=1.391\,\mathrm{counts}$.

\subsection{\GalSim Setup}
\label{sec:galsimsetup}

The above sections detail the generation of all required \GalSim input parameters. With these parameters defined, input catalogues for all regions of interest are constructed. Input catalogues contain positional information alongside all other parametric variables required to construct a \Sersic model for each simulated source (see Section \ref{sec:inputcats} for further details).

We aim to fully explore the accuracy of various background subtraction techniques as a function of field density, source profile shape and background light contamination. In addition to two density regimes (low and high as defined in Section \ref{sec:sample}, termed here as `denlo' and `denhi', respectively) and two profile types ($n=1$ and $n=4$, see Section \ref{sec:sersicindex}), we also define a bright-end flux subset of the data. This allows us to explore the performance of sky subtraction algorithms with a known removal of faint undetected components, the primary contributors to EBL. Three simulated populations roughly segregated by flux are defined in Section \ref{sec:numcounts}: bright mock sources ($m_r<22$ mag), intermediate brightness detected and used sources ($m_r>22$ mag, becoming increasingly incomplete fainter than $m_r=25$ mag) and faint missing sources (i.e., known missing sources in the range $m_r>25$ mag). We define a `bright' sample as the combination of those simulated sources belonging to the first two of these simulated populations, i.e., bright mock sources and intermediate brightness detected and used sources. From the initial $428257$ simulated sources in the low density regime and $636215$ simulated sources in the high density regime, this bright end cut extracts $6597$ low density region sources and $11977$ high density region sources. Taken together, we generate a total of eight simulated regions. These regions are summarised in Table \ref{tab:simtab}.

\begin{table}
    \setlength{\tabcolsep}{5pt}
    \begin{tabular}{ l | l | l | l | l }
    label & based on & flux subset & source density & \Sersic index\\
    \hline
    denlo1a & 8283-38 & all & low & 1\\
    denlo4a & 8283-38 & all & low & 4\\
    denhi1a & 9592-20 & all & high & 1\\
    denhi4a & 9592-20 & all & high & 4\\
    denlo1b & 8283-38 & bright & low & 1\\
    denlo4b & 8283-38 & bright & low & 4\\
    denhi1b & 9592-20 & bright & high & 1\\
    denhi4b & 9592-20 & bright & high & 4\\
    \end{tabular}
    \caption{A summary table providing an overview of our final eight simulated regions. From left to right, columns are: dataset label; the input HSC-SSP PDR1 tract-patch ID upon which this simulated data is based; the flux population subset; the relative source density of the region; and, the \Sersic index assigned to each simulated source in that region. Flux population subset is either `all' (i.e., all simulated sources as outlined in Section \ref{sec:numcounts}) or `bright'. This `bright' flux subset identifier indicates that the `faint missing' sources described in Section \ref{sec:numcounts} are excluded from this simulation, thereby limiting it to the `bright mock' and `intermediate detected and used' simulated sources alone.}
    \label{tab:simtab}
\end{table}

Input catalogues for each simulated frame defined in Table \ref{tab:simtab} are constructed. An associated YAML `feedme' configuration file is also generated specific to each simulated region. This configuration file specifies input and output filenames, defines the PSF and \Sersic components and assigns global parameters. For further information see Section \ref{sec:yamlconfig} and \citet{Rowe2015}.

\subsection{Simulated Fields}
\label{sec:simfields}

The \GalSim software package is run across each simulated field defined in Table \ref{tab:simtab}. Figure \ref{fig:simstampfull} shows the results of these runs. All simulated \Sersic sources are convolved with an empirical PSF, and realistic Poisson noise has been added commensurate with typical HSC-SSP PDR1 data. The simulated background is set to zero in all of these simulations. For each density quartet (low density on the left, high density on the right) the top left panel shows the full simulated field populated exclusively with exponential $n=1$ disk-like components. The top right panel shows the equivalent full simulated field populated exclusively with de Vaucouleurs type $n=4$ spheroid-like components. The bottom left and bottom right panels also show the $n=1$ and $n=4$ fields, respectively, yet only for the bright flux sub-population as defined in Section \ref{sec:galsimsetup} (i.e., bright mocks and intermediate brightness detected and used sources). The scale inset into the top left panel applies equivalently to all panels.

As can be seen, the absence of EBL in the lower panels marks them as visually distinct from their full simulated counterparts in the upper panels, with the background sky in the bright-only panels notably lower. Further, the extended $n=4$ \Sersic profiles appear to visibly contaminate a relatively larger fraction of the background sky than equivalent exponential type sources do. Both such factors are potentially significant contributors to inaccurate background sky estimation.

\begin{figure*}
    \centering
    \includegraphics[width=0.9\textwidth]{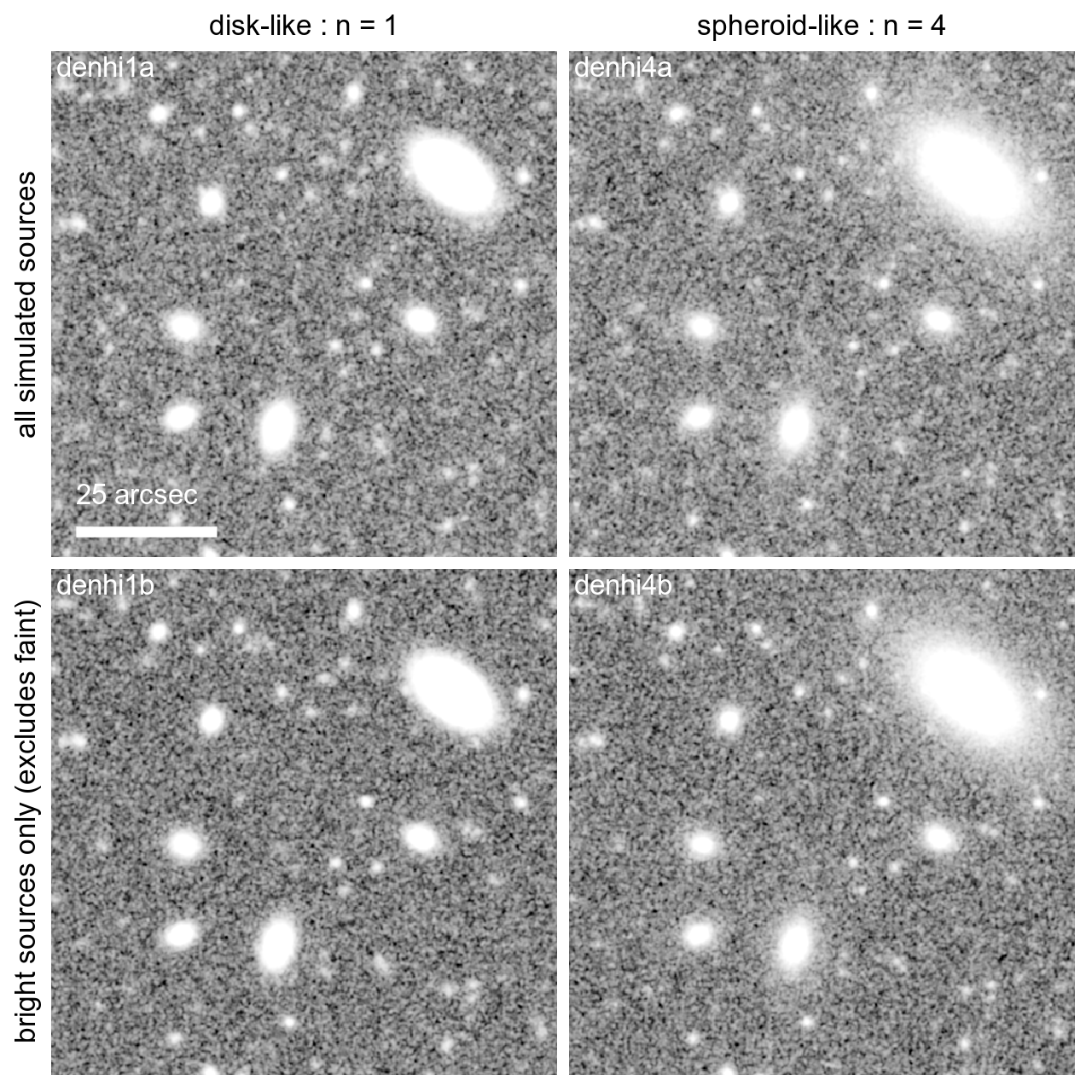}
    \caption{Low density simulated fields. As for the low density data shown in Figure \ref{fig:simstampfull}, but for a zoomed in region as represented by the inset scale. When zoomed in, the presence (top row) or otherwise (bottom row) of faint EBL sources and their impact on the apparent sky level is visibly evident.}
    \label{fig:simstampzoomlo}
\end{figure*}

Figure \ref{fig:simstampzoomlo} shows a zoomed in region from the low density data shown in Figure \ref{fig:simstampfull}. Here it becomes apparent that the brightest of the faint background sources present in the full simulated data appear as faint fuzzy patches scattered across the sky. Such sources are notably missing in the bright only simulations. Their form bears a striking similarity to many such objects recorded throughout the literature, notably: the `small condensed galaxy' in \citet{Arp1965}; the `dwarf galaxies' of \citet{Sandage1984}; the `compact galaxies' in \citet{Guzman1997}; the `little blue fuzzies' of \citet{Brough2011}; the `green peas' of \citet{Cardamone2009,Bauer2013}; the `little blue spheroids' defined in \citet{Kelvin2014a,Kelvin2014b}; `low surface brightness galaxies' in \citet{Williams2016}, and; the `ultra diffuse galaxies' of \citet{vanderBurg2016}, amongst others. Automated classification algorithms are now identifying and characterising such sources on an increasingly regular basis, as in, e.g., \citet{Sreejith2018,Turner2019}, with possible formation mechanisms outlined in \citet{Stinson2007}. As noted above, these systems may potentially significantly affect background sky estimation procedures.

\begin{figure*}
    \centering
    \includegraphics[width=0.9\textwidth]{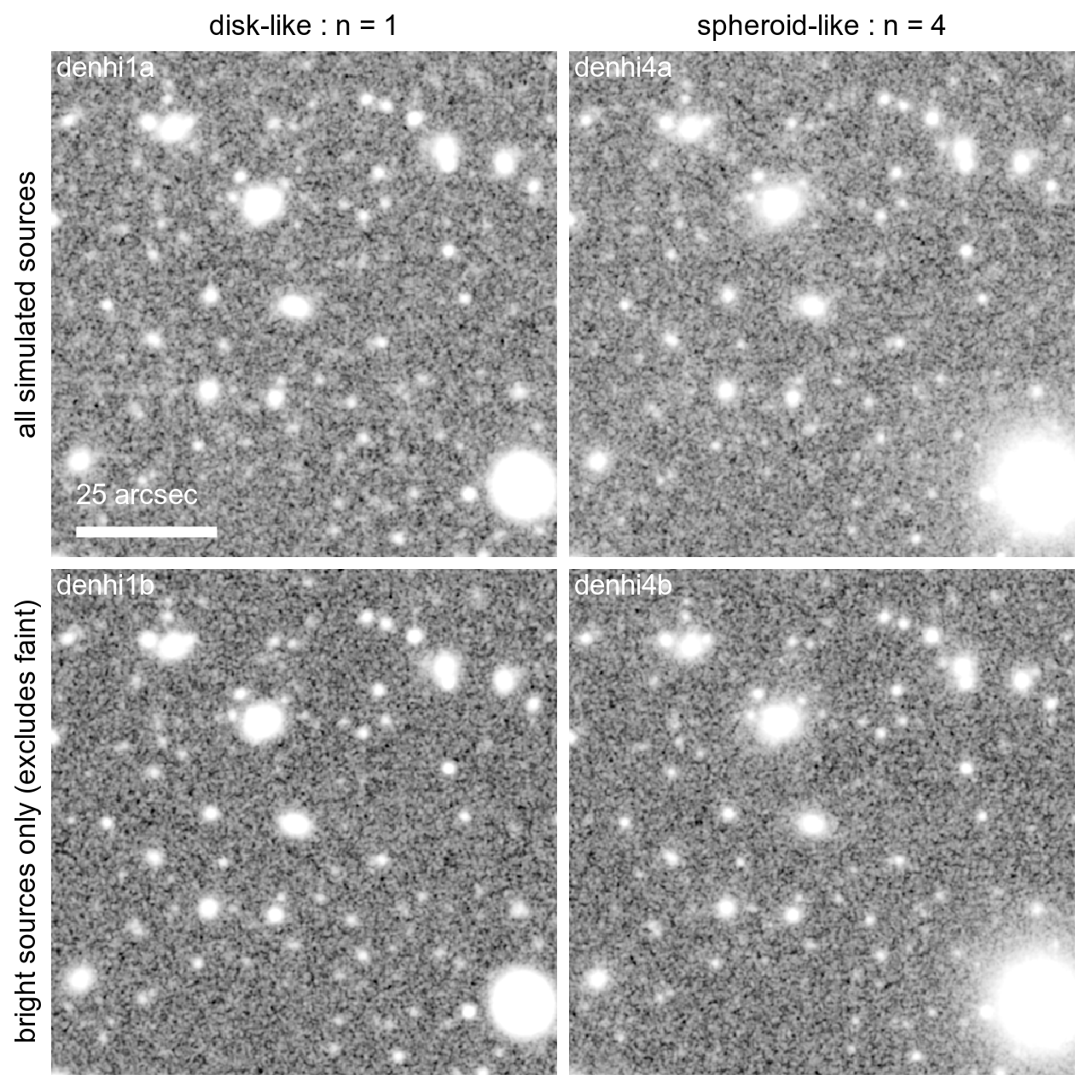}
    \caption{High density simulated fields. As for the high density data shown in Figure \ref{fig:simstampfull}, but for a zoomed in region as represented by the inset scale. As with the low density region, when zoomed in, the presence or otherwise of faint EBL sources and their impact on the apparent sky level is visibly evident.}
    \label{fig:simstampzoomhi}
\end{figure*}

Figure \ref{fig:simstampzoomhi} shows a zoomed in region from the high density data shown in Figure \ref{fig:simstampfull}. The zoom factor in this figure is identical to that used for Figure \ref{fig:simstampzoomlo}, i.e., each panel shown here shows the same solid angle on the sky as their equivalent panels in the low density regime. As noted above, the impact of faint sources is increasingly evident here. Not only do the numerous unresolved faint sources act to increase the global sky pedestal, but they also act to further enhance the wings of intermediate brightness galaxies. Such densely populated regions are becoming increasingly common in an era of low surface brightness astronomy.

The eight simulated fields summarised in Table \ref{tab:simtab} and shown in Figure \ref{fig:simstampfull} constitute our simulated dataset. These simulated images are background subtracted to zero counts (i.e., Poisson noise oscillates about a level of zero counts) and populated with a range of extended PSF-convolved \Sersic profile objects. These simulated data are used for all subsequent analyses throughout the remainder of this study.

\section{Threshold Levels}
\label{sec:thresholds}

To facilitate a fair comparison of detected source catalogues generated from each of our $57$ input sky-subtracted fields, we require a uniform absolute detection threshold to use as an input to \SExtractor. The default operation of \SExtractor identifies sources as contiguous regions of flux lying above some relative threshold level, normally $1.5\sigma$ above the locally determined sky. As sky estimation varies across fields, the resultant threshold level will also vary, therefore precluding our use of the default \SExtractor configuration in this case.

\begin{figure}
    \centering
    \includegraphics[width=\columnwidth]{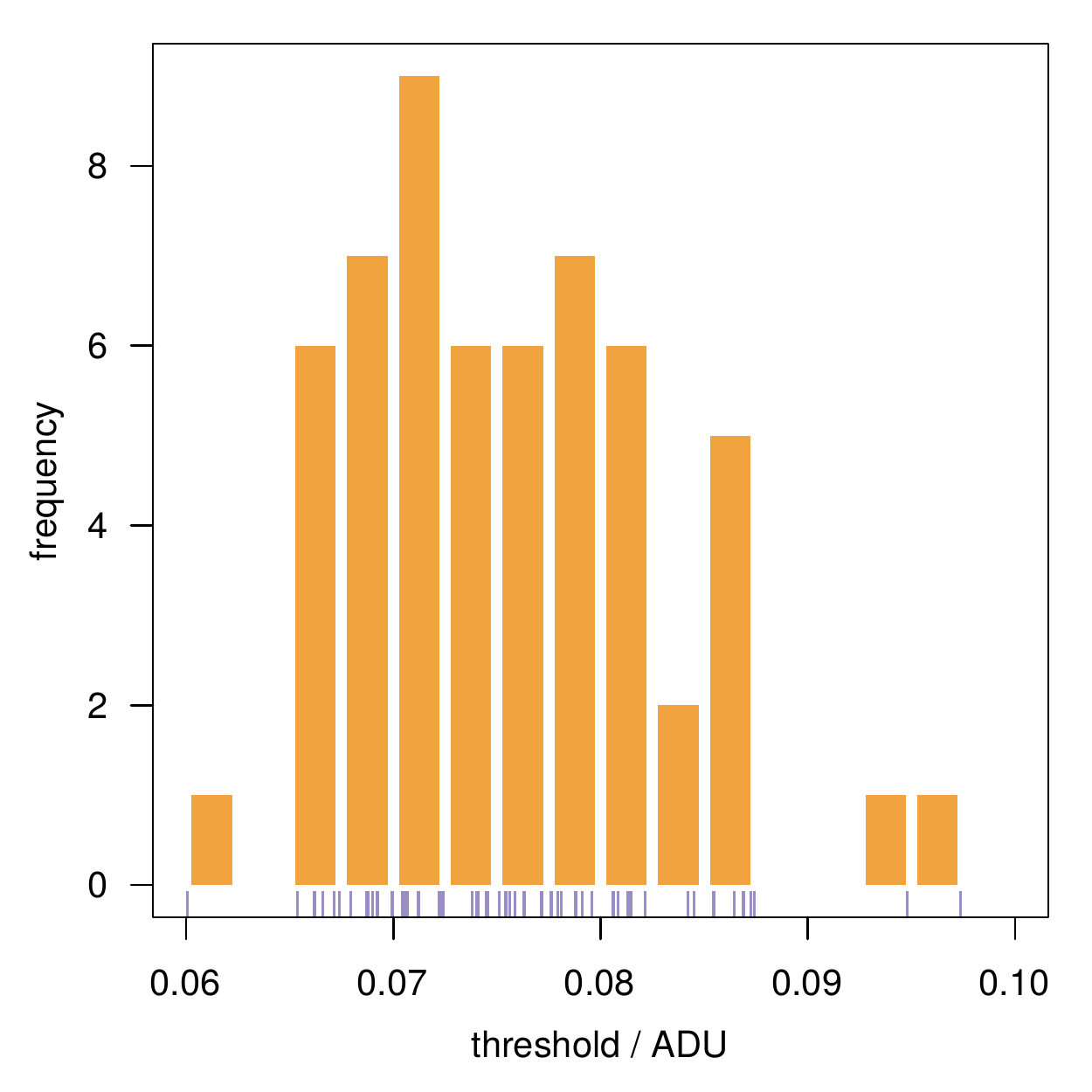}
    \caption{Histogram showing relative $1.5\sigma$ detection threshold levels as output by \SExtractor when operated in default mode upon each of our $57$ input fields. Individual field values are shown in the underlying rug plot. We adopt a threshold level of $0.07$ counts as our global absolute detection threshold level, corresponding to the approximate position of the peak in this figure.}
    \label{fig:threshlvls}
\end{figure}

To estimate a global absolute threshold level, we initially ran \SExtractor in default mode on each of the $57$ fields. The resultant $1.5\sigma$ threshold levels in counts are shown in the histogram in Figure \ref{fig:threshlvls}. Individual field values are shown in the underlying rug plot. As can be seen, our threshold levels vary from $\about0.06$ counts to $\about0.1$ counts, peaking at $\about0.07$ counts. We choose this peak value of $0.07$ counts as our global absolute threshold level with which to subsequently generate our source detection catalogues with the \SExtractor software package.

\section{\GalSim Input Catalogues}
\label{sec:inputcats}

Table \ref{tab:simcat} shows an excerpt from the input catalogue produced for the exponential ($n=1$) simulation of the low density region. Here we show the five brightest sources out of a total of $428257$ objects in this density regime. Columns, from left to right, are: $x$ pixel position in the field; $y$ pixel position in the field; total source flux; half light radius $R_e$; axis ratio ($q=b/a$); position angle; \Sersic index $n$; and, postage stamp size. Associated units (where appropriate) are shown beneath column headers in parentheses. Parameter values are rounded to a varying number of decimal places, as shown.

\begin{table}
    \setlength{\tabcolsep}{2.5pt}
    \begin{tabular}{ r | r | r | r | r | r | r | r }
    \multicolumn{1}{c}{$x$} & \multicolumn{1}{c}{$y$} & \multicolumn{1}{c}{flux} & \multicolumn{1}{c}{$R_e$} & \multicolumn{1}{c}{$q$} & \multicolumn{1}{c}{P.A.} & \multicolumn{1}{c}{$n$} & \multicolumn{1}{c}{stamp size}\\
    \multicolumn{1}{c}{(pix)} & \multicolumn{1}{c}{(pix)} & \multicolumn{1}{c}{(counts)} & \multicolumn{1}{c}{(pix)} & \multicolumn{1}{c}{ } & \multicolumn{1}{c}{(deg.)} & \multicolumn{1}{c}{ } & \multicolumn{1}{c}{(pix)}\\
    \hline
    527.0 & 2786.5 & 11480.40237 & 11.202 & 0.56 & -38.8 & 1 & 263 \\
    387.3 & 454.0 & 7569.54847 & 9.987 & 0.55 & -40.8 & 1 & 235 \\
    2460.5 & 1071.5 & 5751.59779 & 11.058 & 0.82 & -1.70 & 1 & 205 \\
    1471.4 & 1174.9 & 4328.31835 & 7.920 & 0.83 & 39.7 & 1 & 151 \\
    4049.5 & 1094.9 & 3874.05190 & 8.682 & 0.65 & -52.0 & 1 & 183
    \end{tabular}
    \caption{An excerpt from the \GalSim input catalogue relating to our low density regime. This catalogue, and others of a similar format, are used by \GalSim to generate simulated imaging. Shown here are parameters relating to the five brightest simulated sources in the low density region from a total of $428257$ objects. Columns, from left to right, are: $x$ pixel position in the field; $y$ pixel position in the field; total source flux; half light radius $R_e$; axis ratio ($q=b/a$); position angle; \Sersic index $n$; and, postage stamp size. Associated units (where appropriate) are shown beneath column headers in parentheses. An input catalogue of this format is generated for each simulated region.}
    \label{tab:simcat}
\end{table}

\section{\GalSim \bf\texttt{YAML} Configuration Files}
\label{sec:yamlconfig}

\begin{figure}
    \centering
    \includegraphics[width=\columnwidth]{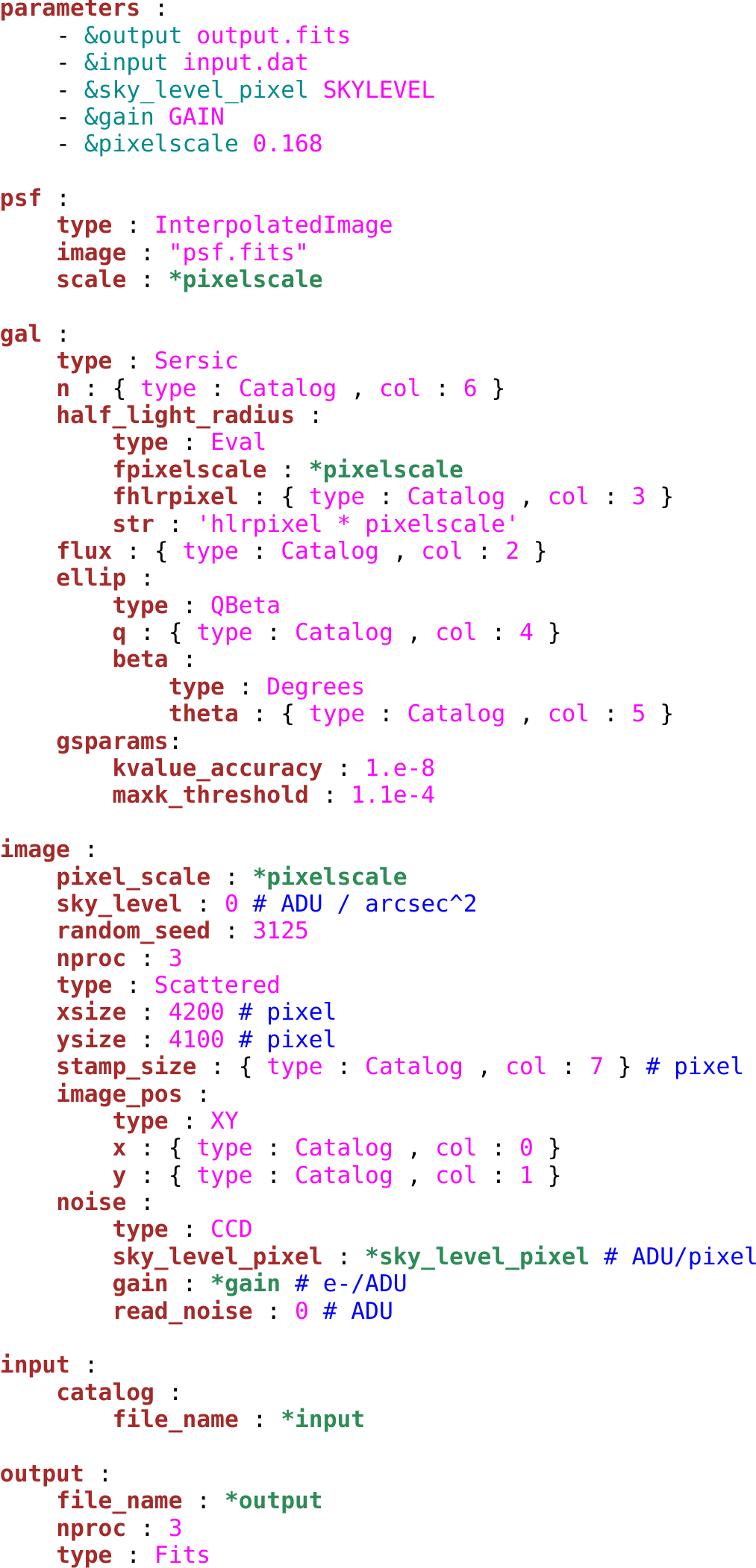}
    \caption{An example \texttt{YAML} configuration file used as an input to \textsc{GalSim}. This configuration file produces a simulated \texttt{output.fits} file based on the \texttt{input.dat} input catalogue. PSF generation is outlined in Section \ref{sec:psf}, whilst estimates of the original sky level pedestal (\texttt{SKYLEVEL}) and averaged detector equivalent gain (\texttt{GAIN}) are detailed in Section \ref{sec:rmsskylevel}. See the text for further information.}
    \label{fig:yamlconfig}
\end{figure}

Figure \ref{fig:yamlconfig} shows an example \texttt{YAML} configuration file used as an input to \textsc{GalSim} (version 2.2.6). The file is split into sections as indicated, namely, from top to bottom: global parameters; PSF information; simulated source profile setup; global image definitions; input file name; and, output processing configuration. PSF estimation and FITS file generation are detailed in Section \ref{sec:psf}. Note that the background sky level is set to zero counts in all output simulated images. Derivation of the global parameters \texttt{SKYLEVEL} and \texttt{GAIN} (an equivalent gain) are discussed in Section \ref{sec:rmsskylevel}. Following initial testing, we opt to modify \texttt{kvalue\_accuracy} and \texttt{maxk\_threshold} from their default values to those shown here in order to minimise limiting artefacts which would otherwise be apparent in our output data. For complete definitions of these and all remaining parameters not discussed here, we refer the reader to \citet{Rowe2015}.

\section{Postage Stamp Sizes}
\label{sec:postagestampsizes}

We use \SExtractor to fit a single \Sersic function to each detected source as given by a prior run of either \SExtractor, \Gnuastro, or the \LSSTPs. As an output, \SExtractor provides a catalogue containing \Sersic component fitting information, and (optionally) an associated check image containing all successfully fitted \Sersic models in 2D space. Initially, we used the \SExtractor check images as the basis for our modelled masks (see Section \ref{sec:modelledmasking}), however, on closer inspection it became clear that these images were inappropriate for our purposes.

\begin{figure*}
    \centering
    \includegraphics[width=0.9\textwidth]{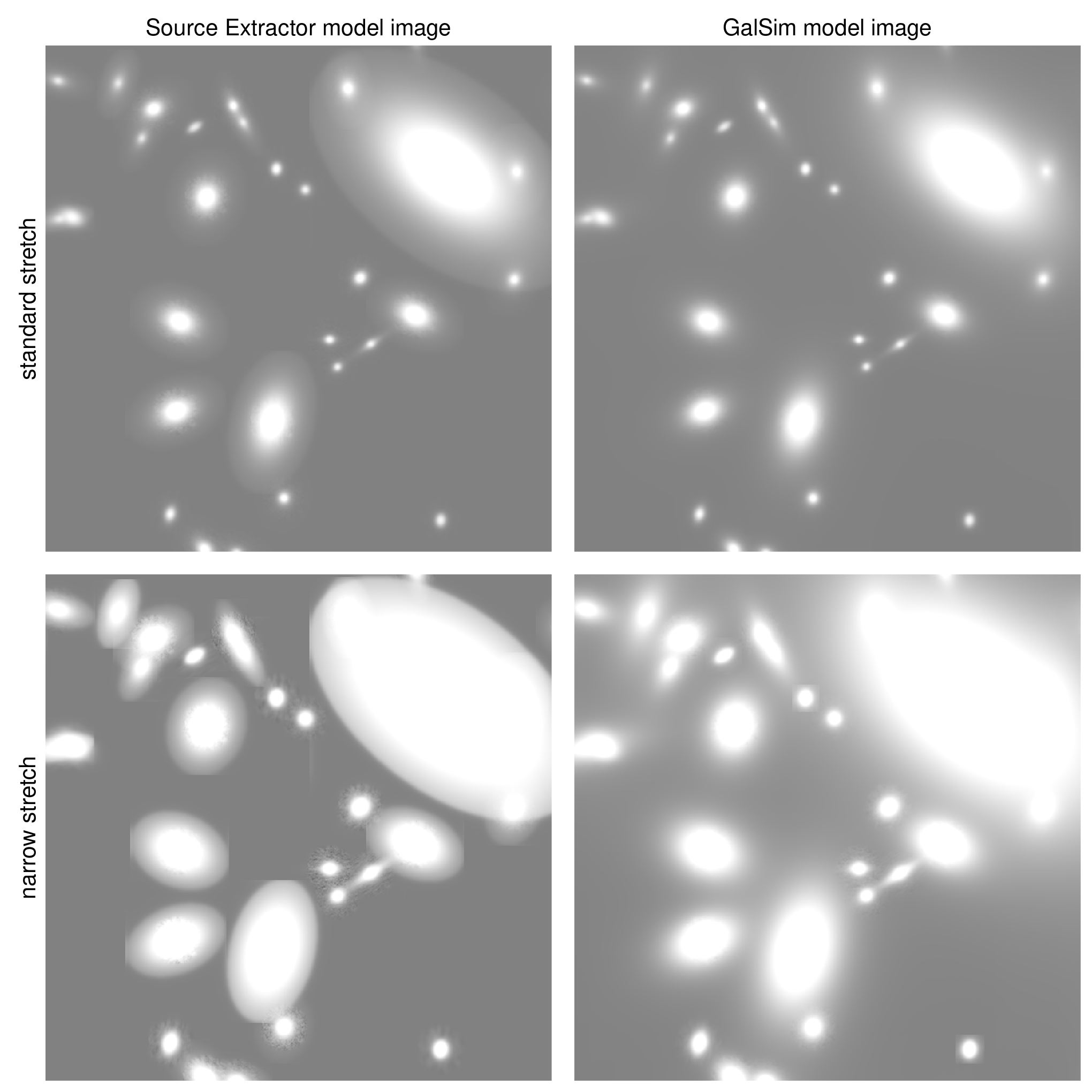}
    \caption{A comparison of initial \Sersic model imaging produced by \SExtractor (left panels) to our final utilised refined \GalSim \Sersic modelled imaging (right panels). Model \Sersic parameters are derived from \SExtractor. The upper two panels show model imaging using a standard arctan $z$-scale stretch, similar to that found elsewhere in this paper. The lower two panels are arctan scaled over a narrower $z$-scale stretch, highlighting LSB flux around the outskirts of sources. Whilst \SExtractor seemingly produces robust \Sersic catalogue data, it does not allow the user to control the postage stamp size out to which \Sersic models are propagated in an accompanying check image. As a result, various modelled source edge effects are visible (see left panels) in output \SExtractor check imaging owing to the fact that the default \SExtractor postage stamp size is too small for the purposes of this study. We use \GalSim to define bespoke postage stamps for each simulated source (see right panels), minimising this effect.}
    \label{fig:modmaskcomp_sex}
\end{figure*}

Figure \ref{fig:modmaskcomp_sex} shows a comparison of \SExtractor model imaging (left panels) to \GalSim model imaging (right panels) using an initial \SExtractor run as the basis for determining which sources should be modelled. The top two panels show these 2D data using a standard arctan stretch, similar to figures found elsewhere in this paper. The bottom two panels show these same data using a narrower stretch, designed to highlight some of the fainter features. As is clear in the lower-left panel, \SExtractor model imaging contains significant artefacts around the edges of \Sersic model sources. The default and immutable \SExtractor postage stamp size is too small, resulting in a visible hard boundary at the edges of most sources. The extent of profile truncation becomes more severe as the source becomes more extended. In addition, notable edge boundary flux-wrapping artefacts are introduced for the brightest of sources, adding a peak of flux to regions where no sources were previously placed. Switching to \GalSim allows us to control the postage stamp box size used for each simulated source, resulting in a smoother and more accurate 2D modelled plane (see lower-right panel in both figures). We opt to define postage stamp boxes for each source such that a source aligned parallel to the $x$-axis would reach a surface brightness limit of $\mu_r=35\,\mathrm{mag}/\mathrm{arcsec}^{-2}$ at the boundary of the postage stamp. This limit was chosen to be sufficiently faint to facilitate exploration of the background sky, yet not too deep to become overly computationally expensive. A minimum postage stamp size of $11\times11$ pixels is enforced at all times.

\label{lastpage}
\end{document}